\documentclass[twocolumn]{aastex701}

\usepackage{mathtools,amsmath}
\usepackage{array}
\usepackage{xcolor}

\newcommand{\rdmp}{\textsc{redMaPPer}}
\newcommand{\metacal}{\textsc{metacalibration}}
\newcommand{\cosmodc}{\texttt{cosmoDC2}}
\newcommand{\skysim}{\texttt{SkySim5000}}

\newcommand{\clmm}{\texttt{CLMM}}

\graphicspath{{./}}

\begin{document}

\title{Constraining Galaxy Cluster Triaxiality via Weak Lensing --- I. Preparation for the Rubin Data Beyond Leading Order}

\author[orcid=0000-0001-5422-1958]{Shenming Fu}
\affiliation{NSF-DOE Vera C. Rubin Observatory / SLAC National Accelerator Laboratory, 2575 Sand Hill Road, Menlo Park, CA 94025, USA}
\affiliation{Kavli Institute for Particle Astrophysics and Cosmology, Stanford University, Stanford, CA 94305, USA}
\email[]{shenming.fu.astro@gmail.com}

\author[orcid=0000-0002-4508-4581]{Radhakrishnan Srinivasan}
\affiliation{Department of Physics and Astronomy, Stony Brook University, Stony Brook, NY 11794, USA}
\email[]{rk29121996@gmail.com}  

\author[orcid=0000-0002-6389-5409]{Tae-hyeon Shin}
\affiliation{Carnegie Mellon University, 5000 Forbes Ave, Pittsburgh, PA 15213, USA}
\email[]{taehyeos@andrew.cmu.edu}

\author[orcid=0000-0003-0693-2469]{Rance Solomon}
\affiliation{Faculty of Physics, Ludwig-Maximilians-Universit\:at, Scheinerstr. 1, 81679 Munich, Germany}
\email[]{rance.solomon@gmail.com}

\author[orcid=0000-0002-5284-1237]{Deric Jones}
\affiliation{University of Michigan, 500 S State St, Ann Arbor, MI 48109, USA}
\email[]{dericj@umich.edu}

\author[orcid=0000-0001-8868-0810]{Camille Avestruz}
\affiliation{University of Michigan, 500 S State St, Ann Arbor, MI 48109, USA}
\email[]{cavestru@umich.edu}

\author[orcid=0000-0001-5969-4631]{Yuanyuan Zhang}
\affiliation{NSF's National Optical-Infrared Astronomy Research Laboratory, 950 N. Cherry Ave., Tucson, AZ 85719, USA}
\email[]{yuanyuanzhang.astro@gmail.com}

\author[orcid=0000-0001-5679-6747]{Michel Aguena}
\affiliation{Istituto Nazionale di Astrofisica, Viale del Parco Mellini No.84 00136 ROMA, Italy}
\email[]{aguena@inaf.it}

\author[orcid=0000-0001-6487-1866]{C\'eline Combet}
\affiliation{Univ. Grenoble Alpes, CNRS, LPSC-IN2P3, 53, Avenue des Martyrs, 38000, Grenoble, France}
\email[]{celine.combet@lpsc.in2p3.fr}

\author[orcid=0000-0003-2314-5336]{Anthony Englert}
\affiliation{Department of Physics, Brown University, Providence, RI 021912, USA}
\email[]{anthony_englert@brown.edu}

\author[orcid=0000-0001-8000-1959]{Benjamin Levine}
\affiliation{Department of Physics and Astronomy, Stony Brook University, Stony Brook, NY 11794, USA}
\email[]{ben@levlin.com}

\author[orcid=0000-0002-8676-1622]{Alex I. Malz}
\affiliation{Space Telescope Science Institute, 3700 San Martin Drive, Baltimore, MD 21218, USA}
\email[]{aimalz@nyu.edu}

\author[orcid=0000-0002-1818-929X]{Constantin Payerne}
\affiliation{Univ. Paris-Saclay, CEA, IRFU, 91191, Gif-sur-Yvette, France}
\email[]{constantin.payerne@gmail.com}

\author[orcid=0000-0002-3645-9652]{Marina Ricci}
\affiliation{Université Paris Cité, CNRS, Astroparticule et Cosmologie, F-75013 Paris, France}
\email[]{ricci@apc.in2p3.fr}

\author[orcid=0000-0002-3881-7724]{Anja von der Linden}
\affiliation{Department of Physics and Astronomy, Stony Brook University, Stony Brook, NY 11794, USA}
\email[]{anja.vonderlinden@stonybrook.edu}

\author{the LSST Dark Energy Science Collaboration}
\email[]{}
\nocollaboration{all}

\correspondingauthor{Radhakrishnan Srinivasan}
\email[show]{rk29121996@gmail.com}

\begin{abstract}

The 3D mass distributions of galaxy clusters are generally triaxial, a geometry that is difficult to constrain from projected observations. In this work, we measure the projected halo shapes of clusters from weak lensing signatures using the Cluster Lensing Mass Modeling software, a tool developed by the Dark Energy Science Collaboration to analyze data from NSF-DOE Rubin Observatory's Legacy Survey of Space and Time (LSST). We measure ensemble halo ellipticity on the plane of the sky via axis-aligned stacking and multipole expansion of weak lensing signatures. We study a precursor dataset --- the \textsc{redMaPPer} cluster catalog, the \textsc{metacalibration} shape catalog, and the Directional Neighborhood Fitting photometric redshift catalog from the Dark Energy Survey Year 3 public data release. We select clusters that have a high centering probability ($>90\,\%$) of the identified central galaxy, and use the satellite galaxy distribution to determine the major-axis orientation for stacking. We extend the analysis to the second order of ellipticity in monopole and quadrupole measurements. The projected ellipticity of the cluster sample is found to be $0.307^{+0.013}_{-0.014}$ (axis ratio $0.530^{+0.016}_{-0.015}$) with extra percent-level systematics introduced mainly by miscentering. The projected cluster ellipticity shows no statistically significant dependence on redshift or richness (mass proxy). We further verify the accuracy of the cluster shape measurement using mock catalogs. This analysis is applicable to datasets from wide-area cosmic surveys such as LSST, Euclid, and the Roman Space Telescope, where larger sample sizes will lead to tighter constraints on cluster shapes.

\end{abstract}



\let\oldmaketitle\maketitle
\renewcommand{\maketitle}{\oldmaketitle\setcounter{footnote}{0}}

\section{Introduction}

Galaxy clusters are the largest gravitationally collapsed structures in the universe, consisting mostly of dark matter ($\sim80\,\%$). 
Clusters are located at the intersections of cosmic filaments in the large-scale structure.
In the bottom-up hierarchical model of cosmology, clusters grow via merging with nearby groups and clusters~\citep{Borgani2011}.

The continuous infall of neighboring structures and the torque generated by the large-scale gravitational field lead to triaxial shapes of clusters rather than spherical.  
In addition, the orientations of components of a cluster (e.g., the central galaxy, distribution of member galaxies, gas, and dark matter) are usually broadly aligned, and the general orientation of the cluster aligns with the neighboring major cosmic filaments, as a result of the long-term evolution under gravity~\citep{Donahue2016,Herbonnet2019,Herbonnet2022,Zhou2023,Fu2024b,Lokken2025}. 

Clusters provide a key probe of cosmology through the halo mass function: the mass number density as a function of the redshift, which is sensitive to $\Omega_{\rm m}$ and $\sigma_8$ at the high-mass end~\citep{Allen2011,DES2025}.  
Accurate mass measurement is the cornerstone of this probe. The mass of a cluster can be inferred from observables such as galaxy number density, X-ray emissions, and the Sunyaev–Zeldovich (SZ) effect through the mass-observable relations~\citep[MOR, e.g.,][]{Bocquet2019,McClintock2019,Chiu2025}. The statistical relation between the mass and an observable can be calibrated by weak lensing (WL), which provides direct measurements of cluster masses~\citep{McClintock2019}. Constraining a statistical relation is particularly useful for low-mass clusters, where the WL signal of a single cluster is noisy, but it becomes statistically significant when averaged over an ensemble of clusters~\citep[especially when the cluster counts are included in the analysis;][]{Lesci2022}. 

The triaxial shapes of clusters can cause a scatter of $\sim20\,\%$ in the WL mass measurement of single clusters, if assuming a spherical density model \citep{beckerkravtsov2011}. 
A common method to reduce the scatter is stacking a large sample of clusters without aligning their axes, but a percent-level bias can still exist after stacking \citep{McClintock2019}. 
Recently, several studies pointed out that triaxiality (with associated filaments) can cause detection/selection bias due to projection effects. When a cluster is oriented along (or perpendicular to) the line of sight (LoS), more (fewer) galaxies are considered as cluster members, and the cluster is easier (harder) to be detected/selected~\citep[][]{Osato2018,Wu2022,Zhang2023,srinivasanBCGellipticity}; a similar case happens, to lesser extent, when clusters are detected through their gas content~\citep{Herbonnet2019,Kafer2019,Saxena2025}. 
Accurately constraining cluster halo shapes (e.g., as a function of cluster properties) is a crucial step towards understanding the WL mass measurement scatter and biases and is thus important for cluster cosmology analysis. 
In addition, measurement of cluster ellipticities as a function of redshift can be used to probe cosmology directly~\citep{kasunevrard2005,Ho2006}.

The shapes of galaxy clusters are also useful for understanding small-scale astrophysical mechanisms. 
For example, simulations show that self-interacting dark matter produces rounder halos, and therefore the shape distribution of relaxed clusters can be used to constrain the self-interaction cross-section of dark matter~\citep{Robertson2019,McDaniel2021,Gonzalez2024}. 
By comparing the distributions of gas, galaxies, and dark matter, one can also study the coevolution of the cluster components with the environment~\citep{Joachimi2015}. 

Observational studies have been made on cluster shapes. 
Noting that observation only sees a 2D projection of the cluster's 3D shape, and that there are multiple definitions of the 2D ellipticity, we convert their ellipticity measurements into axis ratios instead to give an overview. For instance, using data from the Sloan Digital Sky Survey~\citep[SDSS, $i\sim22$ at $5\,\sigma$;][]{Gunn1998}\footnote{\url{https://www.sdss4.org/dr17/imaging/other_info/}. We note that the depth does not change much between phases.} with WL quadrupole measurements,~\citet{Shin2018} found that the projected axis ratio of the ensemble is $\sim0.57$. 
\citet{Oguri2010a,Umetsu2018,Okabe2025} showed that small samples of massive clusters exhibit axis ratios of $\sim0.54-0.67$ using WL.
Mimicking observational WL analysis, \citet{Payerne2023} found a mean axis ratio of $\sim0.59$ for massive clusters in \textsc{The Three Hundred} cosmological hydrodynamical simulation~\citep{Cui2018}.
For comparison, the halo axis ratio of the less massive, luminous red galaxies (LRGs) in the same SDSS-WL dataset, is reported to be $\sim0.78$~\citep{Clampitt2016}.   
Recently,~\citet{Robison2023} showed that the halos of SDSS LRGs have an effective axis ratio of $\sim0.61$. 
In addition, it is worth mentioning that when the X-ray and SZ observations are available, the 3D cluster shape can be directly constrained via combining the lensing signal and the gas information~\citep{Limousin2013,Sereno2018,Gavidia2026}.

The Dark Energy Survey (DES) Y3 \rdmp{} catalog contains the optical detection of clusters across $\sim5000~\deg^2$ sky with $\sim2$ mag deeper observations than SDSS~\citep[$i\sim24$;][]{SevillaNoarbe2021}, which is the largest deep optical cluster catalog to date, but no analysis of the ensemble shape through WL has been performed on those clusters. 
The upcoming NSF-DOE Vera C. Rubin Observatory Legacy Survey of Space and Time (LSST) will have an even larger footprint covering the $\sim18000~\deg^2$ of the southern sky with even deeper observations than DES~\citep[$i\sim27$;][]{Ivezic2019}.  
The LSST Dark Energy Science Collaboration (DESC) has developed various tools to prepare for the analysis of LSST data, including the Cluster Lensing Mass Modeling software~\citep[\clmm;][]{Aguena2021} designed for the WL mass measurement of clusters. 
\clmm{} is able to efficiently and consistently analyze cluster WL data in a large volume. This DESC tool can be applied to any WL cluster dataset, and recently a triaxiality module has been added to perform analysis of cluster shapes~\citep[generally following the procedure of][see also~\clmm v2, The CLMM Team et al., under review]{Shin2018}. 

In this work, we apply \clmm{} to public datasets from DES: the Y3 \rdmp{} cluster catalog, \metacal{} shapes, and Directional Neighborhood Fitting (DNF) photometric redshifts (photo-zs), to perform the first ensemble projected halo ellipticity measurement on the DES Y3 data, with an initial investigation of measurement systematics (more details will be presented in a follow-up paper of this series). Additionally, this work helps demonstrate the performance of \clmm{} on real data. This paper also paves the way for future cluster triaxiality analyses with LSST. 
In summary, the goals of this paper are as follows. 
\begin{itemize}
    \item Measure for the first time the effective projected ellipticity of a large ensemble of clusters from the DES weak lensing data. 
    \item Test the dependence of ellipticity on cluster mass and redshift. 
    \item Perform an initial investigation of systematics contributing to weak lensing triaxiality measurements. 
\end{itemize}

This paper is organized as follows. Section~\ref{sec:data} briefly introduces the datasets used in this work. 
Section~\ref{sec:theory_method} presents the theoretical background and our method. 
We show our results in Section~\ref{sec:results} and discuss our analysis in Section~\ref{sec:discussion}. Finally, we draw conclusions and summarize our work in Section~\ref{sec:conclusion}. 
Throughout the paper, we assume a flat $\Lambda$CDM cosmology with $\Omega_{\rm m}=0.3$ and $H_0=70$~km~s$^{-1}$~Mpc$^{-1}$.

\section{Data}
\label{sec:data}

\subsection{Cluster catalog} \label{sec:cluster_catalog} 
This work uses a sample of galaxy clusters from the public DES Y3 \rdmp{} catalog~\citep{DES2025}\footnote{\url{https://des.ncsa.illinois.edu/releases/y3a2/Y3key-cluster}}, which is based on the galaxy catalog derived from the first three years of DES observations. The \rdmp{} cluster detection technique makes use of the distinctive concentration of bright red-sequence galaxies in  clusters. More details of the \rdmp{} algorithm are presented by~\citet{Rykoff2014}.

The catalog provides five central galaxy (CG) candidates for each galaxy cluster, along with the candidates' probability of being the true CG (\texttt{pcen}). 
We use the highest \texttt{pcen} value among CG candidates ($\texttt{pcen0}\equiv\max\{\texttt{pcen}\}$) in several ways. First, we adopt the most probable candidate as the CG of the cluster, whose position is used as the cluster coordinates in~\rdmp{}~\citep{Rykoff2016}. Second, we use $\texttt{pcen0}$ as a proxy for centering quality and relaxation~\citep{Fu2024a}. 
Specifically, we select the clusters where $\texttt{pcen0}>0.9$ ($55\,\%$ among all clusters). 
A low \texttt{pcen0} suggests that there are multiple similar CG candidates in the cluster, and the determination of true CG is difficult --- this may be caused by perturbed systems, and the CG candidate with low \texttt{pcen0} value
may not represent the center of the galaxy cluster very well, which leads to miscentering issues.
Thus, a high \texttt{pcen0} indicates that the cluster is less likely to be a merging system and more likely to be relaxed.
Note, this is an inference. Spectroscopic redshift (spec-z) measurements can improve the determination of the dynamical state, but currently there is no spec-z survey with sufficient coverage and completeness for cluster members. Future large targeted surveys may provide a better solution.

Following~\citet{McClintock2019,DES2025}, we also select clusters within $0.2<z<0.65$ ($85\,\%$ of the whole sample). The lower limit is to ensure the cluster's photo-z and richness quality~\citep{Rykoff2016}, while the upper end is to ensure the cluster completeness~\citep{McClintock2019}.  
Additionally, we select clusters for which the \metacal{} catalog covers the centers.
In the end, we have selected 9076 (out of 21092) clusters in total, which span richness $20<\lambda<222$ (we symbolize the upper limit with $\infty$ in the following analysis for simplicity). We define this as the \textit{baseline} sample. 

To study how the cluster ellipticity changes with redshift and mass, we divide the clusters into two bins of redshift and richness (mass proxy), respectively, where both divisions are done one at a time. We describe the details of the divisions in Section~\ref{sec:cluster_sample_division}. 

\subsection{Galaxy shape catalog}

For the weak lensing measurements, we use the public DES Y3 shape catalog~\citep{Gatti2021}\footnote{\url{https://des.ncsa.illinois.edu/releases/y3a2/Y3key-catalogs}}. 
This catalog is based on the first three years of DES observations and covers the entire DES footprint.  
The catalog includes shear estimates per object derived with the \metacal{} algorithm~\citep{Huff2017,Sheldon2017}. In brief, \metacal{} performs small linear transformations on the galaxy image (i.e., adds small shear to the image) to derive the shear response --- a ``matrix'' between the measured galaxy shape and the reduced shear. We use the source galaxies selected by this catalog.

\subsection{Photometric redshift catalog}

To distinguish background galaxies from foreground ones for lensing analysis, we use the DES Y3 Directional Neighbourhood Fitting (DNF) photometric redshifts~\citep[photo-z;][]{DeVicente2016}. DNF uses a galaxy's brightness and color to determine its photo-z based on a spectroscopic redshift training set.  
DNF provides two photo-z measurements: \texttt{ZMC\_SOF} (nearest neighbor redshift), which is useful for estimating redshift distributions of ensembles of galaxies, while \texttt{ZMEAN\_SOF} gives the best point redshift estimate.

\subsection{Toy model: mock catalogs} 
\label{sec:mock}

To validate our measurement method, we feed a mock shear catalog, which mimics the lensing signal of a halo with known shape (``truth''), into our halo ellipticity fitting pipeline and compare the output shape with the input. 
The mock catalog is based on a toy model --- it simulates exactly the lensing effects of an elliptical Navarro–Frenk–White~\citep[NFW;][]{Navarro1997} halo. 
We consider a halo that has mass $M_{\rm 200c}=2\times10^{14}M_\odot$  and redshift $z_l=0.4$, which corresponds to a ``typical'' cluster (richness $\lambda\sim32$) for the DES Y3 \rdmp{} sample with the centering probability $>90\,\%$~\citep{DES2025}. The choice of these typical parameters is based on our redshift and richness sample divisions, which are described in Section~\ref{sec:cluster_sample_division}. The $200$c means that the mean overdensity of the enclosed mass is 200 times the critical density of the universe at that redshift.  
We consider common values for the cluster halo concentration $c_{\rm200c}=2,4,6$.  
The source redshift is assumed to be on a plane and chosen to be $z_s=0.8$, which approximates the median of the LSST ``gold'' sample~\citep{LSSTSC2009} and provides a representative high lensing efficiency for the halo. 
The lensing effects are produced by an elliptical NFW surface density/convergence~\citep{Wright2000} in Eq.~\ref{eq:kappa_elp} with an axis ratio of $q=b/a$~\citep[][]{Schramm1990,Keeton2001,Oguri2010a,Oguri2010b}, where $(\theta_1,\theta_2)$ are angular coordinates on the plane of the sky (PoS). We describe the details in Appendix~\ref{app:lensing_elliptical_halo}. 
\begin{equation}
    \kappa (\boldsymbol{\theta}) = \kappa_{\rm NFW} (\sqrt{q\theta_1^2+\theta_2^2/q})
    \label{eq:kappa_elp}
\end{equation}

We compute the lensing shear on a grid without including shape noise or measurement error ($100\times100$ pix over $1000\times1000$ arcsec, $5.4\times5.4$ Mpc). This gives a mock source galaxy density of 36 per arcmin$^{-2}$, which is close to the expected LSST WL sources in the ``gold'' sample~\citep{LSSTSC2009}. Note that the overall variations in observation depth (i.e., differences in grid resolution for the mock) and the number of stacked clusters will only change the uncertainty of the fitting result.  We do not expect these variations to lead to additional biases, and therefore any trends identified in our subsequent analysis should be robust to other choices in depth and number of clusters. 
In addition, we do not consider the contribution of lensing signal from the Large-Scale Structure (LSS) as it is much weaker than that of the 1-halo term~\citep[typically by one order of magnitude;][]{McClintock2019}, and we constrain the analysis within small cluster-centric radii ($<2$ Mpc; Section~\ref{sec:theory_method}). We will further investigate the contributions of LSS/LoS structures using cosmological simulations in future work. 
We generate catalogs corresponding to the lens axis ratios from 0.1 to 1.0 with a step of 0.1 to test the performance and limit of our cluster shape measurement algorithm in Section~\ref{sec:dis_accuracy_taylor}, and we use the mock to test systematic effects in Section~\ref{sec:dis_multi_sys},~\ref{sec:dis_miscenter}, and~\ref{sec:dis_lss}. 
A similar mock has also been used in the \clmm v2 paper to demonstrate the software performance. 

\section{Theory \& Method}
\label{sec:theory_method}

\begin{figure*}[htb!]
    \plotone{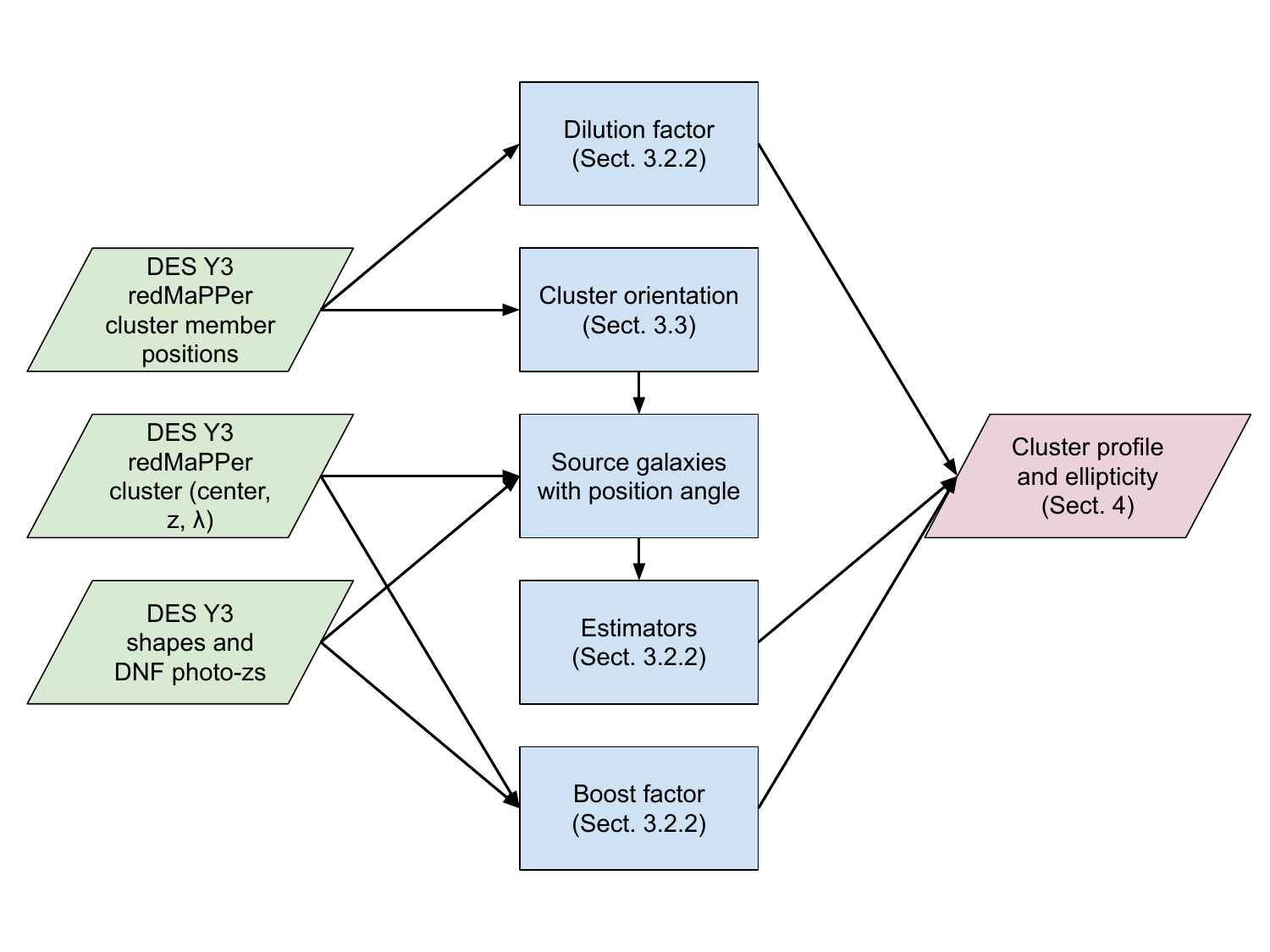}
    \caption{Flowchart of the analysis pipeline. The green parallelograms, blue rectangles, and pink parallelograms show the data, processing steps, and results, respectively. }
    \label{fig:pipeline}
\end{figure*}

We study the cluster triaxiality by measuring the projected cluster ellipticity using WL. 
We use a multipole expansion to efficiently model the lensing effects of an elliptical halo. 
We determine the cluster orientation projected on the sky using the distribution of member galaxies and then stack an ensemble of clusters with their axes aligned to reduce noise.  
Our analysis pipeline is summarized in Figure~\ref{fig:pipeline}. 

\subsection{Cluster weak lensing effects}
\subsubsection{General mass distribution}
A foreground cluster deflects photons from background galaxies via gravitational lensing. 
The convergence ($\kappa$) and the shear ($\boldsymbol{\gamma}$) describe the magnification and the stretch (with two components, $\gamma_1$ and $\gamma_2$, separated by 45 degrees) of the background image, respectively. 
They are connected by derivatives of a lensing potential $\phi$ that satisfies Eq.~\ref{eq:kappa} and Eq.~\ref{eq:gamma} with sky coordinates. 
\begin{equation}
    \kappa = \frac{1}{2}(\partial_{11}\phi+\partial_{22}\phi) = \frac{\Sigma}{\Sigma_{\rm crit}}
    \label{eq:kappa}
\end{equation}
\begin{equation}
    \gamma_1 = \frac{1}{2}(\partial_{11}\phi-\partial_{22}\phi);~ \gamma_2 = \partial_{12}\phi = \partial_{21}\phi
    \label{eq:gamma}
\end{equation}
The convergence is a normalized surface density --- in Eq.~\ref{eq:kappa}, $\Sigma=\Sigma({\boldsymbol{\theta}})$ is the lens surface mass density, and $\Sigma_{\rm crit}$ is the critical surface density satisfying Eq.~\ref{eq:critical_surface_density}, 
\begin{equation}
    \Sigma_{\rm crit} = \frac{v_c^2}{4\pi G} \frac{D_s}{D_l D_{ls}},
    \label{eq:critical_surface_density}
\end{equation}
where $v_c$ is the speed of light, and $D_s$, $D_l$, $D_{ls}$ are the angular diameter distances from the observer to the source, to the lens, and between the source and the lens, respectively.   

For an axisymmetric mass profile, lensing stretches the background source image tangentially with respect to the lens center. It is typical to consider a transformation of the shear along the tangential and cross directions. 
The transformation between shear $\boldsymbol{\gamma}$ in the polar system $\boldsymbol{\gamma}_{\rm p}$ and the Cartesian system $\boldsymbol{\gamma}_{\rm c}$ is $\boldsymbol{\gamma}_{\rm p}=\mathbf{A} \boldsymbol{\gamma}_{\rm c}$ and $\boldsymbol{\gamma}_{\rm c}=\mathbf{A}^{-1} \boldsymbol{\gamma}_{\rm p}$, where 
\begin{equation}
    \boldsymbol{\gamma}_{\rm p}=\begin{pmatrix}
    \gamma_{\rm T} \\  \gamma_{\rm X}
    \end{pmatrix},~\boldsymbol{\gamma}_{\rm c}=\begin{pmatrix}
    \gamma_{\rm 1} \\  \gamma_{\rm 2}
    \end{pmatrix},
\end{equation}
\begin{equation}
\mathbf{A}=
\begin{pmatrix}
-\cos{2\theta} & -\sin{2\theta} \\ 
\sin{2\theta} & -\cos{2\theta}
\end{pmatrix},~\mathbf{A}^{-1}=\begin{pmatrix}
-\cos{2\theta} & \sin{2\theta} \\
-\sin{2\theta} & -\cos{2\theta}
\end{pmatrix}, 
\label{eq:shear_conversion_matrix}
\end{equation}
and $\theta$ is the position angle (counterclockwise from the positive x-axis direction). 
Equivalently in the complex system $\boldsymbol{\gamma}_{\rm p}=-\exp{(-2i\theta)}\boldsymbol{\gamma}_{\rm c}$, with $\boldsymbol{\gamma}_{\rm c}=\gamma_{\rm 1} + i\gamma_{\rm 2}$ and $\boldsymbol{\gamma}_{\rm p}=\gamma_{\rm T} + i\gamma_{\rm X}$.

Along a circular loop with radius $R=D_l\sqrt{\theta_1^2+\theta_2^2}$, the mean tangential shear field is a ratio between the excess surface density $\Delta\Sigma$ and the $\Sigma_{\rm crit}$ at a given source redshift, 
where $\Delta\Sigma$ is the difference between the mean surface density within the loop and the mean surface density along the loop (Eq.~\ref{eq:mean_shear_excess_surface_density}). The surface density profile can be arbitrary~\citep{Schneider2005}.
\begin{equation}
    \langle \gamma_{\rm T}(R)\rangle = \frac{\Delta\Sigma(R)}{\Sigma_{\rm  crit}} = \frac{\bar{\Sigma}(<R)-\langle \Sigma(R) \rangle}{\Sigma_{\rm  crit}}
    \label{eq:mean_shear_excess_surface_density}
\end{equation}

\subsubsection{Elliptical mass distribution}
When describing the surface mass distribution, one can use a multipole decomposition 
starting from a circular component (zeroth-order multipole/monopole) and then add in components to account for the non-circular shape (second-order multipole/quadrupole and higher-order multipoles; Eq.~\ref{eq:multipoles}). 
We consider the elliptical shape of the cluster lens, and thus only use the first two components (monopole and quadrupole) of the expansion, following the procedure of~\citet{Adhikari2015,Clampitt2016,Shin2018}. The higher-order multipoles of the surface density and their lensing effects are much smaller.

\begin{equation}
    \Sigma(R,\theta) = \Sigma_0(R) + \Sigma_2(R)\cos{2\theta} + \cdots
    \label{eq:multipoles}
\end{equation}

In Eq.~\ref{eq:multipoles}, the mass is centered at the origin, $R=D_l\sqrt{\theta_1^2+\theta_2^2}=\sqrt{x^2+y^2}$ is the projected radial distance on the image/lens plane (PoS) with polar angle $\theta$, $\Sigma_0(R)$ is the \textit{monopole}, and $\Sigma_2(R)$ is the \textit{quadrupole}. Here we assume a flat-sky approximation at small angles and take $(x,y)$ as physical coordinates. The major axis is along the x-axis. We assume that the ellipticity is small and nearly fixed along the cluster-centric radius (tested in Section~\ref{sec:galaxy_radial_number_density}), and this requires $\Sigma_2(R)\ll\Sigma_0(R)$ (demonstrated in Section~\ref{sec:results}).

\begin{figure}[!tbp]
    \plotone{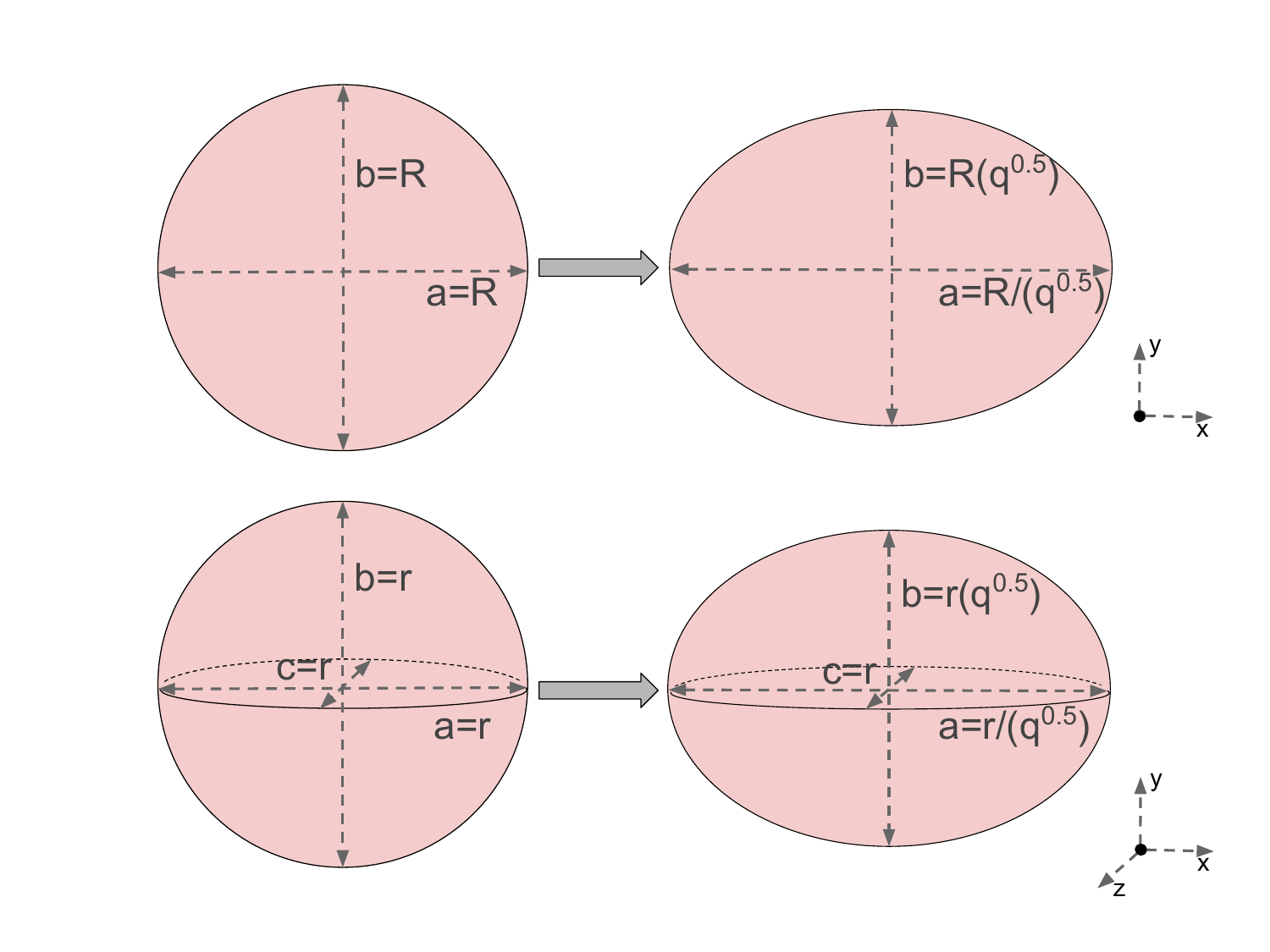}
    \caption{Illustration of the deformation. \textit{Top}: 2D projection with surface density transformation $\Sigma(\sqrt{x^2+y^2})\rightarrow\Sigma(\sqrt{qx^2+y^2/q})$. The iso-density contour changes from a circle to an ellipse with ratio of axes $a:b =1/\sqrt{q}:\sqrt{q}$. For convenience of calculation, we choose the $a$- and $b$-axes to be oriented along the $x$- and $y$-axes, respectively. Since the transformation has a Jacobian determinant of unity, the enclosed area and mass are preserved. 
    \textit{Bottom}: Effective 3D structure with density transformation $\rho(\sqrt{x^2+y^2+z^2})\rightarrow\rho(\sqrt{qx^2+y^2/q+z^2})$. The $c$-axis is aligned with the LoS (i.e., the $z$-axis). The $a$- and $b$-axes are also aligned with the $x$- and $y$-axes, respectively.
    Here, the iso-density shell changes from a sphere to an ellipsoid (triaxial model) with ratio of axes, $ a:b:c = 1/\sqrt{q}:\sqrt{q}:1$. The transformation again has a Jacobian determinant of unity, preserving the enclosed volume and mass. The \textit{Top} structure is the projection of the \textit{Bottom} structure. The mass and shape of the ellipsoid provide an \textit{effective} representation of the halo.
    }
    \label{fig:Illustration}
\end{figure}

To derive the formulas for the monopole and quadrupole, we begin with an arbitrary circular/spherical density profile $\Sigma_{\rm sph}(R)$. 
Next, let us slightly stretch the circular profile (Figure~\ref{fig:Illustration}), so that the surface density has the form $\Sigma(R,\theta)=\Sigma_{\rm sph}(\tilde{R})$, where $\tilde{R}=\sqrt{qx^2+y^2/q}$.  
Here $q$ is the lens axis ratio, and Eq.~\ref{eq:lens_ellipticity} gives the lens ellipticity. 

\begin{equation}
    \epsilon = \frac{1-q}{1+q}
    \label{eq:lens_ellipticity}
\end{equation}

Note, in this work we adopt $q$ instead of $q^2$ in the form of ellipticity, so that it has smoother behavior at high ellipticity (i.e., the ellipticity does not rapidly increase to 1 when $q\rightarrow0$) and thus gives better fitting results --- at a small value of $q$ we can more easily pin down the value of $\epsilon$.

Using Taylor expansion, one can expand the new $\Sigma$ into powers of $\epsilon$, match the form of Eq.~\ref{eq:multipoles}, and then obtain Eq.~\ref{eq:sigma0_sigma2}. 

\begin{equation}
    \left\{
    \begin{aligned}
    \Sigma_0(R) &= \Sigma_{\rm sph}(R) \left[1 + \epsilon^2 \left(\frac{u}{2} + \frac{u^2}{4} + \frac{1}{4}\frac{\mathrm{d} u}{\mathrm{d}\ln R} \right) \right] + \mathcal{O}( \epsilon^4 )\\
    \Sigma_2(R) &= -\epsilon u \Sigma_{\rm sph} (R) + \mathcal{O}( \epsilon^3 )
    \end{aligned}
    \right.\label{eq:sigma0_sigma2}
\end{equation}

In Eq.~\ref{eq:sigma0_sigma2}, we define $u\equiv {\mathrm{d} \ln{\Sigma_{\rm sph}}} / {\mathrm{d}\ln{R}}$, the ``logarithmic slope'' of $\Sigma_{\rm sph}$, for simplicity of the form.\footnote{We note that the minimum of $u$ is a proxy for the ``splashback radius''~\citep{Diemer2014} of the 2D circular profile.} 
We provide the derivation details in Appendix~\ref{app:derive_multipoles_forms}. 
When $\epsilon\rightarrow0$, we have $\Sigma_0\rightarrow\Sigma_{\rm sph}$ and $\Sigma_2\rightarrow0$. 
For an NFW profile as $\Sigma_{\rm sph}$, $u$ is negative and, as $R$ increases, $u$ decreases, making the coefficient of the $\epsilon^2$ term inside the square brackets negative (between $\sim-0.35$ and 0). An interesting detail is that as $R\rightarrow r_{\rm s}$, $u\rightarrow-1.2$, and when the radius is sufficiently large $u\rightarrow-2$ and $\Sigma_0\rightarrow\Sigma_{\rm sph}$~\citep{Wright2000}. 

\clmm v2 uses high-precision numerical differentiation to compute the $u$ term and its derivatives, which makes it flexible to switch between different spherical profile models such as NFW, Einasto~\citep{Einasto1965}, etc. 

Compared to previous studies~\citep[e.g.,][]{Shin2018}, in this work we go further and include a correction term of $\epsilon^{2}$ for the monopole. This second order term contributes to a percent-level correction to the mass profile (Section~\ref{sec:dis_meaning_of_mass}; see also the \clmm v2 paper). 

In theory, a 2-halo term can also be included in $\Sigma_{\rm sph}$, if we assume both the 1-halo and the 2-halo density profiles have the same ellipticity and orientation, and then the effect of LSS on both the monopole and the quadrupole will be taken into account. 
In this work, we focus on the signal within the 1-halo regime (inside a radial distance of 2 Mpc) and thus exclude the 2-halo term in the main analysis, but we discuss its effect on the measurement in Section~\ref{sec:dis_lss}.

\subsection{Lensing analysis}
\label{sec:lensing_analysis}

\subsubsection{Shear measurements}
\label{sec:shear_measurements}

In weak lensing, $\kappa,\gamma\ll1$, and the observable reduced shear $\boldsymbol{g}=\boldsymbol{\gamma}/(1-\kappa)$ satisfies
\begin{equation}
    \langle \boldsymbol{g} \rangle \sim \langle \boldsymbol{\gamma} \rangle \sim  \langle  \mathsf{R}\rangle ^{-1} \langle \boldsymbol{e} \rangle,
\end{equation}
for an ensemble of background galaxies, where $\mathsf{R}$ is the response matrix and $\boldsymbol{e}$ is the galaxy shape. 

The shape catalog provides the shape, shear response, and weight of each galaxy. The weight is based on the shape dispersion and the measurement error. It is an inverse variance weight minimizing the scatter of the shear signal --- it reduces the error bar of the shear estimate, but keeps the mean value almost fixed. In the following text, when we mention a response, we refer to the total response ($\mathsf{R}=\mathsf{R}_{\gamma}+\mathsf{R}_{\rm s}$).
The shear response $\mathsf{R_\gamma}$ is primarily a function of the size and brightness (or signal-to-noise ratio, S/N) of the source galaxy, but also slightly depends on the source redshift and location on the sky~\citep{Gatti2021}. 
The selection applied to the sources (for building a clean sample) can be shear-dependent: the shear affects the measured properties of galaxies (such as size and brightness), which changes the distribution of source properties in the selected sample. Therefore, a selection response $\mathsf{R}_{\rm s}$ needs to be included in the analysis as well~\citep{Sheldon2017} --- it is evaluated by applying the same selection to the (artificially) sheared images and finding the change of the mean shape of corresponding galaxies on the original (unsheared) image. 

Along an annulus at radius $R$, the shear satisfies $\langle g_{\rm T} \Sigma_{\rm crit} \rangle |_R \sim \langle \gamma_{\rm T} \Sigma_{\rm crit} \rangle |_R \sim \Delta\Sigma(R) $ for a sample of sources (Eq.~\ref{eq:mean_shear_excess_surface_density}). The shear depends on the source redshift. 
When determining the mass distribution ($\Delta\Sigma$) in each radial bin $R$, 
we consider two methods below for the estimator. In both methods, we select source galaxies that satisfy $\texttt{ZMEAN\_SOF}>z_l+0.1$. {Note that this cut is used by~\citet{McClintock2019} as well. Applying a redshift cut may lead to a selection bias. However, the photo-zs measured on the artificially sheared images are not publicly available and the computation of the corresponding selection response is beyond the scope of this paper. We expect its effect on the measured halo ellipticity to be minimal, especially because it can be regarded as a multiplicative bias on both the monopole and quadrupole measurements (see also the description for the boost factor below). The detection/deblending could be affected by shear, which can be addressed by \textsc{metadetection}~\citep{Yamamoto2025}; we will test shear measurements other than \metacal{} in future work.} 

\paragraph{Method 1 (\textit{Per-object response estimator})} 
This method generally follows the algorithm of~\citet{McClintock2019}. We integrate the response with $\Sigma_{\rm crit}^{-1}$ in the estimator so that they are summed together (Eq.~\ref{eq:delta_sigma_estimator1}, where $w_i$ is the weight per source galaxy $i$). 
\begin{equation}
    \Delta \hat{\Sigma} = \frac{\sum_i w_i e_{\textrm{T},i}}{\sum_i w_i \mathsf{R}_i \Sigma_{\textrm{crit},i}^{-1}} = \frac{\langle e_{\textrm{T}} \rangle}{\langle \mathsf{R} \Sigma_{\textrm{crit}}^{-1} \rangle}
    \label{eq:delta_sigma_estimator1}
\end{equation}
We use the reciprocal of  $\Sigma_{\rm crit}$ to avoid the singularity when the lens redshift $z_l$ and the source redshift $z_s$ are close (when $D_{ls}\rightarrow 0$, $\Sigma_{\rm crit}\rightarrow\infty$). The weight $w_i$ is the product of another $\Sigma_{\rm crit}^{-1}$ and the shape weight ($w_{\textrm{shape}}$) provided by the catalog, that is,  
\begin{equation}
    w_{i} = w_{\textrm{shape},i}\Sigma_{\textrm{crit},i}^{-1}\;.
\end{equation}
The selection response is estimated by all selected sources, and the shear response along the tangential direction is computed. We use \texttt{ZMEAN\_SOF} for the selection and weight while we use \texttt{ZMC\_SOF} for evaluating the critical surface density and effectively marginalizing photo-z uncertainty, following the strategy of~\citet{McClintock2019}.

\paragraph{Method 2 (\textit{Per-HEALPix response estimator})}  
For each galaxy, we convert the shape into the shear estimate per source $\hat{g}$ using the mean response $\bar{\mathsf{R}}$ of a sufficiently large sky area~\citep[per HEALPix with NSIDE=32
and RING ordering,  $\sim3\deg^2$ per pixel;][]{Gorski2005,Fu2024a}, so that $\hat{g}_{1,2}=e_{1,2}/\bar{\mathsf{R}}$. The weight $w_i$ includes both the shape weight and another $\Sigma_{\textrm{crit}}^{-1}$, and for simplicity, we use $\texttt{ZMEAN\_SOF}$ for $z_s$ in all $\Sigma_{\textrm{crit}}$ (Eq.~\ref{eq:delta_sigma_estimator2}). 
\begin{equation}
    \Delta \hat{\Sigma} = \frac{\sum_i w_i \hat{g}_{\textrm{T},i}} {\sum_i w_i \Sigma_{\textrm{crit},i}^{-1}} = \frac{\langle \hat{g}_{\textrm{T}} \rangle }{ \langle \Sigma_{\textrm{crit}}^{-1} \rangle} =  \frac{\langle e_{\textrm{T}}  \rangle}{\bar{ \mathsf{R} } \langle \Sigma_{\textrm{crit}}^{-1}\rangle} 
    \label{eq:delta_sigma_estimator2}
\end{equation}

The major difference between the two methods is the treatment of the response. 
The second method is more straightforward and easier to compute and thus provides a quick test of physical quantities, but the first method is more accurate. We find that the variation between the two methods is a few percent. 
When we measure the halo shape, this difference can be regarded as a constant factor in both the monopole and the quadrupole, and then this factor can be largely canceled out in the cluster ellipticity measurement (similar to the boost factor; see the following subsections). 
In the following text, we adopt the \textit{first} method in the main analysis and demonstrate the robustness of the second method in Section~\ref{sec:des_y1_rm}.

\subsubsection{Monopole and quadrupole measurements} \label{sec:monopole_and_quadrupole_measurements}
\paragraph{Models}
We first build theoretical models for the cluster halo ellipticity measurement. 
Following~\citet{Adhikari2015}, by putting the multipole expansion of the convergence $\kappa=\sum_{m\geq0} \kappa_m(R) \cos{m\theta}$ into the equation $\kappa=\frac{1}{2}\nabla\phi$ (Eq.~\ref{eq:kappa})  and solving it, one can obtain the multipoles of the lensing potential $\phi$ (in the polar coordinate system),  and then obtain the multipoles of the shear $\gamma$ (for both tangential and cross components). 
Keeping only the $m=0$ and $m=2$ multipoles,  one can get the monopole and quadrupole of shear. The model for the monopole is shown in Eq.~\ref{eq:sigma0_sigma2}. For the quadrupole, \citet{Clampitt2016} transformed the tangential and cross components of the quadrupole into Cartesian components, which can be measured more straightforwardly. 
Then~\citet{Shin2018} re-organized the Cartesian components and built models that separated the internal (integrated up to a radius R) and external (integrated from radius R to $\infty$)
quadrupole signals and maximized the detection significance. Based on that work, we rewrite the models as follows (more details of the derivation are given in Appendix~\ref{app:derive_multipoles_models_general}):

\begin{equation}
    \left\{
    \begin{aligned}
        \Delta\Sigma_{4\theta} &= \frac{1}{2\pi} \int_0^{2\pi} (\Sigma_{\rm crit}\gamma_{1}\cos{4\theta} + \Sigma_{\rm crit}\gamma_{2}\sin{4\theta})\mathrm{d}\theta \\
        &= \frac{\Sigma_2}{2} - L_1\;, \\
        \Delta\Sigma_{\rm const} &= \frac{1}{2\pi} \int_0^{2\pi} \Sigma_{\rm crit}\gamma_{1} \mathrm{d}\theta = \frac{\Sigma_2}{2} - L_2\;, \\
        L_1 &= \frac{3}{R^4}\int_0^R R'^3 \Sigma_2(R')\mathrm{d}R'\;, \\
        L_2 &= \int_R^{\infty} \frac{\Sigma_2(R')}{R'}\mathrm{d}R'\;.
    \end{aligned}
    \right.
    \label{eq:models_main}
\end{equation}

In Eq.~\ref{eq:models_main}, $L_1$ is determined by the quadrupole profile ($\Sigma_2$) inside a circle with radius $R$, while $L_2$ is determined by the one outside $R$. Therefore, $\Delta\Sigma_{4\theta}$ is mainly affected by the structure inside radius $R$, while $\Delta\Sigma_{\rm const}$ is mostly affected by the external structure (outside radius $R$). Also, since $\Sigma_2$ grows with the halo ellipticity and concentration (i.e., steeper slope; Eq.~\ref{eq:sigma0_sigma2}), $\Delta\Sigma_{4\theta}$ and $\Delta\Sigma_{\rm const}$ also grow with them.

\paragraph{Estimators}
Estimators for the monopole are defined in Section~\ref{sec:shear_measurements}. And, following the excess surface density estimator in Eq.~\ref{eq:delta_sigma_estimator1} (\textit{Method 1}), we compute the following quadrupole estimators ($\Delta\hat{\Sigma}_{4\theta},\Delta\hat{\Sigma}_{\rm const} $) using the observed data to match the theoretical models above. We find that in each estimator, projecting the shear response along the numerator direction changes the denominator negligibly; thus, for simplicity, we use the tangential response for all estimators.

\begin{equation}
    \left\{
    \begin{aligned}
    \Delta\hat{\Sigma}_{4\theta} &= \frac{\sum_i  w_i ( e_{1,i}\cos{4\theta_{i}} + e_{2,i}\sin{4\theta_{i}} )}{\sum_i w_{i} \mathsf{R}_{i} \Sigma_{{\rm crit},i}^{-1}} \\
    \Delta\hat{\Sigma}_{\rm const} &= \frac{\sum_i  w_i e_{1,i} }{\sum_i w_{i} \mathsf{R}_{i} \Sigma_{{\rm crit},i}^{-1}}
    \end{aligned}
    \right.
    \label{eq:estimators_main}
\end{equation}
Here, the weight per source $w_i$ is the same as the one in Section~\ref{sec:shear_measurements}.  
We have implemented these quadrupole models and estimators in \clmm v2. The position angle is computed using \texttt{Astropy}. 

To reduce noise, we stack the clusters in a sample when computing the monopole and quadrupole estimators. That is, for each summation term we sum over all source galaxies for all clusters in the sample.
We compute the estimators in logarithmically-spaced radial bins. We estimate the covariance using jackknife resampling; we split the sky footprint of clusters into 100 jackknife patches using the \texttt{kmeans\_radec}\footnote{\url{https://github.com/esheldon/kmeans_radec}} algorithm. The covariance can also be estimated analytically~\citep{Kohlinger2015,Schrabback2026}, and the result is expected to be consistent with the jackknife, especially at the scales we measure in this paper~\citep{McClintock2019}.

\paragraph{Dilution factor for angle misalignment correction}
In practice, the misalignment between the orientations of the member galaxy distribution and the dark matter halo dilutes the triaxial signal, thus leading to a smaller measured halo ellipticity, i.e., the cluster looks rounder~\citep{Shin2018}. 
The dilution effect depends on both the halo ellipticity and the number of member galaxies (and thus richness). 
We use simulations to evaluate the dilution effect under specific ellipticity and richness, and then compute the mean dilution after matching the richness distribution of the simulated clusters to that of the true cluster sample. 
For each cluster sample, we construct a ``dilution factor'' as a function of the true halo ellipticity $D(\epsilon_{\rm true})$ to correct for the dilution effect. We present the details as follows.

When the halo major axis has a position angle of $\theta_0$, the true position angle of a background source galaxy with respect to the major axis changes from $\theta$ to $\theta-\theta_0$. Then, the quadrupole models change from $\Delta\Sigma_{\rm 4\theta,const} $ to $ \Delta\Sigma_{\rm 4\theta,const}\cos{2\theta_0}$ per cluster  (Appendix~\ref{app:derive_dilution_factor}). When stacking an ensemble of clusters, we consider a mean $\langle \cos{(2\theta_0)}\rangle$ for the dilution factor $D$ given an input ellipticity $\epsilon$. This factor is then applied to the quadrupole terms affecting the measured halo ellipticity.  

The misalignment issue has two underlying causes. First, galaxies may not perfectly trace dark matter. To address this, we need tests from cosmological hydrodynamical simulations, which are beyond the scope of this paper, and we postpone this analysis to future work.  
Second, even if the galaxy distribution fully traces dark matter, the limited number of galaxies can lead to noise in the orientation measurement --- a type of Poisson noise. 
In this work, we only consider the latter cause of the misalignment, and we use Monte Carlo sampling to estimate the Poisson noise by randomly drawing a sample of galaxies based on the richness and then measuring the orientation of their distribution.  

Next, we describe the steps of our algorithm. This is similar to the method of~\citet{Shin2018}, where the biases from the Poisson noise, the membership aperture, and the interlopers have been addressed, but we update the method with more detailed modeling. 

1. Take an ellipticity value as input. 

2. Pick one cluster from a cluster sample, select $N_{\rm mem}$ members with membership probability $P_{\rm mem}>0.5$ (the same cut as to compute the projected cluster orientation; Section~\ref{sec:halo_orientation}).

3. Draw (rounded) $\sum P_{\rm mem} - 1$  galaxies from an elliptical NFW distribution with the input ellipticity; the summation is over these selected members with $P_{\rm mem}>0.5$. 
Here, the minus 1 is because the member galaxies include the central galaxy and it is chosen as the origin of the member galaxy distribution. The concentration of the profile is derived from the mass\footnote{We assume the red-sequence member galaxies trace the dark matter distribution~\citep{vanUitert2016,Hennig2017}. We expect the effect of concentration on the orientation to be weak, as the $1/R^2$ weight in constructing the second moments (Eq.~\ref{eq:second_moments}) removes the dependence on galaxy radii from the orientation estimator (Eq.~\ref{eq:position_angle}).}~\citep{Diemer2018,Diemer2019}, while the mass is derived from the richness with scatter included~\citep{DES2025,To2025}. 
In addition, the galaxy positions are constrained within a radius of $R_\lambda = 1.95 (\lambda/100)^{0.45}$~$h^{-1}$Mpc~\citep[percolation radius;][]{DES2025}, which defines the outer boundary of the membership aperture.

4. Draw (rounded) $N_{\rm mem} - \sum P_{\rm mem}$ galaxies from a uniform distribution within $R_\lambda$ to simulate interlopers (i.e. galaxies along the LoS but not associated with the cluster). 

5. Calculate the major axis orientation using the weighted second moments (Section~\ref{sec:halo_orientation}, Eq.~\ref{eq:second_moments}) of all galaxy points generated by Steps 2 to 4. 

6. Go to another cluster and repeat Steps 2 to 5. 

7. Go to another ellipticity value and repeat Steps 1 to 6.

\begin{figure}
    \plotone{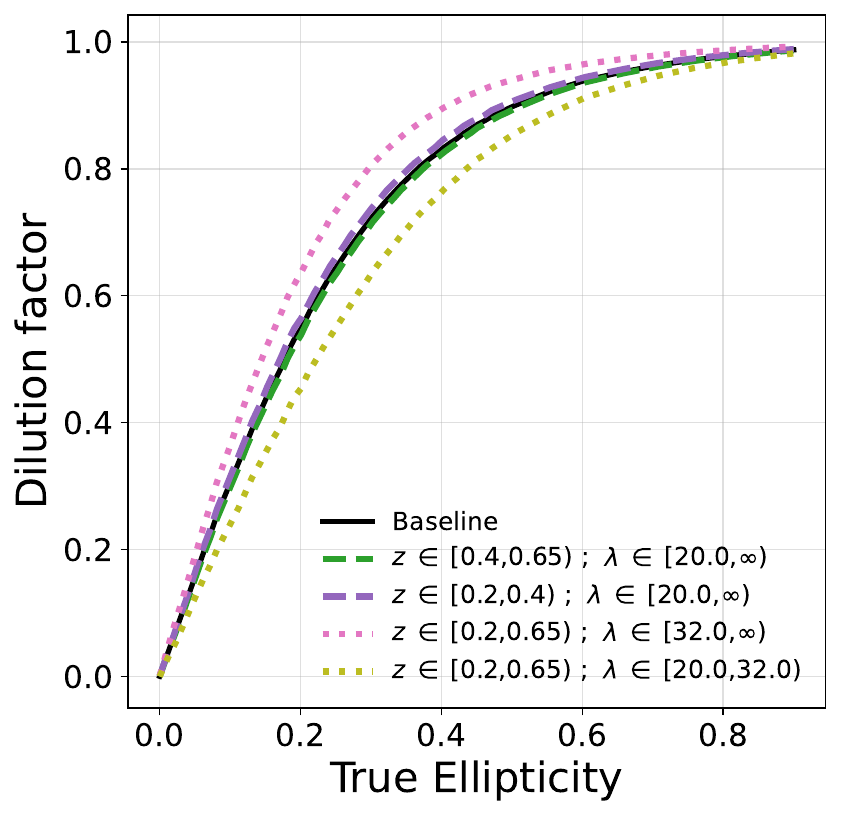}
    \caption{Dilution factor as a function of the \textit{true} ellipticity for the baseline sample (defined in Section~\ref{sec:cluster_catalog}) and for the redshift- and richness-split cluster sub-samples (defined in Section~\ref{sec:cluster_sample_division}).}
    \label{fig:dilution_factor}
\end{figure}

8. Repeat Steps 1 to 7 several times; we choose 32 times and find that it provides sufficient statistics to build a smooth dilution factor curve (Figure~\ref{fig:dilution_factor}).

9. Calculate the mean dilution factor weighted by mass at a given ellipticity: $
\langle \cos{2 \theta_0}\rangle $. Here, the mass is proportional to $\lambda^{1.053}$ (and a weak function of redshift), which is from the mass-richness relation of DES Y3~\citep{DES2025}. We use this weight because more massive clusters contribute more to (quadrupole) lensing signals, and when the radial distance $R$ is sufficiently larger than the scale radius $r_s$, $\Delta\Sigma$ scales approximately with mass at a given $R$. Hence, we obtain an effective dilution factor for the cluster sample: $D_{\rm eff}=\langle \Delta\Sigma_i D_i\rangle/\langle \Delta\Sigma_i \rangle \sim \langle \Delta\Sigma_{i;\,\rm 4\theta,const} D_i\rangle/\langle \Delta\Sigma_{i;\,\rm 4\theta,const} \rangle$; $i$ is the index of each cluster.\footnote{Note that $\Delta\Sigma^{\rm 4\theta,const}\propto\Sigma_2$ and $\Sigma_2\propto\Sigma_{\rm sph}\propto\Sigma_0$, when $\epsilon$ is small and fixed, and when the logarithmic slope of $\Sigma_{\rm sph}$ is fixed (Eq.~\ref{eq:sigma0_sigma2},~\ref{eq:models_main}).
}

10. Build a mapping between input true ellipticities and the mean dilution factor of the cluster sample. 
Figure~\ref{fig:dilution_factor} shows the dilution factor as a function of ellipticity for each cluster sample. It grows with ellipticity. It is sensitive to richness, since clusters with fewer members are more affected by misalignment, but it is not sensitive to redshift.

The derivation of the dilution factor may be improved by using more realistic clusters in N-body cosmological simulations such as~\skysim{}~\citep{DESC2021}, an extension of~\cosmodc{}~\citep{Korytov2019} featuring more comprehensive modeling.  
Using the cosmological simulation, we can find the misalignment between the cluster's 3D major axis projected on the sky (using the particle information) and the major axis of the 2D projected galaxy/subhalo positions on the sky. We show an initial test in Appendix~\ref{app:sim_orientation_comparison}. 

Additionally, it is possible that the misalignment can vary with the radial distance; we assume that the change is small. For a massive cluster, it is also possible to directly fit the orientation angle~\citep{Payerne2023}, informed by the weak-lensing data-vectors, which could be tested in future work. 
Moreover, halo-shear-shear correlation allows halo shapes to be determined without knowledge of major axes~\citep{Adhikari2015,Liu2025}.

\paragraph{Boost factor}
After obtaining the lensing estimate via WL, we also consider a correction of the lensing signal called the ``boost factor'' $\mathcal{B}$. 
Note, this is a general systematic in WL, not just for analyzing the triaxial signal. 
Due to the photo-z error, some (unlensed) cluster member galaxies can be recognized as background galaxies and thus dilute the lensing signal. The correction for this is called the boost factor. The boost factor can be estimated by the number density of galaxies because clusters have high concentrations of galaxies, while background galaxies are generally assumed to be uniformly and randomly distributed. However, lensing magnification/deflection, galaxy obscuration/blending, and masking may also affect the observed number density of source galaxies. Therefore, instead of using the number density, we use the photo-z to estimate the boost factor~\citep{Varga2019}.

We decompose the observed redshift distribution $P(z)$ of galaxies selected in the analysis into two components --- true background source galaxies and cluster member galaxies causing the contamination (Eq.~\ref{eq:boost_factor_decompose}). The background sources produce lensing signals, while the cluster galaxies do not. Here we assume that the intrinsic alignment is small. It is also possible that the contamination has a small dependence on the polar angle; we skip it for now and leave it to future work. 
\begin{equation}
    P(z \mid R) = f_{\rm cl}(R) P_{\rm m}(z) + (1-f_{\rm cl}(R)) P_{\rm bg}(z)
    \label{eq:boost_factor_decompose}
\end{equation}

In Eq.~\ref{eq:boost_factor_decompose}, $P(z \mid R)$ is the observed redshift distribution at the radial bin $R$, $f_{\rm cl}(R)$ is the fraction of cluster galaxies and $(1-f_{\rm cl}(R))$ is the fraction of background galaxies. $P_{\rm m}(z)$ is the redshift distribution of the cluster member galaxies, which is assumed to be a Gaussian with both the mean and the standard deviation to be fitted.\footnote{The mean can be higher than the cluster redshift provided by \rdmp{}, because these cluster galaxies have photo-zs biased high~\citep{Varga2019}.} 
$P_{\rm bg}(z)$ is the redshift distribution of background galaxies, which is determined by galaxies far away from the cluster center on PoS ($9.5-125~h^{-1}$Mpc). Eq.~\ref{eq:boost_factor_decompose} assumes that the redshift distribution of background galaxies is independent of the radius. The lensing magnification can modify the background photo-z distribution, but previous studies have shown that its impact on the inferred boost factor is at the percent level or below~\citep{Gruen2017,Varga2019}. 

In $P(z \mid R)$ and $P_{\rm bg}(z)$, the galaxies are selected by $\texttt{ZMEAN\_SOF}>z_l+0.1$ and weighted by the same weights for constructing $\Delta\hat{\Sigma}$ (Eq.~\ref{eq:delta_sigma_estimator1}, denominator). 
The $f_{\rm cl}$ is estimated in each radial bin $R$ (with the same mean and scatter of the cluster's Gaussian $P_{\rm m}(z)$ jointly fit among all radial bins), and then the boost factor is determined by Eq.~\ref{eq:boost_factor_value}. 
\begin{equation}
    \mathcal{B}(R)=\frac{1}{1-f_{\rm cl}(R)}
    \label{eq:boost_factor_value}
\end{equation}

\begin{figure}
    \plotone{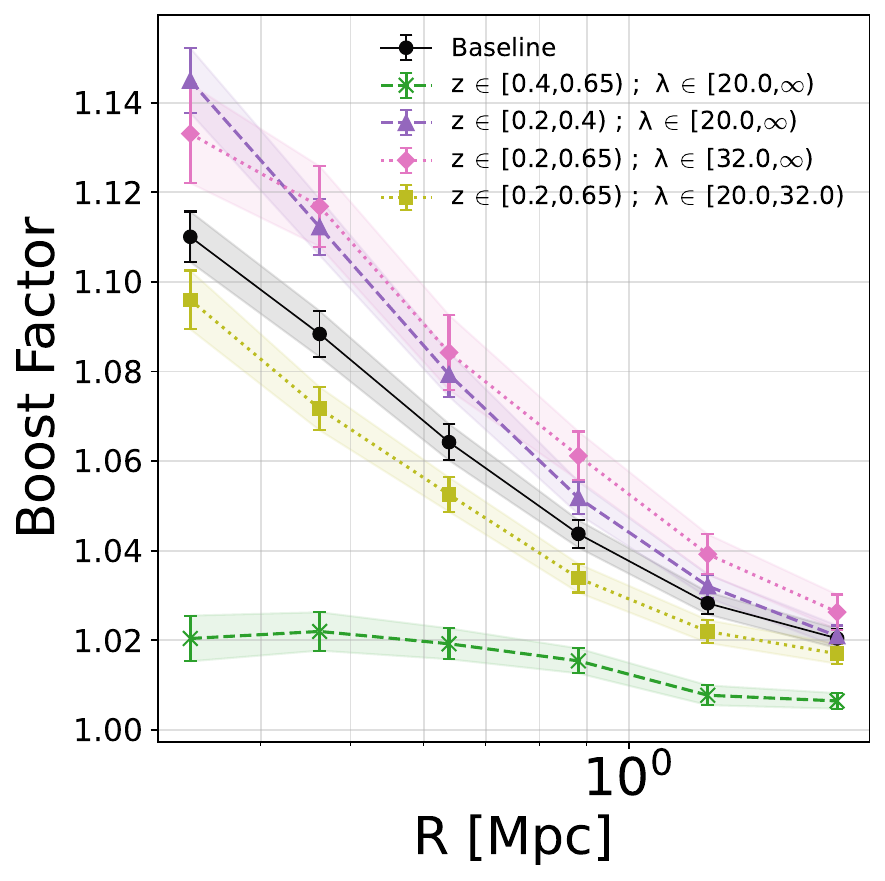}
    \caption{Boost factors computed for the stacked lensing profiles as a function of radial distance. The boost factor for the baseline sample, as defined in Section~\ref{sec:cluster_catalog}, is shown in black (with circular markers and solid lines). The boost factors for other sub-samples (defined in Section~\ref{sec:cluster_sample_division}) are shown in different colors, markers, and line-styles as described in the legend. The shaded area represents the $1\,\sigma$ confidence interval. }
    \label{fig:Y3RM_zAll_lAll_boost_factor}
\end{figure}

We use \texttt{ZMC\_SOF} to construct $P(z\mid R)$ and $P_{\rm bg}(z)$ for $z \in (0.1,1.25)$ with bin width $\Delta z=0.05$. Figure~\ref{fig:Y3RM_zAll_lAll_boost_factor} shows the boost factor for each cluster sample. The error bars are estimated by sky jackknife. As expected, the boost factor grows with richness as more cluster members contaminate the background sample. Also, it decreases with redshift, because at lower redshift the cluster galaxies are relatively redder (more quenched) making it more difficult to distinguish cluster members from the background galaxies. 

Note that we apply boost factors to both the monopole ($\Delta\Sigma_0$) and the quadrupoles ($\Delta\Sigma_{\rm 4\theta,const}$). 
This factor will propagate to $\Sigma_2$ via Eq.~\ref{eq:models_main}, and to $\Sigma_0$ via $\Delta\Sigma_0$. Therefore, $\Sigma_0$ and $\Sigma_2$ will be multiplied by the same factor at a given radius. 
Then, Eq.~\ref{eq:sigma0_sigma2} shows that this generally affects the amplitude of $\Sigma_{\rm sph}$ without affecting its logarithmic slope $u$ and the halo shape $\epsilon$, as the $\epsilon^2$ term is small. 
Thus, even if we do not consider the boost factor, the measured halo ellipticity will not greatly change; we keep it here for accuracy. 
Other types of multiplicative biases will produce similar effects.


\paragraph{Fitting}

After correction for the dilution factor and the boost factor $\mathcal{B}(R)$, our final data vector and model for a cluster sample is shown in Eq.~\ref{eq:final_model}: the left hand is the estimator derived from the data; the right hand is the model. 
We use the NFW profile for the spherical mass profile $\Sigma_{\rm sph}(R)$, whose mass and concentration will be fitted. 
\begin{equation}
    \left\{
    \begin{aligned}
        \text{Data} &\leftrightarrow \text{Model}\\
        \mathcal{B}\Delta\hat{\Sigma} &\leftrightarrow\Delta\Sigma_{\rm final;\,mono} =  \Delta\Sigma_{\rm 0}\\
        \mathcal{B}\Delta\hat{\Sigma}_{\rm 4\theta} &\leftrightarrow\Delta\Sigma_{\rm final;\,\rm 4\theta} =  \Delta\Sigma_{\rm 4\theta}(\epsilon_{\rm true})D(\epsilon_{\rm true})\\
        \mathcal{B}\Delta\hat{\Sigma}_{\rm const} &\leftrightarrow\Delta\Sigma_{\rm final;\,\rm const} =  \Delta\Sigma_{\rm const}(\epsilon_{\rm true})D(\epsilon_{\rm true})
    \end{aligned}
    \right.
    \label{eq:final_model} 
\end{equation}

To measure the cluster halo ellipticity, we first evaluate the estimators $ \Delta\hat{\Sigma}$, $\Delta\hat{\Sigma}_{\rm 4\theta}$, $\Delta\hat{\Sigma}_{\rm const}$ (Eq.~\ref{eq:delta_sigma_estimator1}, Eq.~\ref{eq:estimators_main}), and then fit the model parameters  (Eq.~\ref{eq:final_model}). 
We simultaneously fit the monopole and quadrupole lensing signals, specifically $\Delta \Sigma_{\rm final;\,mono}$ to $\mathcal{B}\Delta \hat{\Sigma}$, $\Delta \Sigma_{\rm final;\,4\theta}$ to $\mathcal{B}\Delta \hat{\Sigma}_{4\theta}$, and $\Delta \Sigma_{\rm final;\,\rm const}$ to $\mathcal{B}\Delta \hat{\Sigma}_{\rm const}$ (Eq.~\ref{eq:final_model}). 
We adopt a Gaussian likelihood (Eq.~\ref{eq:likelihood}) and uniform priors on the mass $[10^{13},\,5.0\times10^{15}]$, concentration $[1.0,\,20.0]$, and ellipticity $[0.01,\,0.9]$. 

We denote the three free model parameters by $\mathbf{p}\equiv(M_{\rm 200c},c_{\rm 200c},\epsilon)$. We combine the three estimators and their corresponding model profiles into the data vector $\hat{\mathbf{d}}$ and model vector $\mathbf{m}$, respectively, and define the likelihood in Eq.~\ref{eq:likelihood}, where $\mathcal{N}( \hat{\mathbf{d}} \mid \mathbf{m}, \mathbf{C} )$ denotes a Gaussian probability for the data $\hat{\mathbf{d}}$, centered on the model prediction $\mathbf{m}$, with covariance matrix $\mathbf{C}$.

\begin{equation}
    \left\{
    \begin{aligned}
    \hat{\mathbf{d}} &= \left( \mathcal{B}\Delta\hat{\Sigma},\,\mathcal{B}\Delta\hat{\Sigma}_{\rm 4\theta},\,\mathcal{B}\Delta\hat{\Sigma}_{\rm const} \right)^{\rm T}\\
    \mathbf{m}(\mathbf{p}) &= \left( \Delta\Sigma_{\rm final;\,mono},\,\Delta\Sigma_{\rm final;\,4\theta},\,\Delta\Sigma_{\rm final;\, const} \right )^{\rm T}\\
    \mathcal{L}(\mathbf{p}) &= \mathcal{N}\left[\hat{\mathbf{d}} \mid \mathbf{m}(\mathbf{p}), \mathbf{C} \right]
    \end{aligned}
    \right.
    \label{eq:likelihood}
\end{equation}

\begin{figure}[htb!]
    \plotone{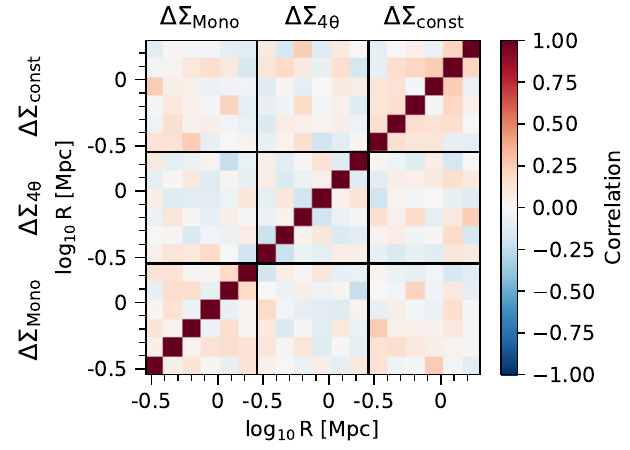}
    \caption{Jackknife-estimated correlation matrix of the boost-factor corrected profiles of the baseline sample. The radial distance bins are consistent with Figure~\ref{fig:Y3RM_zAll_lAll}.}
    \label{fig:correlation_boost}
\end{figure}

The covariance matrix $\mathbf{C}$ is computed by sky jackknife and includes the covariance between radial bins as well as that among the monopole and the two quadrupole profiles (Figure~\ref{fig:correlation_boost}). We find that most off-diagonal terms of the covariance of each profile are much smaller than the diagonal ones. Also, there is no clear sign of correlation between the profiles. Nonetheless, for accuracy, we perform the fitting using the full covariance matrix rather than only its diagonal terms or the covariances of individual profiles. 

We use \textsc{emcee}~\citep{Foreman-Mackey-D-2013:emcee}\footnote{\url{https://emcee.readthedocs.io}} to implement Markov Chain Monte Carlo (MCMC) for the fitting of the three free parameters: mass, concentration, and ellipticity. 
The fitting results of different cluster samples are presented in Section~\ref{sec:results}.

\subsubsection{Other systematics in lensing measurements}

Here we describe several other systematics; they are general in WL, not just in the multipole-expansion analysis. Since they affect both the monopole and quadrupole measurements, we expect their effects on the halo ellipticity measurement to be subdominant (Section~\ref{sec:monopole_and_quadrupole_measurements}).

\paragraph{Blending}
We note that the blending between cluster galaxies and background galaxies can affect the shear measurements. There are more member galaxies near the cluster center, and they are also brighter and larger, and thus their blending effect on the background source galaxies is stronger near the center.
On average, the  blending makes the source galaxies look ``rounder'' at percent levels relative to the shear~\citep[][]{HernandezMartin2020,Fu2021,MacCrann2022,Ramel2026}. 
Although we expect it to be a small effect, we exclude sources within 300 kpc on the sky around the cluster center, which also avoids the effects of strong lensing, intra-cluster light, etc. We further test the effect of blending on the halo shape measurement in Section~\ref{sec:dis_multi_sys}. 

\paragraph{Reduced shear}
Similarly, we do not consider the factor $1/(1-\kappa)$ for reduced shear $g$, because we exclude small radii (Appendix~\ref{app:derive_reduced_shear}). We test its effect on our measurement in Section~\ref{sec:dis_multi_sys}. 

\paragraph{Miscentering}
Likewise, we expect the effect of the cluster center offset (miscentering) to be small because of the $\texttt{pcen0}>0.9$ cut and the radial cut. We test its effect on our measurement in Section~\ref{sec:dis_miscenter}.

\subsection{Halo orientation determination}\label{sec:halo_orientation}

In this work, we stack the galaxy clusters along their 2D projected major axis to measure the monopole and quadrupole lensing signals. Consequently, we need to determine the orientation of the major axis of each cluster and rotate the WL shear estimates accordingly. To measure the orientation, we select cluster members with membership probability $P_{\rm mem}>0.5$ to reduce interlopers, and compute the second moments of the satellite galaxy distribution \citep[Eq.~\ref{eq:second_moments};][]{Shin2018}. 

\begin{equation}
    \begin{aligned}
    I_{xx} &= \frac{\sum_{i}x_{i}^{2}\omega_{i}}{\sum_{i}\omega_{i}}, & I_{yy} &= \frac{\sum_{i}y_{i}^{2}\omega_{i}}{\sum_{i}\omega_{i}}, & I_{xy} &= \frac{\sum_{i}x_{i}y_{i}\omega_{i}}{\sum_{i}\omega_{i}}.
    \end{aligned}
    \label{eq:second_moments}
\end{equation}

In Eq.~\ref{eq:second_moments}, CG is at the origin, $x_{i}$ and $y_{i}$ are the satellite galaxy coordinates, and $\omega_{i} = 1/(x_{i}^{2} + y_{i}^{2})$ is the weight per galaxy, which largely cancels the radial dependence of the major-axis angle estimator below. This weight takes into account the decrease of $P_{\rm mem}$ when the radius increases~\citep{Rykoff2014,Rozo2015,Rykoff2016}, and this weight can be mimicked in the Monte Carlo analysis of the dilution factor (Section~\ref{sec:monopole_and_quadrupole_measurements}). 
For each galaxy, both the position angle and the separation with respect to the origin are computed by \texttt{Astropy} to derive $x_i$ and $y_i$. The orientation of the major axis, $\theta_{\rm major}$, is then given as:

\begin{equation}
    \tan(2\theta_{\rm major}) = \frac{2I_{xy}}{I_{xx} - I_{yy}}.
    \label{eq:position_angle}
\end{equation}

Next, we rotate the ellipticity $\boldsymbol{e}$ by the orientation angle of the major axis. Using complex notation, we denote the ellipticity components as $\boldsymbol{e}= e_{1} + ie_{2}$, 
then the rotated ellipticity is
\begin{equation}
    \begin{aligned}
    \boldsymbol{e}^{\rm rot}&= e_{1}^{\rm rot} + ie_{2}^{\rm rot}\\
    &= \boldsymbol{e} \exp(-2i\theta_{\rm major})\\
    &= (e_{1}\cos2\theta_{\rm major} + e_{2}\sin2\theta_{\rm major}) + \\
    &\qquad i(-e_{1}\sin2\theta_{\rm major}+e_{2}\cos2\theta_{\rm major}).
    \end{aligned}
\end{equation}

Though we use the galaxy distribution's major axis orientation as the cluster orientation on PoS, the CG's orientation can also be used as a proxy of the cluster orientation~\citep{Shin2018,Fu2024a,Fu2024b}. Additionally, we expect a misalignment between the galaxy distribution and the true halo orientation. This can result in a dilution of the stacked cluster ellipticity measurement. We use Monte Carlo simulation to estimate this dilution factor as described in Section~\ref{sec:monopole_and_quadrupole_measurements}. We also test the misalignment between the galaxy distribution and the true halo orientation in the cosmological simulation \skysim{}~(Appendix~\ref{app:sim_orientation_comparison}).

\subsection{Cluster sample division}\label{sec:cluster_sample_division}

The DES Y3 cluster sample (i.e., the baseline sample) is described in Section~\ref{sec:cluster_catalog}. We further divide the baseline sample by redshift and richness. The motivation is to study how the projected halo shape changes with redshift and richness (a proxy for halo mass). The clusters span a range of redshifts in [0.2, 0.65) and richness in [20, $\infty$). We split the whole sample once at a redshift of $z_{\rm split} = 0.4$, and once at a richness value of $\lambda_{\rm split} = 32$. The split locations are chosen to achieve approximately equal S/N of the stacked WL measurements in both bins~\citep{Fu2024a}; the sums of the richness in the two richness bins are comparable. We have 3231 (5845) clusters in the low (high) redshift bins and 5968 (3108) clusters in the low (high) richness bins. The halo ellipticity measurements for the cluster sub-samples are presented in Section~\ref{sec:e_z_lambda}. 

\section{Results}
\label{sec:results}

\subsection{Ellipticity of the baseline cluster sample (DES Y3 \rdmp{})}
\label{sec:ellipticity_baseline}

\begin{figure*}[htb!]
    \plotone{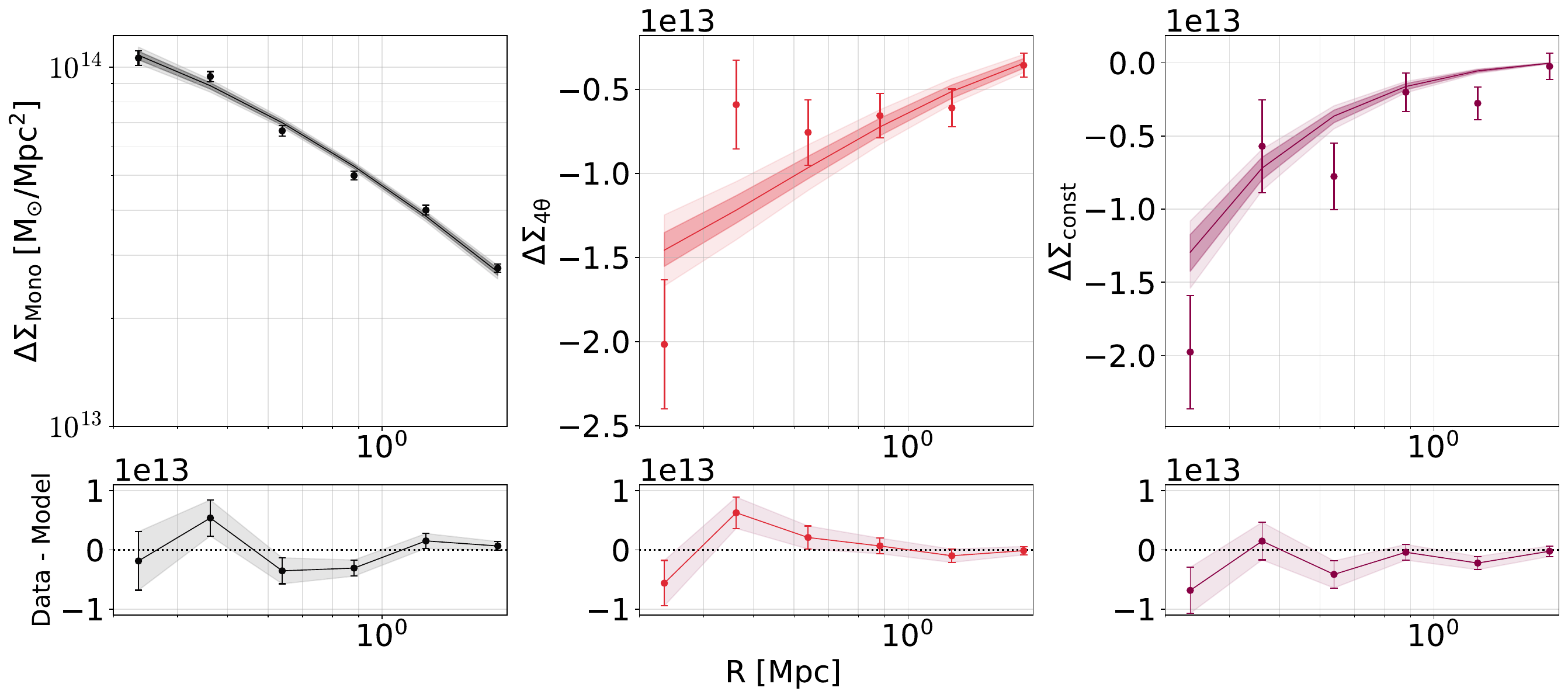}
    \caption{Fitting results for the stacked lensing monopole (\textit{Left}) and quadrupole (\textit{Middle} and \textit{Right}) profiles of the baseline cluster sample (selected by $\texttt{pcen0}>0.9$ and $0.2<z<0.65$ in DES Y3 \rdmp). 
    The fitted parameters are shown in Figure~\ref{fig:Y3RM_zAll_lAll_posterior}. The projected ellipticity of the cluster sample is $0.307^{+0.013}_{-0.014}$.
    \textit{Top}: Points with error bars represent the measured data (boost-factor corrected). The solid curves show the best-fit model predictions (dilution factor included). The shaded regions indicate the $1\,\sigma$ (dark) and $2\,\sigma$ (light) confidence intervals. \textit{Bottom}: Residuals between the data and model predictions. 
    }
    \label{fig:Y3RM_zAll_lAll}
\end{figure*}

\begin{figure*}[htb!]
    \plotone{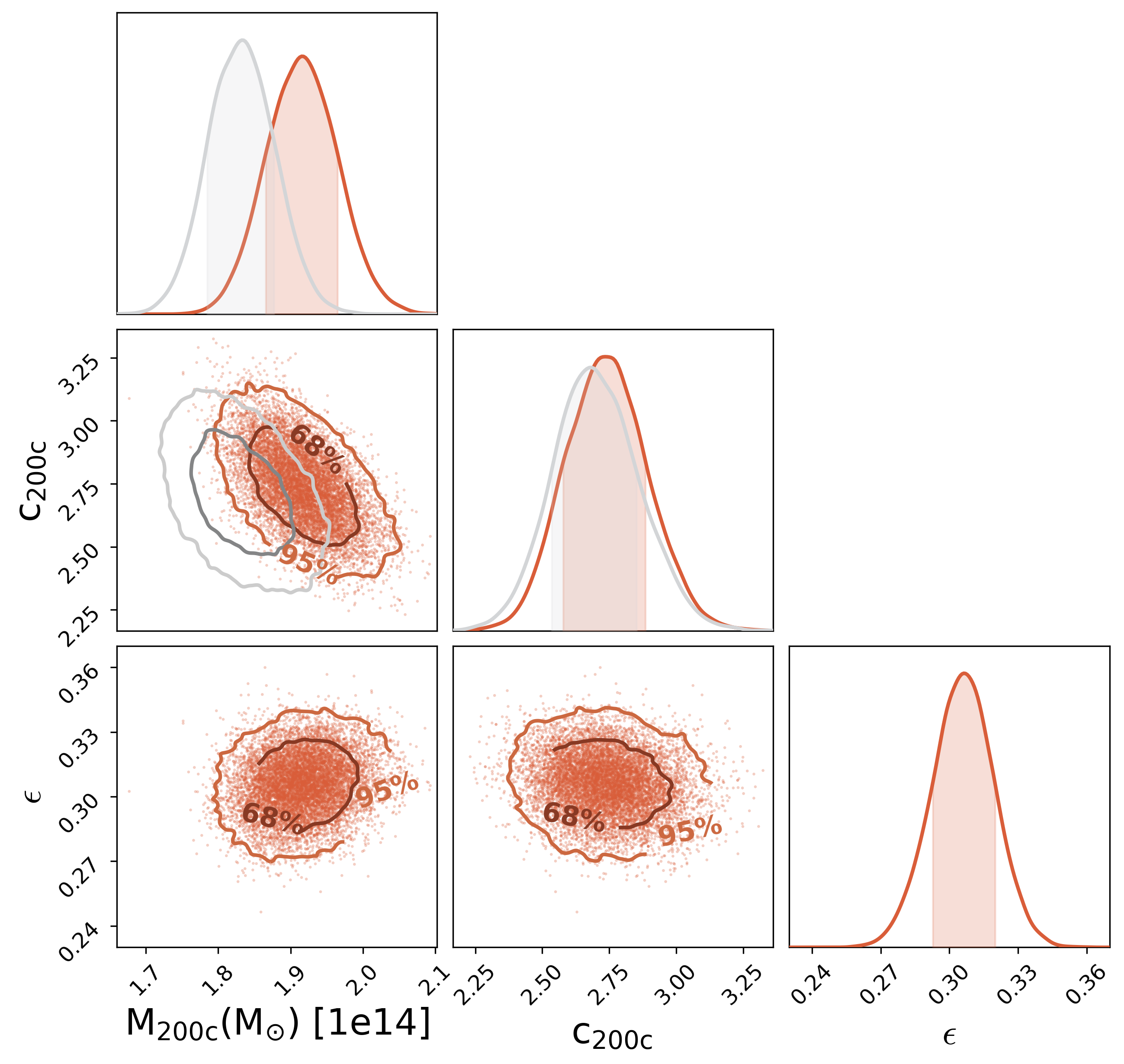}
    \caption{
    Posterior of the stacked lensing monopole and quadrupole simultaneous fit shown in Figure~\ref{fig:Y3RM_zAll_lAll}. The model parameters are mass ($M_{\rm 200c}$ in the unit of $M_{\odot}$), concentration ($c_{\rm 200c}$), and ellipticity ($\epsilon$). The marginalized distribution of each parameter is given, with the shaded area showing $1\,\sigma$. In gray we also show the fit to the mass and concentration assuming only a spherical NFW model. The mass change is $-4.4\,\%$ (see Section~\ref{sec:dis_meaning_of_mass} for discussions).}
    \label{fig:Y3RM_zAll_lAll_posterior}
\end{figure*}

We have described the DES Y3 \rdmp{} cluster sample in Section~\ref{sec:data}. For the baseline sample, we stack all the clusters. We fit the quadrupole and monopole excess surface density profiles simultaneously using the likelihood in Eq.~\ref{eq:likelihood}. 
We obtain the best-fit mass ${M_{\rm 200c}} = 1.915^{+0.048}_{-0.051}\times10^{14} \rm M_{\odot}$,  concentration $c_{\rm 200c} = 2.74^{+0.16}_{-0.16}$, and halo ellipticity $\epsilon = 0.307^{+0.013}_{-0.014}$ ($\sim22\,\sigma$ detection of a non-zero ellipticity) or axis ratio $q = 0.530^{ + 0.016}_{ - 0.015}$ for the baseline sample. Here we report the mean of the 1-D marginalized posterior distribution as the best-fit parameter value, and we use the 16th, 84th percentiles to calculate the $1\,\sigma$ uncertainties. 
Figure~\ref{fig:Y3RM_zAll_lAll} shows the best-fitting results for $\Delta\Sigma_{\rm mono}$, $\Delta\Sigma_{4\theta}$, and $\Delta\Sigma_{\rm const}$ with the $1\,\sigma$ and $2\,\sigma$ confidence intervals. We find a good fit to the data for monopole and quadrupole profiles in the 1-halo regime between 0.3 and 2.0~Mpc, divided into 6 logarithmic bins. The confidence bands are computed using the posterior samples shown in Figure~\ref{fig:Y3RM_zAll_lAll_posterior}.\footnote{In Figure~\ref{fig:Y3RM_zAll_lAll_posterior}, the $1\,\sigma$ ($68\,\%$) region in the 2D posterior scatter plot is slightly wider than the one of the 1D marginalized distribution; this is caused by the marginalization. 
}
That figure also shows the fit to the mass and concentration with only a spherical model,  and the result is comparable to previous studies of similar samples~\citep{Medezinski2018}.
We note that if we do not consider the dilution factor, the best-fit result yields an axis ratio of $q\sim0.6$. 
To investigate these results further, we fit the cluster samples when splitting them by cluster properties such as redshift and richness.

\subsection{Ellipticity as a function of cluster properties}\label{sec:e_z_lambda}

\begin{figure*}[htb!]
    \plotone{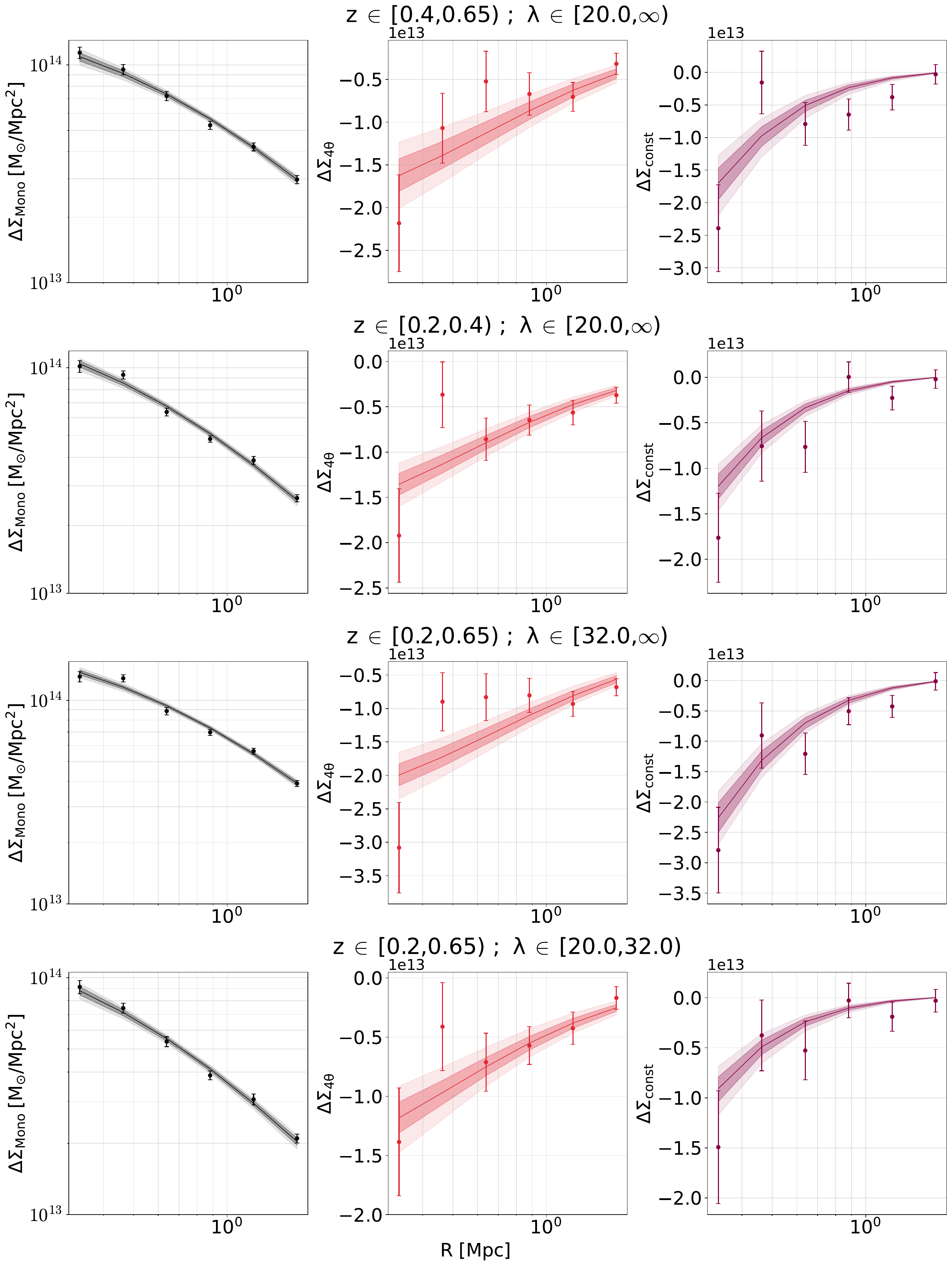}
    \caption{Fitted stacked profiles of the selected DES Y3 \rdmp{} clusters when split by redshift or richness. The dark (light) shaded area shows the $1\,\sigma$ ($2\,\sigma$) of the confidence interval. The data have been corrected for boost factors and the models include dilution factors. \textit{Top two}: High and Low redshift bins. \textit{Bottom two}: High and Low richness bins. 
    The fitted ellipticities are shown in Table~\ref{tab:Y3RM_e_summary}. 
    }
    \label{fig:Y3RM_profile_vs_lz}
\end{figure*}

We study the dependence of halo ellipticity on redshift and richness (mass proxy) by splitting each cluster property into two bins (high and low). These samples are defined in Section~\ref{sec:cluster_sample_division}. We use the same fitting strategy as the one for the baseline sample. The results are shown in Figure~\ref{fig:Y3RM_profile_vs_lz}. The data are well described by the best-fit model profiles across all the cluster sub-samples. The mean and uncertainties of the derived ellipticities are presented in Table~\ref{tab:Y3RM_e_summary}. 

\begin{figure}[htb!]
    \plotone{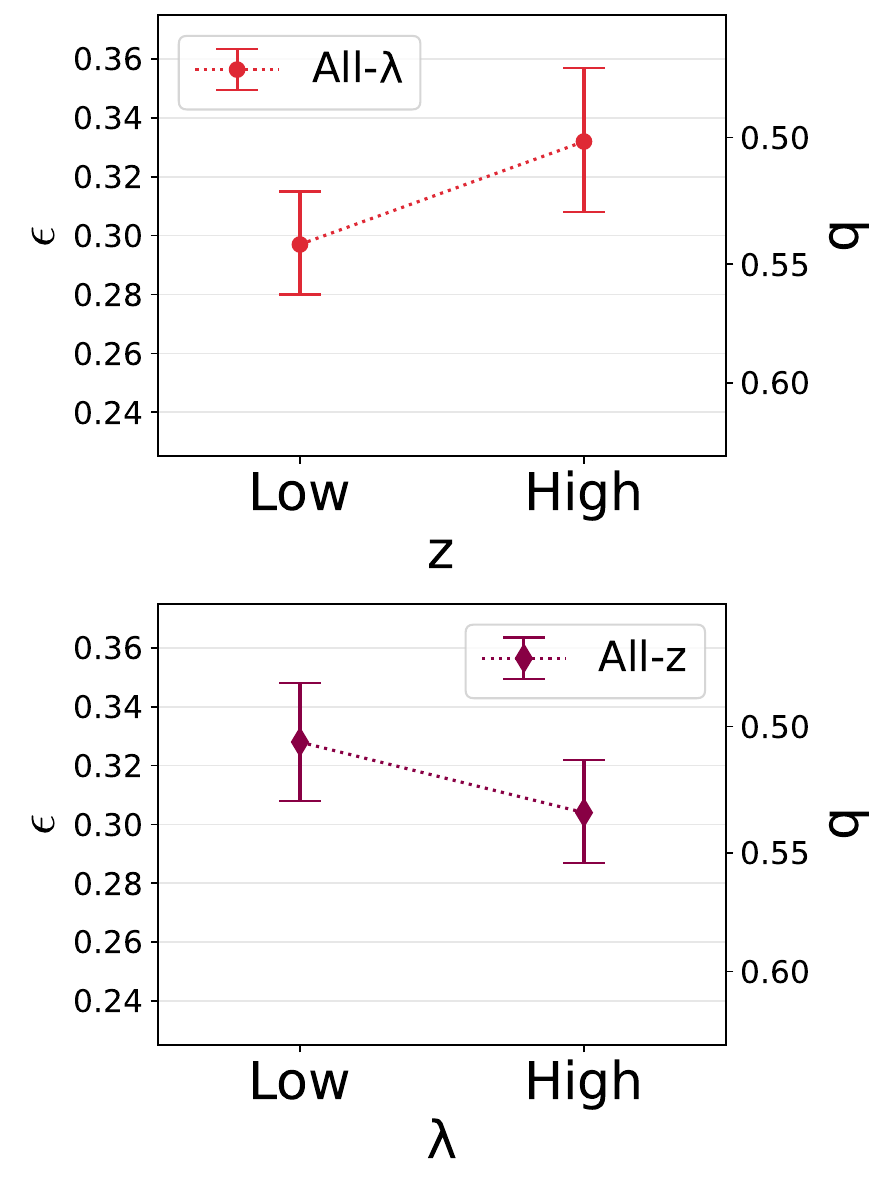}
    \caption{Halo ellipticity ($\epsilon$) and axis ratio ($q$) as a function of redshift (\textit{Top}) or richness (mass proxy) (\textit{Bottom}) for the selected DES Y3 \rdmp{} clusters.}
    \label{fig:Y3RM_e_vs_lz}
\end{figure}

To aid visualization, we also summarize the halo ellipticities of those bins in Figure~\ref{fig:Y3RM_e_vs_lz}. In this figure, the top panel shows the low- and high-redshift cluster samples, where we observe that the ellipticities show no clear trend in redshift (difference $\sim1.2\,\sigma$). This indicates that the redshift evolution of the halo ellipticity is minimal. In the bottom panel, between the richness-split samples, we observe a small difference as well ($\sim0.9\,\sigma$), which suggests that the halo ellipticity has no strong dependence on richness (a proxy for halo mass).  
With forthcoming larger datasets and refined modeling of systematics, we will be able to test these results with higher statistical significance. We further discuss these results in Section~\ref{sec:comparison_studies}.

\begin{table}[htb!]
    \centering
    \begin{tabular}
    {c|c|c}
    Sample & Ellipticity $\epsilon$ & Axis ratio $q$ \\
    \hline
    Baseline & $0.307^{ + 0.013}_{ - 0.014}$ & $0.530^{ + 0.016}_{ - 0.015}$ \\
    High redshift ($z > 0.4$) & $0.332^{ + 0.025}_{ - 0.024}$ & $0.502^{ + 0.027}_{ - 0.029}$ \\
    Low redshift ($z < 0.4$) & $0.297^{ + 0.018}_{ - 0.017}$ & $0.543^{ + 0.021}_{ - 0.021}$ \\
    High richness ($\lambda > 32$) & $0.304^{ + 0.018}_{ - 0.017}$ & $0.534^{ + 0.020}_{ - 0.021}$ \\
    Low richness ($\lambda < 32$) & $0.328^{ + 0.020}_{ - 0.020}$ & $0.506^{ + 0.022}_{ - 0.023}$ \\
    \end{tabular}
    \caption{Summary of ellipticity measurements of different cluster sample divisions. The central value of each $q$ is derived from the mean of the corresponding $\epsilon$, while the error bar of $q$ is computed from the MCMC posterior samples.
    }
    \label{tab:Y3RM_e_summary}
\end{table}

\subsection{Tests on DES Y1 \rdmp{} }\label{sec:des_y1_rm}

\begin{figure*}[htb!]
    \centering
    \includegraphics[width=.8\linewidth]{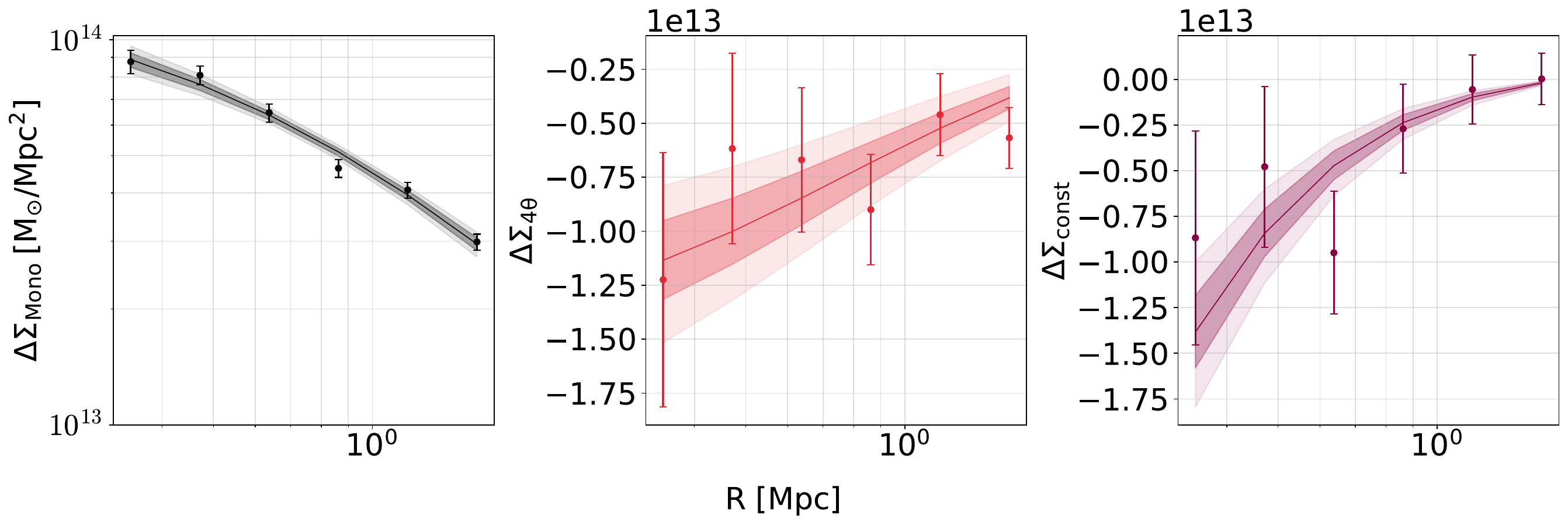}
    \caption{Fitting results for the stacked lensing monopole and quadrupole profiles of clusters with $\texttt{pcen0}>0.9$ and $0.2<z<0.65$ in DES Y1 \rdmp{}. The points with error bars represent the data. The solid curves show the best-fit model predictions. The shaded regions indicate the $1\,\sigma$ (dark) and $2\,\sigma$ (light) confidence intervals. Like Figure~\ref{fig:Y3RM_zAll_lAll} (baseline sample), we take into account the dilution factor. 
    }
    \label{fig:Y1RM_zAll_lAll}
\end{figure*}

As a validation test for the robustness of our measurement, we study the ensemble ellipticity of the DES Y1 \rdmp{} clusters~\citep{McClintock2019}\footnote{\url{https://des.ncsa.illinois.edu/releases/y1a1/key-catalogs/key-redmapper}} using the Y3 shapes and photo-zs (same as those for Y3 clusters above), and compare our mass estimates when binned in redshift and richness to the results of~\citet{McClintock2019}. Here, we also employ the per-HEALpix Response method (i.e., the second method described in Section~\ref{sec:shear_measurements}) and test its robustness when compared to the more accurate but compute-heavy, per-object response performed on the DES Y3 \rdmp{} clusters. We use the method in Table 7 of~\citet{Rykoff2016} to compute $P_{\rm mem}$ for determining the major-axis orientation and the dilution factor. We select clusters with $\texttt{pcen0}>0.9$ and $0.2<z<0.65$ (the default $\lambda>20$) in Y1 \rdmp{}, which gives 3233 clusters out of 6729, and we stack their lensing data. We show the fits to the monopole and quadrupole measurements in Figure~\ref{fig:Y1RM_zAll_lAll}; for simplicity, the error bars of individual profiles are estimated by bootstrapping. We find a stacked cluster ellipticity of $\epsilon = 0.309^{+0.027}_{-0.027}$ ($q=0.529^{+0.031}_{-0.031}$) after taking into account the dilution factor, 
which is consistent with our findings from the Y3 \rdmp{} clusters, albeit we have not taken into account the boost factor here. The boost factor affects both the monopole and quadrupole and thus has a minor impact on the ellipticity measurement (Section~\ref{sec:monopole_and_quadrupole_measurements}); we skip it in this demonstration. Additionally, the different measurement methods of richness and redshift in Y1 may also affect this result~\citep{DES2025}. Nonetheless, we find that the uncertainty of the Y1 \rdmp{} cluster ellipticity is $\sim\sqrt{3}$ times that of Y3 (Section~\ref{sec:ellipticity_baseline}); this is consistent with the number of clusters (the ratio between Y3 and Y1 is $\sim3$), and the footprint of Y3 is $\sim3$ times that of Y1 with similar depth. 

Furthermore, we split the clusters into redshift and richness bins following~\citet{McClintock2019}. This allows us to compare the mass estimate from the triaxial-model fitting to the spherical-model fit results. We caution the readers that the mass reported in this work is the mass within a 3D iso-density ellipsoidal envelope centered at the cluster (Figure~\ref{fig:Illustration}) and is not directly comparable to a spherical overdensity mass as reported by~\citet{McClintock2019}; in Section~\ref{sec:dis_meaning_of_mass} we further discuss the mass measurements. Nevertheless, we perform this analysis to confirm the robustness of our models. We measure a fractional mass difference $(M_{\rm 200m, \,this\; work} - M_{\rm 200m,  \,M19})/M_{\rm 200m, \,M19}$ in each redshift-richness bin and report it in Table~\ref{tab:Y3RM_McClintock_mass_bias}. Here, ``$\rm{200m}$'' means that the mean overdensity of the enclosed mass is 200 times the mean matter density of the universe at that redshift. The error bars in the lensing profiles are computed using bootstrapping and the uncertainties in the fitted parameters (NFW mass and concentration) are inferred from the MCMC chains. Then, we use error propagation to compute the uncertainties in this table. 
The mass difference is at the order of $\sim10\,\%$. We do not draw strong conclusions as we have not yet considered many of the systematic biases that \citet{McClintock2019} have taken into account, e.g., boost factor, miscentering, 2-halo modeling, etc. However, this test is sufficient to validate the calculations presented in this work.

\begin{table*}[htb!]
    \centering
    
    \begin{tabular}{c|c|c|c}
       Mass difference ($\%$) & $z \in [0.2,0.35)$ &$z \in [0.35,0.5)$ & $z \in [0.5,0.65)$ \\
     \hline
     $\lambda \in [20,30)$ & $-0.5\pm 11.2$ & $-5.1 \pm 8.4$ & $17.8 \pm 10.2$ \\
     $\lambda \in[30,45)$ & $-7.5\pm 7.2$ & $-8.4 \pm 8.3$ & $-12.8 \pm 11.9$ \\
     $\lambda \in[45,60)$ & $-5.1\pm 9.9$ & $-3.3 \pm 9.7$ & $-15.2 \pm 14.5$ \\
     $\lambda \in[60,\infty)$ & $-1.9\pm 9.9$ & $-8.8 \pm 8.9$ & $-6.4 \pm 14.7$ \\
    \end{tabular}
    \caption{Mass difference tests of the DES Y1 \rdmp{} cluster samples (\texttt{pcen0} $> 0.9$) split according to~\citet[][M19]{McClintock2019}. The mass difference percentage is calculated as ${(M_{200m, \rm \,this\; work} - M_{200m, \rm \,M19})}/{M_{200m, \rm \,M19}}\times100$.}
    \label{tab:Y3RM_McClintock_mass_bias}
\end{table*}

\section{Discussion}
\label{sec:discussion}

In this section, we first verify our analysis, then discuss the insights that the above results can offer, and finally give an outlook for upcoming datasets. 

\subsection{Ellipticity measurement bias}\label{sec:cluster_ellipticity_measurement_bias}
 
\subsubsection{Accuracy of the Taylor expansion}\label{sec:dis_accuracy_taylor}

\begin{figure*}[htb!]
    \plotone{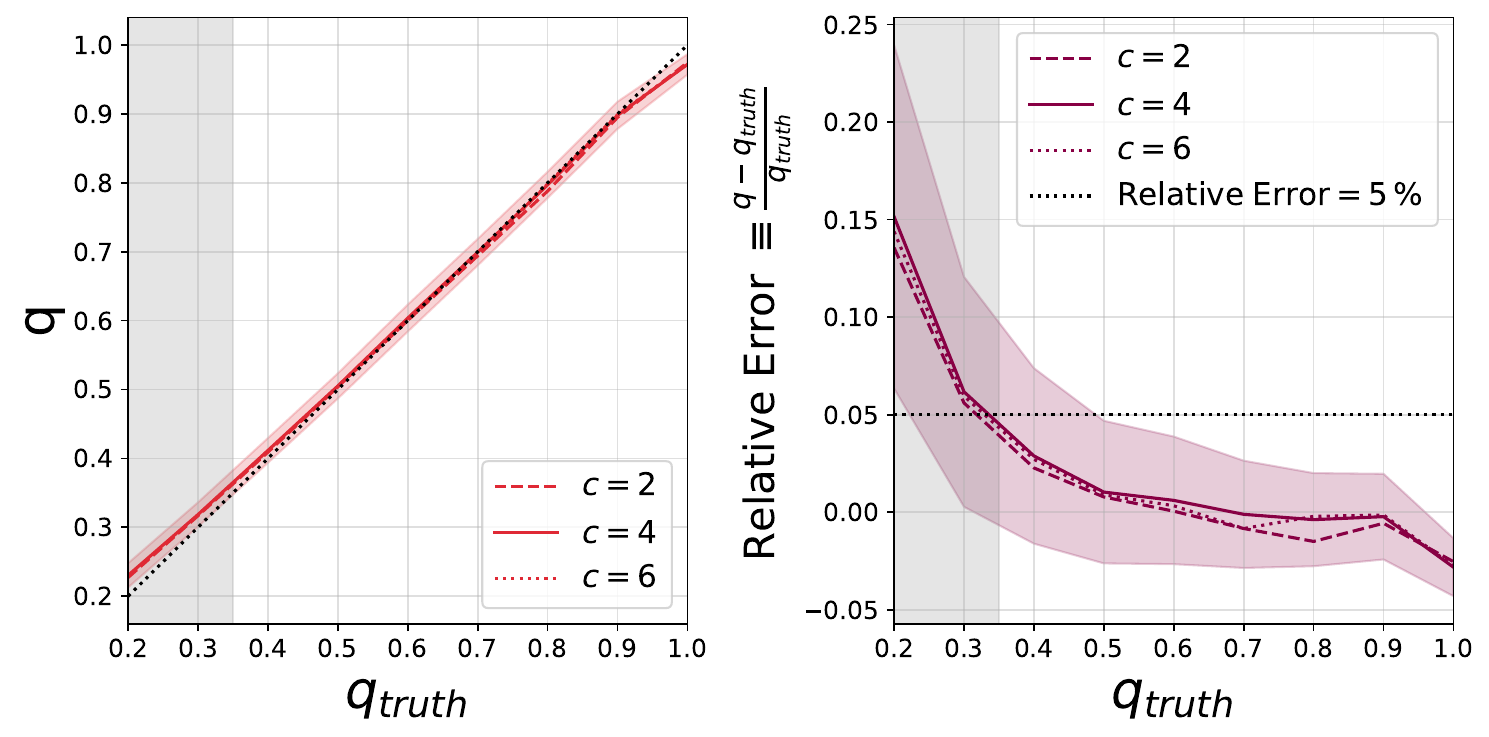}
    \caption{Comparison between the true and fitted axis ratios derived from mock catalogs with different concentration values. The red-dashed, solid, and dotted lines correspond to $c_{\rm 200c}=2$, 4, and 6, respectively.
    \textit{Left}: Fitted axis ratio ($q$) vs. true value ($q_{\rm truth}$). \textit{Right}: Relative error ($\Delta q/q_{\rm truth}$) vs. $q_{\rm truth}$. The shaded areas around the lines represent the $1\,\sigma$ confidence interval for the mock clusters with the concentration 4. The gray bands mark the cases when the recovered axis ratio has a bias $\gtrsim 5\,\%$ compared to the truth (i.e., $q_{\rm truth} < 0.35$).} 
    \label{fig:q_true_vs_measure}
\end{figure*}

We begin by testing the accuracy of our ellipticity measurement. 
We build \textit{mock} shear catalogs of elliptical halos with known shapes (Section~\ref{sec:mock}) and feed them into our analysis pipeline, so that we can directly compare the input and output halo ellipticities. The expansion in Eq.~\ref{eq:sigma0_sigma2} is expected to accumulate errors when the ellipticity is large or the axis ratio is small. 
As mentioned in Section~\ref{sec:mock}, we generate a set of mock clusters with different axis ratios, with a step size of 0.1 in the range of [0.1, 1.0], to determine the bias in the measurement as a function of the input $q$. 

We evaluate estimators using the mock data (replacing $e$ with $\gamma$ in the estimators without weights) and then fit them with the same model (Section~\ref{sec:lensing_analysis}). We use the same radial range as in the observational analysis above, but adopt narrower radial bins (14 bins) given the higher source number density and the absence of shape noise in the mock data.

The result is presented in Figure~\ref{fig:q_true_vs_measure}. 
In the left panel, we show a direct comparison between the true axis ratio and the fitted result. 
In the right panel, we show the relative error of the axis ratio as a function of the true value. We find that the bias is $\lesssim 5\,\%$ when $q_{\rm truth} > 0.35$. For all of the cluster samples presented in this work, the recovered axis ratio is well above this limit. 
Also, we do not see clear differences among the results of concentration from 2 to 6, indicating that this analysis is not sensitive to halo concentration. The fitted mass is also consistent with the input (difference $<1\,\%$ when $q>0.4$). 
In addition, we note that when $q_{\rm truth}\in[0.6,\,1)$ the error is $<1\,\%$, but the offset becomes larger at $q_{\rm truth}=1$. This is caused by the prior $q\in[0,1]$ that we set to avoid unphysical values. The posterior distribution of $q$ peaks at 1, but the tail below 1 leads to the mean below 1. Using a higher resolution grid in the mock can make the posterior distribution tighter and thus reduce this effect, but cannot eliminate it. 

\subsubsection{Meaning of the mass measurement}\label{sec:dis_meaning_of_mass}

Next, we interpret the meaning of the mass measurement in our analysis and the insight it provides us. Note that the mass measurement is not the focus of this paper. 
As mentioned earlier, our fitted mass corresponds to the mass for the underlying spherical profile $\Sigma_{\rm sph}$ in Eq.~\ref{eq:sigma0_sigma2}, and it is the same as the mass of the 3D ellipsoid after the volume-preserving deformation, which corresponds to the effective 3D shape of the cluster (Figure~\ref{fig:Illustration}). The fitted concentration corresponds to the concentration of $\Sigma_{\rm sph}$ before the deformation.

\begin{figure}[htb!]
    \plotone{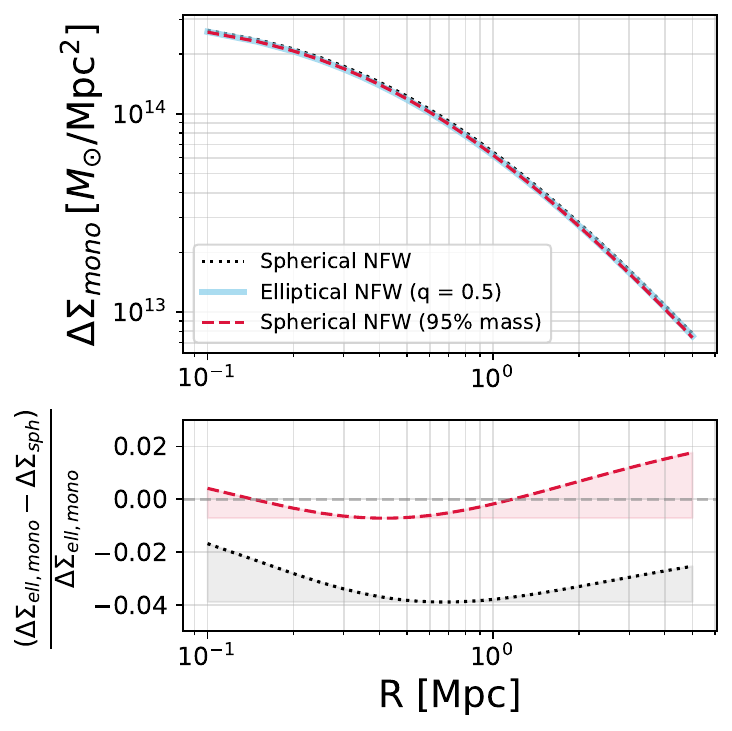}
    \caption{\textit{Top}: Monopole comparison between the elliptical (blue solid) and spherical NFW profiles (black dotted) of clusters with the same mass, $M_{\rm 200c} = 2 \times 10^{14}M_{\odot}$ and concentration, $c_{\rm 200c} = 4$ at redshift $z = 0.4$. 
    We also show the profile of a spherical cluster with $95\,\%$ of the mass  (crimson dashed), which mostly overlaps with the blue curve. \textit{Bottom}: Fractional difference between the monopoles of the elliptical and the two spherical profiles. We find that the elliptical NFW profile's monopole (black dotted) is $\sim3\,\%$ lower when compared to the spherical NFW profile. Equivalently, this translates to a $\sim-5\,\%$ difference in the mass of the spherical NFW profile (crimson dashed).}
    \label{fig:monopole_comparison}
\end{figure}

When we compare the $\Sigma_{\rm sph}$ term and the monopole $\Sigma_0$ in Eq.~\ref{eq:sigma0_sigma2}, 
we find that accounting for the triaxial shape reduces the WL mass by a few percent because of the $\epsilon^2$ term 
--- In Figure~\ref{fig:monopole_comparison}, we show the comparison between the monopole signals when measured using the spherical and elliptical models for a representative elliptical NFW cluster at redshift 0.4 with axis ratio $q=0.5$, mass $M_{\rm 200c} = 2 \times 10^{14}M_{\odot}$, and concentration $c_{\rm 200c} = 4$. As shown in the residuals in the bottom panel of Figure~\ref{fig:monopole_comparison}, the difference between the spherical NFW lensing profile and the elliptical NFW's monopole is on average $\sim3\,\%$. Furthermore, we show that a spherical cluster with $\sim5\,\%$ less mass agrees with the elliptical cluster's monopole profile. In spite of the simplicity of this model, we similarly find the mass difference is $-4.4\,\%$ in the observation (Figure~\ref{fig:Y3RM_zAll_lAll_posterior}).

When clusters are randomly stacked without their axes aligned, the quadrupoles (and higher-order multipoles) are averaged out and only the monopole $\Sigma_{0}$ remains.
If we consider $\epsilon$ as a measured effective ellipticity, then it becomes zero after the random stacking (only a circular profile remains), and then $\Sigma_0=\Sigma_{\rm sph}$. 
But if we think about a collection of clusters with the same shape and the intrinsic ellipticity is $\epsilon_{\rm int}$, then after the random stacking an $\epsilon_{\rm int}^2$ term is still in $\Sigma_0$. However, in that case, the mass in $\Sigma_{\rm sph}$ and $\epsilon_{\rm int}$ will be degenerate and difficult to constrain independently from the monopole alone. Thus, we can only obtain an effective $\Sigma_{\rm sph}$ without considering the intrinsic ellipticity, i.e., compared to the underlying truth, $\Sigma_{\rm sph}$ is biased by $\mathcal{O}(\epsilon_{\rm int}^2)$.

In cluster cosmology, the mass function of clusters uses a mass within a shell where the density is above a certain threshold, and the shell is usually assumed to be a sphere. Using an ellipsoidal shell to account for the triaxiality of clusters can be more realistic and potentially give better measurement.

\subsubsection{Effect of multiplicative systematics}
\label{sec:dis_multi_sys}

From this point on, we discuss the systematic effects on cluster ellipticity measurements. We note that systematics have been studied in cluster mass measurements~\citep[e.g.,][]{Applegate2014,McClintock2019}, but their impact on cluster ellipticity measurements remains to be further explored. 
Here we run tests on the mock (Section~\ref{sec:mock} at $c_{\rm 200c}=4$) to show the robustness of our measurement methods and results. We use a higher resolution mock ($5\times5$ of the original) to reduce the statistical uncertainty and to clarify the trend.

We first test the effect of the reduced shear factor $1/(1-\kappa)$, which leads to a $\lesssim4\,\%$ difference in the mock profiles within the radial range for analysis. We do this by replacing $\gamma_{1,2}$ with $g_{1,2}$ in the fitting input. The results in Table~\ref{tab:systematics} show that the bias in ellipticity when the data vector includes the effect of reduced shear is $0.5\,\%$ and the bias in axis ratio is $-0.4\,\%$. We conclude that the systematics caused by reduced shear are subdominant to the statistical uncertainties in the measurement.

Next, we test the effect of blending by including a multiplicative bias of shear ($-1\,\%$ and $-10\,\%$ for typical values) in both $\gamma_1$ and $\gamma_2$ in the fitting input. The result shows that the bias in ellipticity is less than $2\,\%$ in amplitude for both cases. Most of the bias is absorbed by the fitted mass and concentration, leaving the ellipticity relatively unaffected. 
Additionally, image simulation shows a $-2\,\%$ multiplicative bias for the shear catalog~\citep{Gatti2021,MacCrann2022}, and redshift systematics may also lead to a sub-percent level multiplicative bias~\citep{McClintock2019}. We similarly expect their affects on the ellipticity measurement to be minimal given the tests above.

\subsubsection{Effect of miscentering}
\label{sec:dis_miscenter}

\begin{figure}[htb!]
    \plotone{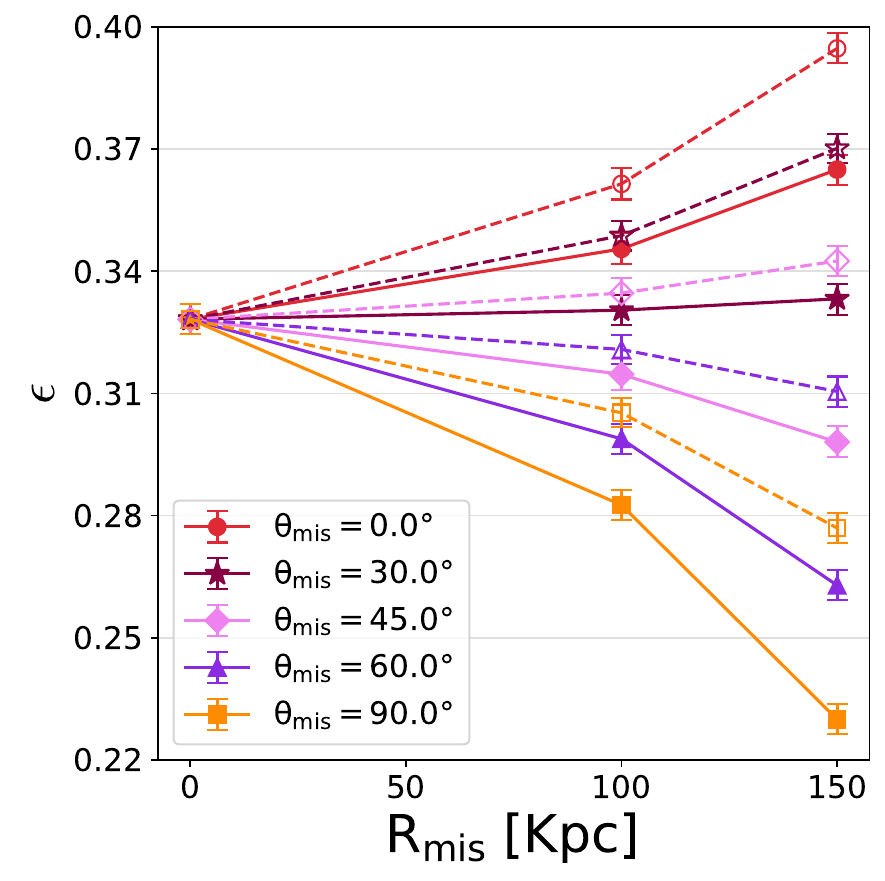}
    \caption{
    Fitted ellipticity for a miscentered elliptical cluster. We use the same mock cluster (Section~\ref{sec:mock}) with concentration $c_{\rm 200c} = 4$ but with a higher resolution  ($5\times5$ of the original) for illustrating the trend. The major axis of the cluster is aligned with the x-axis, whose positive direction corresponds to $\theta_{\rm mis} = 0$; the true center is at the origin.
    The solid curves show the cases when the measurement center is located on a circle: $(x_{\rm mis}, y_{\rm mis})=(R_{\rm mis}\cos\theta_{\rm mis}, R_{\rm mis}\sin{\theta_{\rm mis}})$. The dashed curves show the cases when the measurement center lies on an ellipse: $(x_{\rm mis}, y_{\rm mis})=(R_{\rm mis}/\sqrt{q_{\rm mis}}\cos\theta_{\rm mis},R_{\rm mis}\sqrt{q_{\rm mis}}\sin{\theta_{\rm mis}})$ with $q_{\rm mis} = 0.5$. 
    }
    \label{fig:efit_circular_miscentering}
\end{figure}

To test the effect of miscentering, we set the measurement center at a given offset from the true center and repeat the measurement. We consider typical offsets~\citep[$R_{\rm mis}$;][]{Zhang2019,Kelly2024} at various position angles ($\theta_{\rm mis}$); given the symmetry, we only consider the first quadrant. 
The result (Figure~\ref{fig:efit_circular_miscentering}) shows that the bias grows with $R_{\rm mis}$ when $\theta_{\rm mis}$ is fixed. On the other hand, when $R_{\rm mis}$ is fixed, the measured ellipticity is biased high when the offset is along the x-axis ($\theta_{\rm mis}=0^\circ$) and biased low when the offset is along the y-axis ($\theta_{\rm mis}=90^\circ$); the bias is between those two extreme values when $\theta_{\rm mis}$ is between $0^\circ$ and $90^\circ$. Interestingly, the biases show a symmetry with respect to the $\theta_{\rm mis}=45^\circ$ case when $R_{\rm mis}$ is fixed. Later in Appendix~\ref{app:derive_miscentering} we prove that the bias primarily depends on $\cos{2\theta_{\rm mis}}$, which explains the symmetry about $45^\circ$. 
Going further, in our stacked cluster lensing measurements, the measurement centers may be symmetrically distributed along a circle or ellipse centered at the true center. When the measurement centers are distributed along a circle after the axis-aligned stacking, our derivation shows that the mean bias should be close to the $45^\circ$ case ($-5.5\,\%$ at $R_{\rm mis}=100$~kpc; see Table~\ref{tab:systematics}). If after stacking the measurement centers are distributed along an ellipse aligning with the halo, the amplitude of the mean effect would be smaller (see also Appendix~\ref{app:derive_miscentering}). 
We expect this effect to be even smaller after our \texttt{pcen0} selection.

\subsubsection{Effect of LSS}\label{sec:dis_lss}

To test the effect of LSS, we include a 2-halo model from~\citet{Oguri2011}, which has been implemented in \clmm v2, to mimic the effects of correlated structures such as neighboring filaments and halos. The halo bias is chosen to be 3.63 corresponding to the halo mass for the mock~\citep{Tinker_2010,Diemer2018}.
Note, the effect of uncorrelated LSS has been addressed when we construct the covariance via sky jackknife in the real data analysis.

To simplify the modeling, we analytically compute the lensing effects of the 2-halo component via multipole expansion; we replace $\Sigma_{\rm 2h}(R)$ with $\Sigma_{\rm sph}(R)$ in Eq.~\ref{eq:sigma0_sigma2}, assuming that the 2-halo component has the same ellipticity as the 1-halo component.
We then update the shear and convergence from the original mock: the convergence is increased by $\Sigma_{\rm 2h}(\sqrt{qx^2+y^2/q})/\Sigma_{\rm crit}$; the shear is increased by the 2-halo's contribution of
$\gamma_1(R,\theta)=[-\Delta\Sigma_0(R)\cos{2\theta}+\Delta\Sigma_{4\theta}(R)\cos{4\theta}+\Delta\Sigma_{\rm const}(R)]/\Sigma_{\rm crit}$ and $\gamma_2(R,\theta)=[-\Delta\Sigma_0(R)\sin{2\theta}+\Delta\Sigma_{4\theta}(R)\sin{4\theta}]/\Sigma_{\rm crit}$
(Eq.~\ref{eq:models_12}).

We redo the fitting with the updated lensing profiles.
We find that the 2-halo term mainly affects $\Delta\Sigma_{\rm const}$, which is sensitive to external structure. As the fit to ellipticity requires constraints from both quadrupole estimators, the change on the ellipticity is small; the result shows the ellipticity bias is at the sub-percent level.

\begin{figure*}[htb!]
    \centering
    \plotone{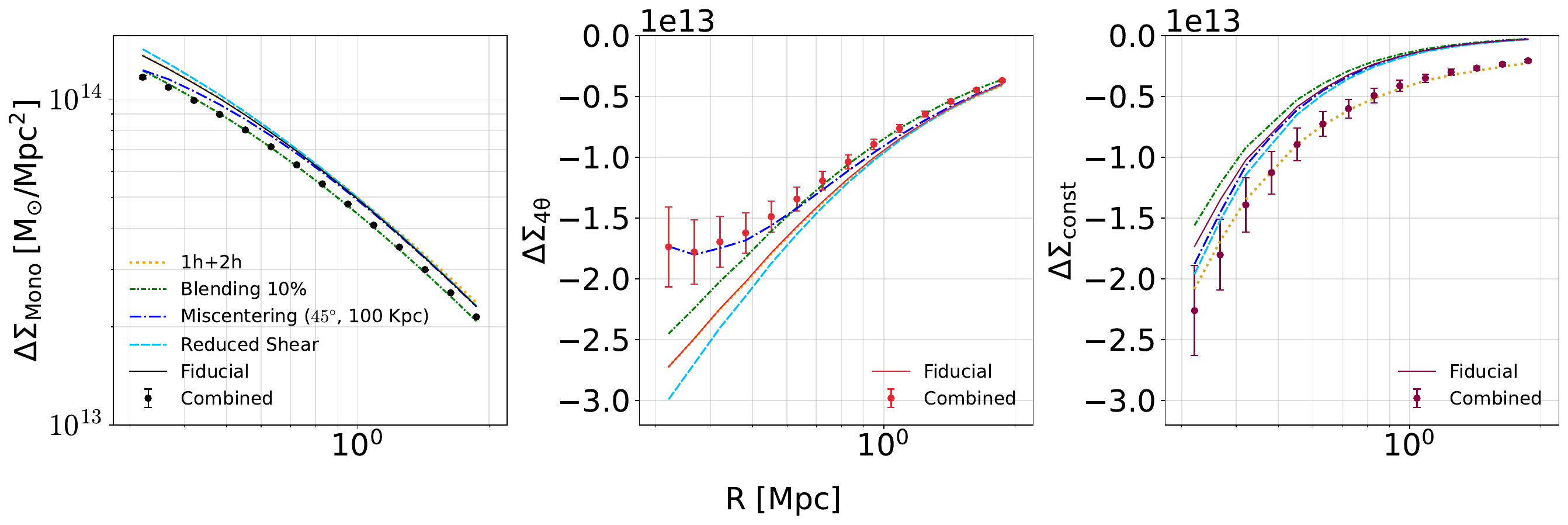}
    \caption{Monopole and quadrupole profiles of the mock after including the systematics. In each panel, the solid line shows the fiducial case, while the orange dotted line, the green short-dash-dot line, the blue long-dash-dot line, and the light blue dashed line, show the cases when individual systematics are included --- 2-halo term, blending, miscentering, and reduced shear, respectively; the points show the case when including all those systematics. 
    For visualization, we only keep the error bars of the ``Combined'' case here.
    }
    \label{fig:profile_systematics}
\end{figure*}

The systematic effects are summarized in Table~\ref{tab:systematics}. 
Most systematics are $\lesssim 1\,\%$ in amplitude, and only the miscentering causes a significant bias (on average $-5.5\,\%$ in ellipticity or $4.2\,\%$ in axis ratio).
We list the results of the ``fiducial'' run as well for reference, where the same high resolution mock has been used but no extra systematic effect has been introduced to the mock.  
The systematic bias in the fiducial comes from the Taylor expansion, the multipole expansion, and the radial binning, and it is included in all systematic estimates. 
Combining the effects of blending (10\,\%), reduced shear, miscentering $(45^{\circ}, 100\,{\rm kpc})$, and 2-halo, yields a $\sim -3\,\%$ systematic bias of ellipticity ($\sim 2\,\%$ for axis ratio), when they are all included in the mock (Figure~\ref{fig:profile_systematics}). 
A more detailed assessment requires cosmological simulations and will be addressed in future work (Section~\ref{sec:future_work}).

\begin{table*}[htb!]
    \centering
    \begin{tabular}{c|c|c|c|c}
        Systematics & $\epsilon$ & $ \epsilon/\epsilon_{\rm truth} -1$ & $q$ & $q/q_{\rm truth}-1$  \\
        \hline
        Fiducial & $0.328\pm0.004$ & $-0.016\pm0.012$ & $0.506\pm0.005$ & $0.012\pm0.010$\\
        Blending 1\,\% & $0.328\pm0.004$ & $-0.016\pm0.012$ & $0.506\pm0.005$ & $0.012\pm0.010$\\
        Blending 10\,\% & $0.329\pm0.004$ & $-0.013\pm0.012$ & $0.505\pm0.005$ & $0.010\pm0.010$\\
        Reduced shear & $0.335\pm0.004$ & $0.005\pm0.012$ & $0.498\pm0.004$ & $-0.004\pm0.008$\\
        Miscentering $(0^{\circ},100\,{\rm kpc})$ & $0.345\pm0.004$ & $0.035\pm0.012$ & $0.487\pm0.004$ & $-0.026\pm0.008$\\
        Miscentering $ (45^{\circ}, 100\,{\rm kpc})$ & $0.315\pm0.004$ & $-0.055\pm0.012$ & $0.521\pm0.005$ & $0.042\pm0.010$\\
        Miscentering $ (90^{\circ}, 100\,{\rm kpc})$ & $0.283\pm0.004$ & $-0.151\pm0.012$ & $0.559\pm0.005$ & $0.118\pm0.010$\\
        2-halo & $0.331\pm0.004$ & $-0.007\pm0.012$ & $0.503\pm0.005$ & $0.006\pm0.010$\\
        *Combined & $0.324\pm0.004$ & $-0.028\pm0.012$ & $0.511\pm0.005$ & $0.022\pm0.010 $\\
    \end{tabular}
    \caption{Summary of systematic effects on the halo shape measurement (see also Figure~\ref{fig:profile_systematics}). *Combined: Result from including the systematic effects of blending (10\,\%), reduced shear, miscentering $(45^{\circ}, 100\,{\rm kpc})$, and 2-halo in the mock.}
    \label{tab:systematics}
\end{table*}

\subsubsection{Null tests}\label{sec:dis_null_test}

To further assess potential systematics in the measurement, we perform null tests by building estimators for $\Delta\Sigma_{\rm x}$, $\Delta\Sigma_{\rm 4\theta;x}$, $\Delta\Sigma_{\rm const;x}$ (Appendix~\ref{app:null_test}) with the same treatment of weight, response, and critical surface density in Section~\ref{sec:shear_measurements},~\ref{sec:monopole_and_quadrupole_measurements}; boost factors are no longer needed for them. 
We compute the  $\chi^2$ values using the data vector and the covariance matrix of the baseline cluster sample (Section~\ref{sec:ellipticity_baseline}). The three profiles have $\chi^2/\mathrm{dof}=2.9/6,\, 8.1/6,\, 6.6/6$, corresponding to $p$-values of 0.83, 0.23, 0.36, respectively, which are all consistent with the null hypothesis. Combining the three profiles and using the full covariance, we obtain $\chi^2/\mathrm{dof}=20.8/18$ ($p=0.29$), which indicates no significant deviation from zero.

\subsection{Ellipticity as a function of radial distance}\label{sec:galaxy_radial_number_density}

When we perform the multipole expansion and the Taylor expansion, we assume that the ellipticity is nearly constant along the radius. This has been tested in simulation~\citep[e.g.,][]{Adhikari2015}. In observation, if we assume that galaxies trace the underlying dark matter distribution, we can test whether the ellipticity of the galaxy number density contour is constant across different radii. It is also useful to compute an ellipticity from the galaxy distribution to compare that with the ellipticity derived from the lensing measurement. The \rdmp{} catalog provides a member galaxy catalog, but it only includes members within a radius of $R_{\lambda}$. If we use this catalog, we would likely produce a biased ellipticity measurement since cluster member  galaxies lying on the outskirts of the elliptical distribution are lost due to the circular cut.

Therefore, we select the \textsc{MagLim} galaxy catalog from DES Y3 to test the number density of galaxies~\citep{Porredon2021}. 
This catalog consists of galaxies selected by a general cut on the $i$-band magnitude using the DNF photo-z, $i < 4\,z_{\rm DNF} + 18$. And, each galaxy has a weight that accounts for the survey variance. 
This catalog does not include any specific cuts on color and has been previously used in galaxy clustering and galaxy-galaxy lensing studies~\citep{Porredon2022}. 
For clusters, the \textsc{MagLim} catalog consists of both the foreground/background galaxies and the cluster galaxies. After stacking a large ensemble of clusters, we expect the foreground/background galaxies to follow a constant number density, 
while the cluster-associated galaxies would exhibit a concentration around the cluster center. 

Using the Y3 \rdmp{} cluster sample with the centering probability cut and the redshift cut as mentioned in Section~\ref{sec:cluster_catalog}, we look for galaxies up to 15 Mpc away from the center of each cluster on the PoS. Then we rotate their coordinates by the major axis angle (derived from the member galaxy distribution; Section~\ref{sec:halo_orientation}), so that the major axis aligns with the new x-axis after the rotation.\footnote{As noted by~\citet{Lokken2025}, the \textsc{MagLim} catalog can also be used to determine the cluster/filament orientation.}
We omit the $P_{\rm mem}$ cut for simplicity, as its effect on the result is minor.
Next, we stack the rotated coordinates from different clusters together, and we compute the weighted number count in each pixel of a grid applied to a  $10\times10$~Mpc window. 
 
\begin{figure*}[htb!]
    \plottwo{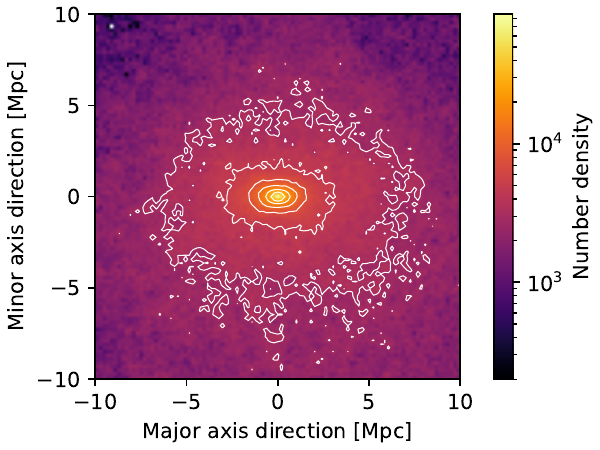}{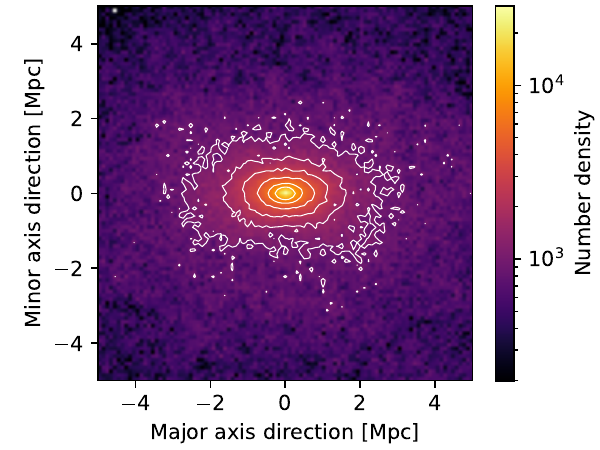}
    \plottwo{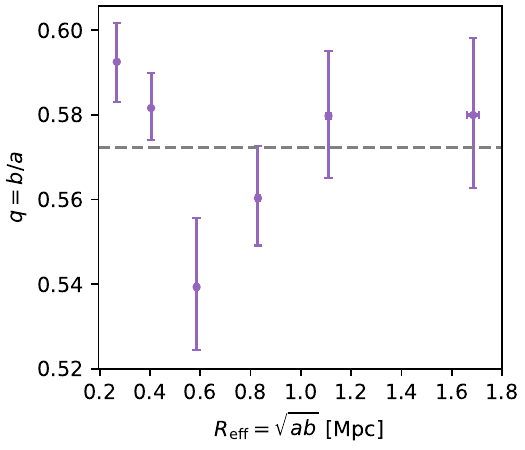}{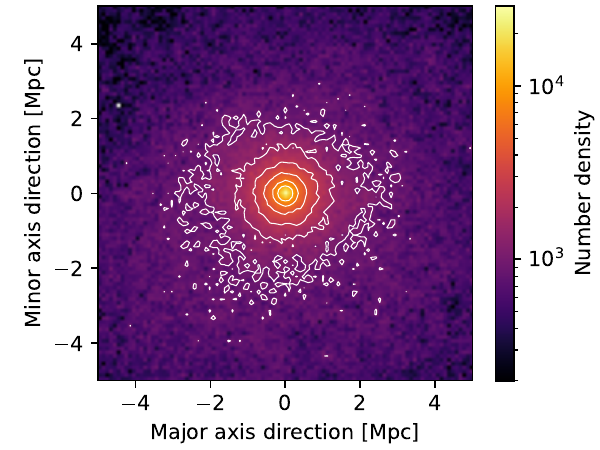}
    \caption{\textit{Top}: \textsc{MagLim} galaxy number density stacked along cluster major axes. 
    The contours are set at equal steps in logarithmic scale.
    The \textit{top right} panel shows the central half region of the \textit{top left} panel. 
    \textit{Bottom left}: Axis ratio as a function of effective radius. Density levels are sampled uniformly in logarithmic space. At each level, pixels with densities within 5\,\% are selected, and an ellipse is fitted to the corresponding coordinates to derive $R_{\rm eff}$ and $q$.
    The points show the median of the bootstrap distributions, and the error bars denote the 16th and 84th percentiles for both x-axis and y-axis. The horizontal dashed line shows the mean of $q$ among the points.
    \textit{Bottom right}: stacked number density without axes aligned (null test).
    }
    \label{fig:rad_dist_num_dens}
\end{figure*}

Figure~\ref{fig:rad_dist_num_dens} shows the galaxy distribution after the axis-aligned stacking. We use color scaling to show the number density after subtracting the mean foreground/background density (approximated by the minimum on the grid). From the contours, we see that the stacked galaxy distribution shows an elliptical shape with its major axis aligned with the x-axis. The axis ratio of the galaxy number density is nearly constant at small radii ($\lesssim 2$~Mpc) and becomes noisy at large radii. 

Next, we measure the ellipticity of the galaxy distribution as a function of radius. At each density level, we fit an ellipse to the corresponding pixel coordinates, allowing a density scatter of 5\,\% and record the best-fit major- and minor-axis lengths. 
We find $q\sim0.6$, which is close to the WL halo ellipticity without the dilution correction (Section~\ref{sec:ellipticity_baseline}). This means that the stacked galaxy distribution without the misalignment correction has a  shape similar to the stacked halo (without the dilution correction).
In addition, we find that $q$ changes only by $\sim10\,\%$ within an effective radius of $R_{\rm eff}=\sqrt{ab}\sim2$~Mpc. Hence, we conclude that the galaxy distribution ellipticity is generally fixed within the cluster scale. This suggests weak radial variation in halo ellipticity, assuming galaxies trace the mass distribution. 

As a null test, we also stack clusters without aligning their axes. The resulting number density distribution is almost circular without showing significant systematics.  

It is worth pointing out that there are other ways to describe the cluster triaxiality without explicitly measuring the projected ellipticity. For example, one can directly compare the matter distribution along the major and minor axes of axis-aligned stacked clusters, and the greater the difference between these two, the more elliptical its shape. The matter-distribution profiles can be inferred from the galaxy number density or lensing signature in observations~\citep{Fu2024a}, or the particle number density or projected mass in simulations (Jones et al. in preparation). 
\citet{Fu2024a} find that the ratio between the major- and minor-axis profiles of galaxy number density is almost constant within $\sim10$~Mpc, and the difference between the two profiles corresponds to 
an axis ratio of $\sim0.7$ (larger than the result here) with the CG as the cluster orientation proxy, suggesting that the galaxy distribution may provide a better proxy for the cluster orientation than the CG~\citep{Shin2018,Fu2024b}.

\subsection{Comparison with similar studies}

\label{sec:comparison_studies}
\begin{figure*}[htb!]
    \plotone{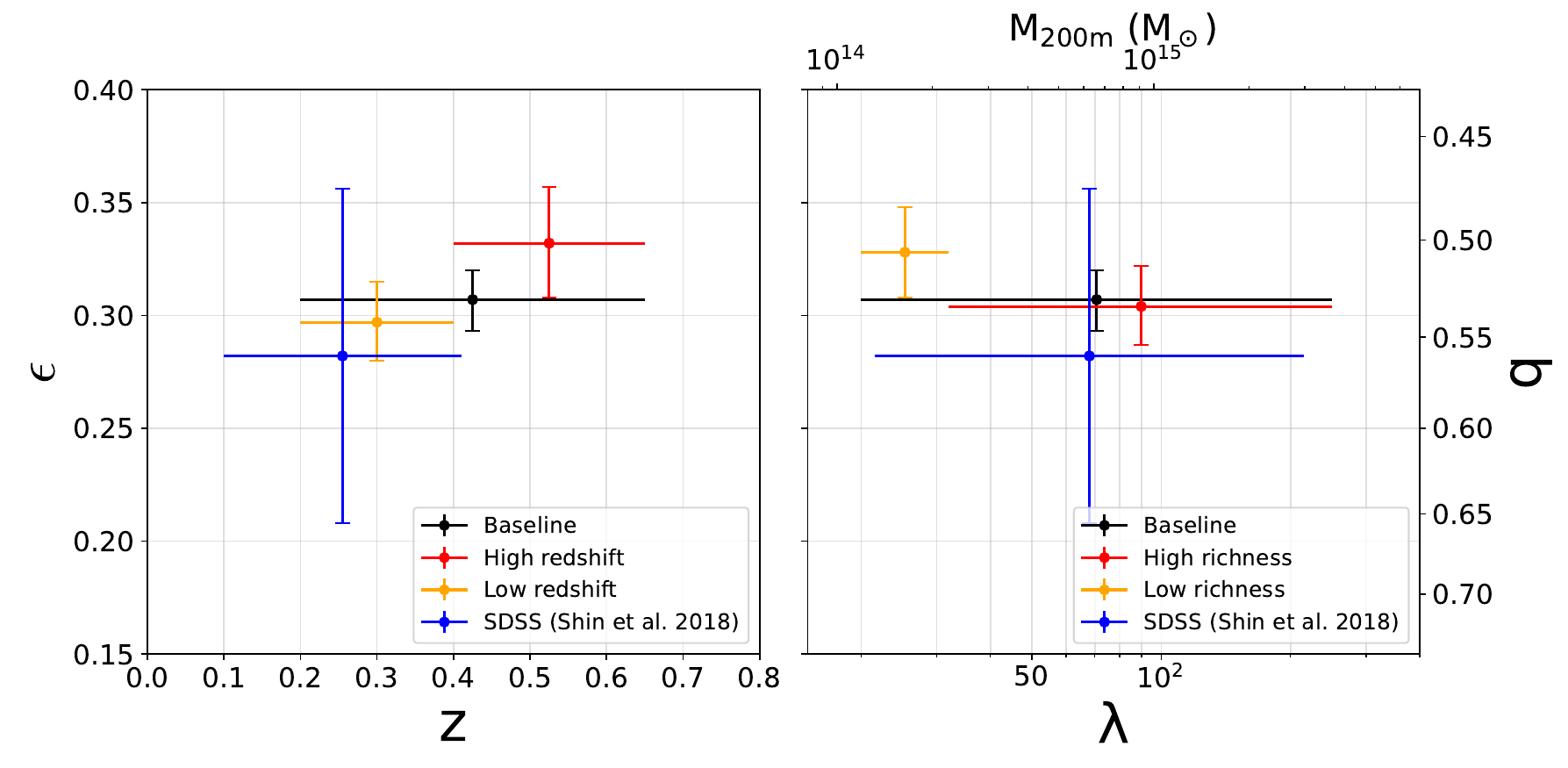}

    \caption{Comparison of projected halo ellipticity measurements for the selected DES Y3 and SDSS \rdmp{} clusters. \textit{Left}: Ellipticity as a function of redshift. We indicate the measurement from \citet{Shin2018} in blue, the baseline sample in black, the low-redshift sample in orange, and the high-redshift sample in red. The horizontal error bar indicates the range of redshifts for the respective cluster sample while the vertical error bar indicates the $1\,\sigma$ uncertainty in $\epsilon$. The bin location for each sample spanning a redshift range of [$z_{\rm min}$, $z_{\rm max}$) is computed as $(z_{\rm min}+z_{\rm max})/2$. \textit{Right}: Ellipticity as a function of richness. We define the bin location for each sample spanning a richness range of [$\lambda_{\rm min}$, $\lambda_{\rm max}$) as $\sqrt{\lambda_{\rm min}\times\lambda_{\rm max}}$. We also compute the mass $M_{\rm 200m}$ using the richness-mass scaling relation provided by~\citet{McClintock2019} at a fixed redshift of $z=0.4$ and show it in the top axis for reference. Note that there are small differences between Y1 and Y3 \rdmp{} clusters' redshifts, richness, and scaling relations~\citep{McClintock2019,DES2025}, but they are not expected to strongly affect the demonstration here.}
    \label{fig:e_vs_z}
\end{figure*}

Comparing our fitted ellipticity for the baseline cluster sample to the measurement made by~\citet{Shin2018} for SDSS clusters, we find that we are in excellent agreement (within $\sim0.3\,\sigma$, measured by combining the uncertainties in quadrature). The uncertainties in the work of~\citet{Shin2018} are clearly larger, although their number of clusters (10428) and the richness range are comparable to this work. The reason could be that the SDSS photometric dataset used in their work is shallower than DES Y3, which leads to fewer background sources for lensing analysis. It also leads to lower redshifts of both lenses and sources, which affects the lensing efficiency (via the angular distance ratio in the critical surface density). 

We show the comparison in Figure~\ref{fig:e_vs_z}, along with the redshift- and richness-split cluster samples. We also convert the SDSS richness from \citet{Shin2018} to DES-Y1-like richness using Eq. 67 in the work of~\citet{McClintock2019}. We find that clusters, when split by redshift (left panel) or richness (right panel), do not show statistically significant trends and are consistent with the measurement of \citet{Shin2018}. 

However, when selecting clusters by richness, we also need to consider projection effects. Clusters aligned along the LoS will be preferentially detected due to a boost in their richness measurements from projection; also, they appear rounder on the sky. 
Conversely, clusters aligned perpendicular to the LoS (along the PoS) will have a lower measured richness value (more difficult to be detected) and appear to be more elliptical. 
Hence, we predict that for an observed cluster sample in a given richness bin, the overall measured ellipticity may be biased low and may not be representative of the true ellipticity of this sample. 
Also, LoS structures can affect cluster detection and halo ellipticity measurement. We speculate that associated filaments may increase the ellipticity because of their alignment with the halo, and unassociated structures may do the opposite because they are randomly distributed.

We can account for these biases using cosmological simulations --- selecting clusters by their true richness (number of satellite galaxies or subhalos), computing their projected ellipticity on the PoS, and then comparing the richness-ellipticity relation in the simulation with the one in the observation. The difference between the simulation and observation results can be used to quantify selection/projection effects. 
For example, simulations show that massive clusters tend to be more elliptical~\citep{Hopkins2005,Allgood2006,Despali2017}, and from observation we have not seen a clear trend in the projected halo ellipticity as a function of cluster richness, which could be caused by selection/projection effects.
We leave more detailed comparisons for a follow-up paper (Section~\ref{sec:future_work}).

\subsection{Outlook for future surveys}

Here we present a rough estimate on how future wide-area deep surveys will improve the statistics of this study. 
The cluster redshift range between 0.2 and 0.65 ensures that the baseline sample is volume-limited~\citep[cluster galaxies used for determining richness are all beyond a certain luminosity threshold;][]{Rykoff2016,McClintock2019,DES2025}, and therefore we expect the sample to have high completeness. Extending to higher redshifts will require deeper optical observations or other probes such as the SZ signal~\citep{Planck2016, Bleem2020, ACTDESHSC2025}. 

Consequently, when discussing the upcoming surveys, we assume the same cluster selection as the baseline sample, with surveys that are deeper than DES Y3. Then the number of clusters detected is roughly proportional to the footprint area $A$ (in the same redshift range). 
The depth mainly affects the number density of background sources $n$. 
We focus on shape noise as the main source of uncertainty, as it is much higher than other sources~\citep{McClintock2019}. 
Then we can estimate that the error bar of the halo ellipticity measurement will change by a factor of $\sqrt{({A_0}/{A})({n_0}/{n})}$, where the subscript 0 denotes the value for the fiducial sample. 
In addition, the number density is approximately proportional to an exponential function of the $i$-band magnitude, $n\propto10^{0.31i}$~\citep{LSSTSC2009}, and then the factor can be rewritten as $\sqrt{({A_0}/{A})10^{0.31(i_0-i)}}\sim 1.43^{i_0-i}\sqrt{A_0/A}$. {Note that depth also slowly changes the source redshift distribution~\citep[by $\sim0.1$ per $i$-band magnitude;][]{LSSTSC2009} and thus the overall lensing signal; we skip its effect on the halo ellipticity measurement for now since it is small.} 

Next, given the footprint and depth of DES Y3~\citep[$\sim5$k~$\deg^2$, $i\sim23$ for $10\,\sigma$;][]{SevillaNoarbe2021},
LSST Y10~\citep[$\sim18$k~$\deg^2$, $i\sim26$;][]{Ivezic2019}, 
Euclid Y6~\citep[$\sim15$k~$\deg^2$, $i\sim25.5$;][]{EuclidCollaboration2022}, 
we obtain the factors $\sim0.18$ (i.e., 82\,\% reduction in the error bar) for LSST Y10 and $\sim0.24$ (76\,\% reduction in the error bar) for Euclid Y6, assuming that the source selection for the WL analysis covers a similar fraction in each survey.  
For comparison, the survey metrics of the Hyper Suprime-Cam Subaru Strategic Program (HSC-SSP), a deep precursor survey of LSST, are $\sim1.4$k~$\deg^2$ and $i\sim25$~\citep{Aihara2018}, which leads to a factor of $\sim0.92$ (the small improvement is caused by the relatively small footprint).
For Roman's High Latitude Wide Area Survey (HLWAS), it covers $\sim2.4$k~deg$^2$ with $Y,J,H\sim26$ in the Medium tier plus $\sim2.7$k~deg$^2$ with $H\sim25.5$ in the Wide tier, where the respective source densities reach $\sim41$ arcmin$^{-2}$ and $\sim27$ arcmin$^{-2}$~\citep[S/N$>18$;][]{Roman2025}\footnote{\url{https://roman.ipac.caltech.edu/page/core-community-surveys}}, and the factor will be $\sim0.47$ (53\,\% reduction) in total. 
We summarize the estimated performances of different surveys in Table~\ref{tab:forecast}.
\begin{table*}[htb!]
    \centering
    \begin{tabular}{c|c|c|c}
    Survey & Area ($\deg^2$) & Depth ($i_{10\sigma}$) & Improvement (\%) \\
    \hline
   LSST Y10 & 18k & 26 & 82 \\
   Euclid Y6 & 15k & 25.5 & 76 \\
   HSC SSP & 1.4k & 25 & 8 \\
   Roman HLWAS & 5.1k & \dots & $53$ \\
    \end{tabular}
    \caption{Estimated performance  improvements of different surveys relative to DES Y3.}
    \label{tab:forecast}
\end{table*}

We also expect improvement in the Point Spread Function (PSF) size of certain instruments, especially space telescopes (e.g., Euclid and Roman),  will further improve the shape measurement and source detection, which potentially leads to smaller WL uncertainties. When the contribution of shape noise becomes significantly smaller after stacking, small systematic effects will be probed more sensitively such as LSS noise and the shift of source redshift distribution~\citep{EuclidCollaboration2024}.
Deeper data will result in more complete detection of high-z clusters, which will further improve the S/N of stacked cluster measurements. 
In fact, the significant reduction of the error bar in LSST/Euclid/Roman will also allow more bins to better constrain the halo shape dependence on mass/redshift. 


\section{Conclusion}
\label{sec:conclusion}

\subsection{Summary of  analysis and findings}

In this work, we study the triaxiality of DES Y3 \rdmp{} galaxy clusters through a WL multipole expansion, making use of the associated source shape catalog and DNF photo-z catalog. 
The clusters are selected with high centering probability ($\texttt{pcen0}>0.9$) for an unambiguous center and a likely less disturbed state, and with redshift $0.2<z_\lambda<0.65$ and richness $\lambda>20$ for high completeness. 
We expand the halo surface density to the second order of the projected halo ellipticity to obtain the monopole and quadrupole terms, and we construct corresponding lensing models for these terms, which efficiently and accurately constrain the projected ellipticity of the cluster sample. 
The estimators associated with the models are computed by stacking the cluster ensemble with their axes aligned, and the orientation of each cluster is determined by the distribution of its satellite member galaxies. 
We obtain an effective ellipticity of $0.307^{+0.013}_{-0.014}$, which corresponds to an axis ratio of $0.530^{+0.016}_{-0.015}$. 
We estimate an extra systematic bias to be a few percent in ellipticity, mostly caused by miscentering.
This is the first time that a WL multipole expansion analysis has gone beyond the leading order in ellipticity, and this is also the first direct ellipticity measurement of this cluster sample. 

In the analysis, we correct for systematics such as the contamination of lensed background galaxies with unlensed cluster galaxies (boost factor) and the misalignment between the orientations of the cluster halo and the satellite distribution (dilution factor). 
We divide the sample by redshift and richness, finding no statistically significant dependence of ellipticity measurements on these quantities; these statistics will be further improved by upcoming wide-area deep surveys. 
We validate our analysis using various datasets from observations (the DES Y1 \rdmp{} catalog) and simulations (mock catalogs with the elliptical NFW model and \skysim{}). 
Our multipole expansion algorithm has been integrated into the \clmm{}, a DESC tool for cluster lensing mass modeling.\footnote{The code of this analysis and the data generated in our results are available from the corresponding author upon reasonable request. The code of \clmm{} is available at \url{https://github.com/LSSTDESC/CLMM}. The data for figures are available on Figshare under a Creative Commons Attribution license: 
\dataset[doi:10.6084/m9.figshare.33480964]{https://doi.org/10.6084/m9.figshare.33480964}.}

\subsection{Future work}\label{sec:future_work}

This work is the beginning of a series that studies cluster triaxiality measurement. We are planning for these following up papers.

The Paper II will focus on cosmological simulations to improve our understanding of systematics. 
The monopole-quadrupole analysis can be tested on the simulation's shear catalog derived from ray tracing to compare the truth with the output. 
We will be able to directly test miscentering --- the offset between the halo center and the identified cluster CG.
Simulations will improve the determination of the dilution factor (Appendix~\ref{app:sim_orientation_comparison}), and we will be able to compare the halo projected ellipticity, the satellite/subhalo distribution ellipticity, and the halo triaxiality. 
Going further, we can test the halo shape dependence on cluster properties such as mass, concentration, redshift, and dynamical state (or central galaxy properties). 
Additionally, since simulation catalogs are free from selection biases and projection effects, they will help quantify the corresponding systematics in observations~\citep{Sunayama2020,Payerne2025,Vecchi2025} by directly comparing the characteristics of simulated clusters to those of the detected 2D clusters based on those simulations.
Moreover, we can study the cluster neighborhood to improve our model at large radii using simulations --- we can compare the orientation/shape of the cluster halo with that of the neighboring filaments. For modeling, we can include a 2-halo term in $\Sigma_{\rm sph}$ (Eq.~\ref{eq:sigma0_sigma2}) as a first step, assuming that the halo aligns with the nearby major filament and shares the ``effective'' ellipticity --- this will be particularly useful for measuring $\Delta\Sigma_{\rm const}$, which is affected by external matter (Eq.~\ref{eq:models_main}). However, the halo ellipticity and the surrounding structure's ellipticity may differ~\citep{Gonzalez2022}. The anisotropic matter distribution around a cluster can have ellipticity that varies with radius and extends to tens of Mpc due to the large-scale tidal field~\citep{Osato2018}, which may require more detailed modeling in our analysis.  

The Paper III will follow the improved modeling of the Paper II and apply that to observational data, especially the Rubin LSST Data Preview 2 (DP2) or Data Release 1 (DR1) once they are available. Those datasets will cover wide sky areas and thus allow a cluster ensemble analysis similar to the one in this paper, but with the new, deeper observations from LSST. The overlapping clusters between DES and LSST will allow for a consistency check. 
Another item that we will test is the direct elliptical NFW fitting to the WL signal~\citep{Oguri2010a,Umetsu2018}, which has been applied to small samples of massive clusters. Instead, we will apply this method to stacked, less-massive clusters (i.e., combined shape catalogs). This method potentially gives a more accurate measurement of the cluster shape, but requires more computing resources than the monopole-quadrupole approximation presented here, as evaluating shear at a given point requires several integrals. 
We may also test the effect of miscentering on the triaxiality analysis in the Rubin data, especially for low $\texttt{pcen0}$ clusters, which are likely to be perturbed systems. 
Additionally, we may consider a wider range of object types and apply the methodology to galaxy groups or LRGs to constrain halo shapes at smaller scales~\citep{Clampitt2016,vanUitert2017,Robison2023}. 


Finally, other probes, such as the gas information derived from cluster X-ray emissions and SZ effects, will provide further constraints on cluster morphology. 
Spectroscopic datasets, especially the future 4MOST~\citep{deJong2019} data covering the southern sky, will provide detailed spectroscopic information of cluster members to mitigate projection effects and interlopers, and help calibrate the redshifts of background galaxies; this will be particularly useful for the triaxiality analysis when combined with the LSST data. 

\begin{acknowledgments}

This paper has undergone internal review in the LSST Dark Energy Science Collaboration. We thank the internal reviewers Anthony Englert and Benjamin Levine for their detailed and constructive comments. 
We thank Shun-Sheng Li, Grzegorz Madejski, Eli Rykoff, Eric Charles, John Franklin Crenshaw, Ian Dell'Antonio, and Patricia Burchat for helpful comments, questions, and support.

The DESC acknowledges ongoing support from the Institut National de 
Physique Nucl\'eaire et de Physique des Particules in France; the 
Science \& Technology Facilities Council in the United Kingdom; and the Department of Energy and the LSST Discovery Alliancein the United States.  DESC uses resources of the IN2P3 Computing Center (CC-IN2P3--Lyon/Villeurbanne - France) funded by the Centre National de la Recherche Scientifique; the National Energy Research Scientific Computing Center, a DOE Office of Science User Facility supported by the Office of Science of the U.S.\ Department of Energy under Contract No.\ DE-AC02-05CH11231; STFC DiRAC HPC Facilities, funded by UK BEIS National E-infrastructure capital grants; and the UK particle physics grid, supported by the GridPP Collaboration.  This work was performed in part under DOE Contract DE-AC02-76SF00515.

This project used public archival data from the Dark Energy Survey (DES). Funding for the DES Projects has been provided by the U.S. Department of Energy, the U.S. National Science Foundation, the Ministry of Science and Education of Spain, the Science and Technology Facilities Council of the United Kingdom, the Higher Education Funding Council for England, the National Center for Supercomputing Applications at the University of Illinois at Urbana-Champaign, the Kavli Institute of Cosmological Physics at the University of Chicago, the Center for Cosmology and Astro-Particle Physics at the Ohio State University, the Mitchell Institute for Fundamental Physics and Astronomy at Texas A\&M University, Financiadora de Estudos e Projetos, Funda{\c c}{\~a}o Carlos Chagas Filho de Amparo {\`a} Pesquisa do Estado do Rio de Janeiro, Conselho Nacional de Desenvolvimento Cient{\'i}fico e Tecnol{\'o}gico and the Minist{\'e}rio da Ci{\^e}ncia, Tecnologia e Inova{\c c}{\~a}o, the Deutsche Forschungsgemeinschaft, and the Collaborating Institutions in the Dark Energy Survey.
The Collaborating Institutions are Argonne National Laboratory, the University of California at Santa Cruz, the University of Cambridge, Centro de Investigaciones Energ{\'e}ticas, Medioambientales y Tecnol{\'o}gicas-Madrid, the University of Chicago, University College London, the DES-Brazil Consortium, the University of Edinburgh, the Eidgen{\"o}ssische Technische Hochschule (ETH) Z{\"u}rich,  Fermi National Accelerator Laboratory, the University of Illinois at Urbana-Champaign, the Institut de Ci{\`e}ncies de l'Espai (IEEC/CSIC), the Institut de F{\'i}sica d'Altes Energies, Lawrence Berkeley National Laboratory, the Ludwig-Maximilians Universit{\"a}t M{\"u}nchen and the associated Excellence Cluster Universe, the University of Michigan, the National Optical Astronomy Observatory, the University of Nottingham, The Ohio State University, the OzDES Membership Consortium, the University of Pennsylvania, the University of Portsmouth, SLAC National Accelerator Laboratory, Stanford University, the University of Sussex, and Texas A\&M University.
Based in part on observations at Cerro Tololo Inter-American Observatory, National Optical Astronomy Observatory, which is operated by the Association of Universities for Research in Astronomy (AURA) under a cooperative agreement with the National Science Foundation.

This research used resources of the National Energy Research Scientific Computing Center (NERSC), a Department of Energy User Facility. 

Part of this research was conducted using computational resources and services at the Center for Computation and Visualization, Brown University.

\end{acknowledgments}

\begin{contribution}
Shenming Fu: Conceptualization, data curation, formal analysis, investigation, methodology, project administration, software, supervision, validation, visualization, writing - original draft. \\
Radhakrishnan Srinivasan: Conceptualization, formal analysis, investigation, project administration, software, supervision, validation, visualization, writing - original draft. \\
Tae-hyeon Shin: Theoretical formalism development, data vector generation, calculation of systematic corrections (dilution factor, boost factor), writing. \\
Rance Solomon: Conceptualization, methodology, software, writing - original draft. \\
Deric Jones: Visualization, validation, feedback, writing and editing. \\
Camille Avestruz: \clmm{} code guidance, feedback, writing, scientific mentorship. \\ 
Yuanyuan Zhang: Feedback, writing, scientific mentorship. \\
Michel Aguena: \clmm{} overview and code review. \\
C\'eline Combet: \clmm{} code review, feedback on original draft. \\
Anthony Englert: DESC Internal Review. \\
Benjamin Levine: DESC Internal Review. \\
Alex I. Malz: Foundational contributions to \clmm{} codebase.\\
Constantin Payerne: Feedback on original draft. \\
Marina Ricci: Feedback on original draft, \clmm{} overview and code review. \\
Anja von der Linden: Scientific mentorship, feedback.

\end{contribution}

\vspace{5mm}

\software{
\texttt{Astropy}~\citep{Astropy2013,Astropy2018,Astropy2022}, 
\texttt{Corner}~\citep{Foreman-Mackey2016}, 
\texttt{H5py}~\citep{Collette2013}, 
\texttt{Healpy}~\citep{Zonca2019}, 
\texttt{Matplotlib}~\citep{Hunter2007}, 
\texttt{Numpy}~\citep{Harris2020}, \texttt{Scipy}~\citep{Virtanen2020}, 
\texttt{Sympy}~\citep{Meurer2017}.
}

\appendix

\section{Lensing effects of an elliptical halo}
\label{app:lensing_elliptical_halo}

Here we summarize the analytical form of
convergence and shear of a 2D elliptical mass distribution\footnote{\url{https://github.com/oguri/glafic2/tree/main/manual}}~\citep[][]{Schramm1990,Keeton2001,Oguri2010a,Oguri2010b}. 

To compute the lensing effects of the surface density described by Eq.~\ref{eq:kappa_elp}, we use these parameters: physical coordinates $(x,y)$, axis ratio $q$,  a radial symmetric function $\kappa(R)$, and its derivative $\kappa'(R)$. 
We consider a surface density $\kappa_{\rm new}(x,y) \equiv \kappa(\sqrt{qx^2+y^2/q})$, an ``area-preserved'' transformation from the circular distribution.

The lensing potential can be calculated by solving the Poisson's equation (Eq.~\ref{eq:poisson}); note this is different from the form in the main text ($\phi D_l^2=\psi$; $\partial_{1,2}=D_l \partial_{x,y}$; Eq.~\ref{eq:kappa},~\ref{eq:gamma}). 
\begin{equation}
    \psi_{xx} + \psi_{yy} = 2\kappa_{\rm new}(x,y)
    \label{eq:poisson}
\end{equation}

Next, to obtain the lensing effects of that surface density, one can compute the second derivatives of that lensing potential (Eq.~\ref{eq:p2d_1},~\ref{eq:p2d_2},~\ref{eq:p2d_3},~\ref{eq:p2d_t}). 

\begin{equation}
    \psi_{xx} = 2x^2q^2 \int_0^1\frac{v\kappa'(t(v))}{2[1-(1-q^2)v]^{1/2}t(v)}dv + q \int_0^1\frac{\kappa(t(v))}{[1-(1-q^2)v]^{1/2}}dv
    \label{eq:p2d_1}
\end{equation}

\begin{equation}
    \psi_{yy} = 2y^2q^2 \int_0^1\frac{v\kappa'(t(v))}{2[1-(1-q^2)v]^{5/2}t(v)}dv + q \int_0^1\frac{\kappa(t(v))}{[1-(1-q^2)v]^{3/2}}dv
    \label{eq:p2d_2}
\end{equation}

\begin{equation}
    \psi_{xy} = 2xyq^2 \int_0^1\frac{v\kappa'(t(v))}{2[1-(1-q^2)v]^{3/2}t(v)}dv
    \label{eq:p2d_3}
\end{equation}

\begin{equation}
    t(v) = \sqrt{vq\left[x^2+\frac{y^2}{1-(1-q^2)v}\right]}
    \label{eq:p2d_t}
\end{equation}

Then, the shear is given by Eq.~\ref{eq:app_gamma}, and it can be verified that the convergence satisfies Eq.~\ref{eq:poisson}. 
\begin{equation}
    \gamma_1=(\psi_{xx} - \psi_{yy})/2; \gamma_2=\psi_{xy}
    \label{eq:app_gamma}
\end{equation}

For an NFW profile, $\kappa(R)=b\hat{\kappa}(\hat{R})$, where $\hat{R}=R/r_s$, $b=4  \rho_{\rm crit} \delta_{\rm c}  r_s / \Sigma_{\rm crit} $, $\rho_{\rm crit}$ is the critical density of the universe and $\delta_{\rm c}$ is the characteristic overdensity~\citep{Wright2000}, and $\hat{\kappa}(\hat{R})$ satisfies Eq.~\ref{eq:k_}, where $F(\hat{R})$ satisfies Eq.~\ref{eq:Fu}. 
Then the derivative $\kappa'(R)=b\hat{\kappa}'(\hat{R})/r_s$,
and $\hat{\kappa}'(\hat{R})$ satisfies Eq.~\ref{eq:kd_}. 
\begin{equation}
    \hat{\kappa}(\hat{R}) =
\begin{cases}
\dfrac{1-F(\hat{R})}{2(\hat{R}^{2}-1)} & (\hat{R}\neq 1)\\
\dfrac{1}{6} & (\hat{R}=1)
\end{cases}
    \label{eq:k_}
\end{equation}

\begin{equation}
    F(\hat{R}) =
\begin{cases}
\dfrac{\mathrm{arctanh}\left(\sqrt{1-\hat{R}^{2}}\right)}
{\sqrt{1-\hat{R}^{2}}} & (\hat{R}<1)\\
1 & (\hat{R}=1)\\
\dfrac{\arctan\left(\sqrt{\hat{R}^{2}-1}\right)}
{\sqrt{\hat{R}^{2}-1}} & (\hat{R}>1)
\end{cases} 
    \label{eq:Fu}
\end{equation}

\begin{equation}
    \hat{\kappa}'(\hat{R}) =
\begin{cases}
\dfrac{-1-2\hat{R}^{2}+3\hat{R}^{2}F(\hat{R})}
{2\hat{R}(\hat{R}^{2}-1)^{2}} & (\hat{R}\neq 1)\\
-\dfrac{1}{5} & (\hat{R}=1)
\end{cases} 
    \label{eq:kd_}
\end{equation}

Currently, we need to evaluate the above integrals at every point, and it is considerably time-consuming. To speed up the calculation, one can use a grid and then interpolate the value at locations in between the grid points. Another possible method is to use tabulated data for the integrals and then rescale the coordinates. Additionally, \citet{Oguri2021} uses expansions to approximate the convergence, so that the second derivatives of the lensing potential have analytical expressions without integrations.   

\section{True halo orientation vs. galaxy distribution orientation in \skysim{}}
\label{app:sim_orientation_comparison}

Using cosmological simulations, one can directly test whether the halo's 3D major axis orientation projected on the sky is consistent with the major axis orientation of the galaxy distribution on PoS. This provides a more realistic estimate of the misalignment (dilution factor) than using Monte Carlo realizations, which is presented in Section~\ref{sec:monopole_and_quadrupole_measurements}. We rely on a simulation dataset that has been widely used in DESC — the \skysim{} ($5000 \deg^2$ footprint area), an extended and improved version of the \cosmodc{}~\citep{Korytov2019}, for this test. 

The \cosmodc{} was the starting point of the Second Data Challenge in DESC.
Built from the Outer Rim run~\citep{Heitmann2019}, a trillion-particle, (4.225~Gpc)$^3$ box cosmological N-body simulation, \cosmodc{} covers 440~$\deg^2$ of sky up to a redshift of $z = 3$.
The simulation is built from a hybrid approach of both data-driven empirical relations and semi-analytic galaxy modeling.
The empirical approach was used to populate halos with galaxies and to provide the necessary statistics for galaxy clusters. It also assigns basic quantities such as colors to the individual galaxies.
The galaxies are then matched to a smaller scale, semi-analytic simulated galaxy catalog~\citep[based on \textsc{Galacticus};][]{Benson2012} in order to provide a more robust set of realistic galaxy properties.

The \skysim{} catalog is built from the same simulation as \cosmodc{} but with a few differences in the final catalog production.\footnote{\url{https://lsstdesc.org/DC2-analysis/tutorials/extragalactic_halo_quantities.html}}
In \cosmodc{}, the galaxies are pasted isotropically into a halo without following the morphology of the halo. In \skysim{}, the galaxies follow an elliptical NFW distribution, which mimics the halo triaxiality and follows the tidal field orientation, and thus the galaxy distribution is more realistic than \cosmodc{} and more comparable to observations. 
Also, the \skysim{} has higher resolution in ray tracing for WL and produces a more accurate modeling of WL shear, which enables multipole analysis based on this dataset as well. 
Moreover, \skysim{} has a \rdmp{} catalog. Directly connecting simulation predictions with observational results is beyond the scope of this paper, but in future work we plan to use \rdmp{} detections and \rdmp{} members from simulations to test for selection and systematic effects (e.g., projection effects and miscentering) in the observations.

\begin{figure}[!tbp]
    \centering
    \includegraphics[width=0.5\linewidth]{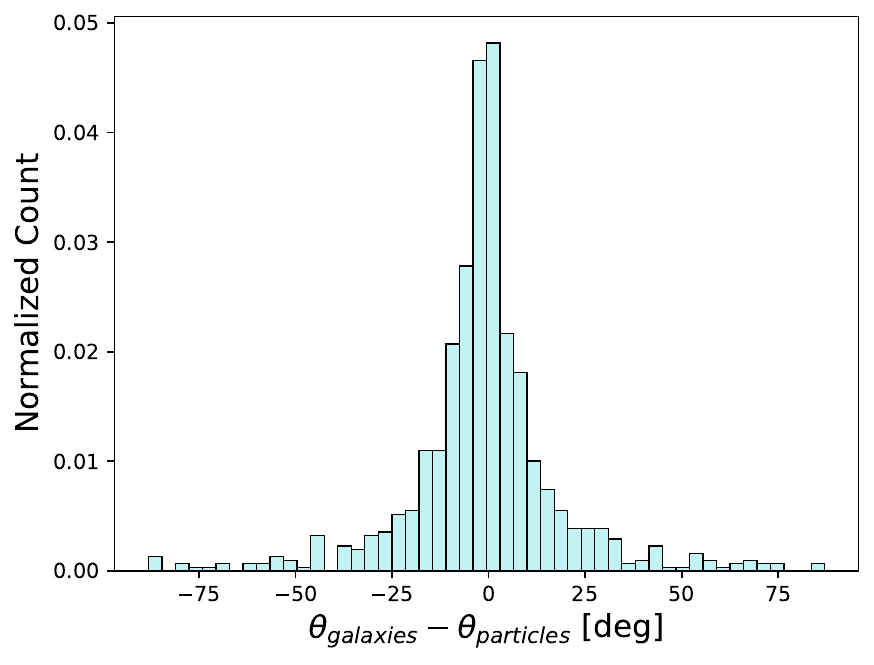}
    \caption{Histogram of the comparison between the major-axis angle of each cluster halo measured from the satellite galaxy distribution vs. the one measured from the particle distribution in \skysim{}.
    }
    \label{fig:Major_axis_orientation}
\end{figure}

We use \skysim{} to compare the true halo orientation derived from particle distribution to the cluster orientation derived from the galaxy distribution as follows.
We consider that the centroid location of a cluster is $\mathbf{x}=(x,y,z)$ and the 3D orientation vector (derived from the inertia tensor of simulation particles) is $\mathbf{a}=(a_x,a_y,a_z)$, and the observer is at the origin $(0,0,0)$. We convert both $\mathbf{x}$ and $\mathbf{a}$ into angular coordinates (in RA, DEC) $\mathbf{\theta_x}$ and $\mathbf{\theta_a}$, using the HEALPix function \texttt{vec2ang}. 
Then, the orientation of the halo projected on PoS can be determined by the position angle (azimuth) of $\mathbf{\theta_a}$ relative to  $\mathbf{\theta_x}$ (computed by \texttt{Astropy}). 
Note, this position angle is measured on the plane tangent to the sphere at the cluster location. 
Then, we compare this angle with the orientation angle derived from the galaxy distribution using the method explained in Section~\ref{sec:halo_orientation}. 
Finally, we fold their difference  by the $180^\circ$ symmetry --- this gives the misalignment angle. We present the result in Figure~\ref{fig:Major_axis_orientation}, which shows consistency  between the two orientations with a sharp peak around 0. 

\section{Theory of monopole and quadrupole}\label{app:derive_multipoles}

\subsection{Expansions}
\label{app:derive_multipoles_forms}

Following the description in Section~\ref{sec:theory_method}, we give the details of the derivation of the monopole and the quadrupole. 

First, we have the axis ratio $q$ and these ellipticity quantities for an ellipse. 
\begin{equation}
    \left\{
    \begin{aligned}
    \epsilon &= \frac{1-q}{1+q} \\
    \chi &= \frac{1-q^2}{1+q^2} 
    \end{aligned}
\right.
\label{eq:ellipticity_definition}
\end{equation}

In this work, we focus on $\epsilon$ for ellipticity measurement. When the ellipse is close to a circle, the ellipticity is small, i.e., when $q\rightarrow1$, both $\epsilon,\chi\rightarrow0$. 

\begin{equation}
    \left\{
    \begin{aligned}
    \chi &= \frac{2\epsilon}{1+\epsilon^2} = 2\epsilon + \mathcal{O}(\epsilon^3)  \\
    \epsilon &= \frac{1 - \sqrt{1 - \chi^2}}{\chi} = \frac{1}{2}\chi + \mathcal{O}(\chi^3) 
    \end{aligned}
\right.
\label{eq:ellipticity_limit}
\end{equation}

Under polar coordinate system, $x=R\cos{\theta},~y=R\sin{\theta}$. We define a deformed radius based on the ellipse:
\begin{equation}
    \tilde{R} \equiv \sqrt{qx^2 + \frac{y^2}{q}} = R\sqrt{q}\sqrt{\cos^2{\theta} + \frac{\sin^2{\theta}}{q^2}}.
    \label{eq:deformed_radius_definition}
\end{equation}
Note that
\begin{equation}
    \begin{aligned}
    \cos^2{\theta} + \frac{\sin^2{\theta}}{q^2} &= \frac{1+\cos{2\theta}}{2} + \frac{1-\cos{2\theta}}{2q^2} = \frac{1}{2}\left[1+\frac{1}{q^2}+\left(1-\frac{1}{q^2}\right)\cos{2\theta}\right] \\
    &= \frac{1}{2}\left(1+\frac{1}{q^2}\right)(1-\chi \cos{2\theta}).
    \end{aligned}
\end{equation}
Thus, 
\begin{equation}
    \tilde{R} = R\sqrt{\frac{1}{2}\left(q+\frac{1}{q}\right)(1-\chi\cos{2\theta})}.
\end{equation}
Note that 
\begin{equation}
    \begin{aligned}
    \frac{1}{2}\left(q+\frac{1}{q}\right) &= \frac{1}{2}\left(\frac{\sqrt{1-\chi}}{\sqrt{1+\chi}} + \frac{\sqrt{1+\chi}}{\sqrt{1-\chi}}\right) 
    = \frac{1}{\sqrt{1-\chi^2}} \\
    &= 1 + \frac{1}{2}\chi^2 + \mathcal{O}(\chi^4).
    \end{aligned}
\end{equation}
Thus, 
\begin{equation}
    \begin{aligned}
    \frac{\tilde{R}}{R} &=  
    \sqrt{1+\frac{1}{2}\chi^2+\mathcal{O}(\chi^4)}\sqrt{1-\chi\cos{2\theta}} = \left(1+\frac{1}{4}\chi^2+\mathcal{O}(\chi^4)\right)\left(1-\frac{\cos{2\theta}}{2}\chi - \frac{\cos^2{2\theta}}{8}\chi^2+\mathcal{O}(\chi^3)\right) \\
    &= 1 - \frac{\cos{2\theta}}{2}\chi + \frac{3-\cos{4\theta}}{16}\chi^2 + \mathcal{O}(\chi^3) = 1 - \epsilon\cos{2\theta} + \frac{3-\cos{4\theta}}{4}\epsilon^2 + \mathcal{O}(\epsilon^3).
    \end{aligned}
    \label{eq:deformed_radius_expansion}
\end{equation}
The last step uses Eq.~\ref{eq:ellipticity_limit}.

We consider an arbitrary axisymmetric distribution and write it in a traditional way, $\Sigma_{\rm sph}(R)=R^h$. This form is general, where $h(R)$ is a function setting the falloff of the profile. 
We have
\begin{equation}
    h \equiv \frac{\ln{\Sigma_{\rm sph}(R)}}{\ln{R}} = h(R).
    \label{eq:power_definition}
\end{equation}
Note, a constant factor $C$ can be absorbed into the power. For example, $C=R^h$ corresponds to $h=\ln{C}/\ln{R}$. 

For an arbitrary function $f=f(R)$, we have 
\begin{equation}
    \left\{
    \begin{aligned}
        \frac{\mathrm{d}f}{\mathrm{d}R} &= \frac{\mathrm{d}f}{\mathrm{d}\ln{R}}\frac{\mathrm{d}\ln{R}}{\mathrm{d}R} = \frac{1}{R}\frac{\mathrm{d}f}{\mathrm{d}\ln{R}}, \\
        \frac{\mathrm{d^2}f}{\mathrm{d}R^2} &= \frac{\mathrm{d}}{\mathrm{d}R}\left(\frac{\mathrm{d}f}{\mathrm{d}R}\right) = -\frac{1}{R^2}\frac{\mathrm{d}f}{\mathrm{d}\ln{R}} + \frac{1}{R^2}\frac{\mathrm{d}}{\mathrm{d}\ln{R}}\left(\frac{\mathrm{d}f}{\mathrm{d}\ln{R}}\right).
    \end{aligned}
    \right.
    \label{eq:general_derivatives}
\end{equation}

Also, based on the definition of $h$ (Eq.~\ref{eq:power_definition}), we have
\begin{equation}
    \frac{\mathrm{d}h}{\mathrm{d}R} = \frac{1}{\ln{R}}\frac{\mathrm{d}\ln{\Sigma_{\rm sph}}}{\mathrm{d}R} - \frac{\ln{\Sigma_{\rm sph}}}{(\ln{R})^2}\frac{1}{R} = \frac{1}{R\ln{R}}\frac{\mathrm{d}\ln{\Sigma_{\rm sph}}}{\mathrm{d}\ln{R}} - \frac{1}{R\ln{R}}\frac{\ln{\Sigma_{\rm sph}}}{\ln{R}}.
    \label{eq:power_derivative1}
\end{equation}
For simplicity, we define
\begin{equation}
    u(R)\equiv\frac{\mathrm{d}\ln{\Sigma_{\rm sph}}}{\mathrm{d}\ln{R}},
\end{equation}
and then from Eq.~\ref{eq:power_derivative1} (and the first line of Eq.~\ref{eq:general_derivatives}) we have
\begin{equation}
    \frac{\mathrm{d}h}{\mathrm{d}R} = \frac{u-h}{R\ln{R}};~\frac{\mathrm{d}h}{\mathrm{d}\ln{R}} = \frac{u-h}{\ln{R}}.
    \label{eq:power_derivative2}
\end{equation}

Next, from \ref{eq:power_derivative2}, 
\begin{equation}
    \frac{\mathrm{d}}{\mathrm{d}\ln{R}}\left(\frac{\mathrm{d}h}{\mathrm{d}\ln{R}}\right) = -\frac{u-h}{(\ln{R})^2} + \frac{1}{\ln{R}}\left(\frac{\mathrm{d}u}{\mathrm{d}\ln{R}} - \frac{\mathrm{d}h}{\mathrm{d}\ln{R}}\right) = -\frac{2(u-h)}{(\ln{R})^2} + \frac{1}{\ln{R}}\frac{\mathrm{d}u}{\mathrm{d}\ln{R}}, 
    \label{eq:power_second_derivative}
\end{equation}
and from the second line of Eq.~\ref{eq:general_derivatives},  Eq.~\ref{eq:power_derivative2}, and Eq.~\ref{eq:power_second_derivative}, 
\begin{equation}
    \frac{\mathrm{d}^2h}{\mathrm{d}R^2} = -\frac{1}{R^2}\frac{u-h}{\ln{R}} + \frac{1}{R^2}\left[ -\frac{2(u-h)}{(\ln{R})^2} + \frac{1}{\ln{R}}\frac{\mathrm{d}u}{\mathrm{d}\ln{R}} \right]
    \label{eq:power_second_derivative2}
\end{equation}

Next consider a Taylor expansion of $h$, and from Eq.~\ref{eq:power_derivative2}, Eq.~\ref{eq:power_second_derivative2}, and then Eq.~\ref{eq:deformed_radius_expansion}, we have
\begin{equation}
    \begin{aligned}
    h(\tilde{R}) &= h(R) + \frac{\mathrm{d}h}{\mathrm{d}R}(\tilde{R} - R) + \frac{\mathrm{d}^2h}{\mathrm{d}R^2}\frac{(\tilde{R} - R)^2}{2} +\cdots \\
    &= h(R) + \frac{u-h}{\ln{R}}\left(\frac{\tilde{R}}{R}-1\right) + \left[-\frac{u-h}{2\ln{R}} - \frac{u-h}{(\ln{R})^2}+\frac{1}{2\ln{R}}\frac{\mathrm{d}u}{\mathrm{d}\ln{R}}\right]\left(\frac{\tilde{R}}{R}-1\right)^2+\cdots \\
    &= h(R) + \frac{u-h}{\ln{R}}\left(-\epsilon\cos{2\theta}+\frac{3-\cos{4\theta}}{4}\epsilon^2\right) + \left[-\frac{u-h}{2\ln{R}} - \frac{u-h}{(\ln{R})^2}+\frac{1}{2\ln{R}}\frac{\mathrm{d}u}{\mathrm{d}\ln{R}}\right]\epsilon^2\cos^2{2\theta} + \mathcal{O}(\epsilon^3) \\
    &= h(R) + \frac{u-h}{\ln{R}}\left(-\epsilon\cos{2\theta}+\frac{3-\cos{4\theta}}{4}\epsilon^2\right) + \left[-\frac{u-h}{2\ln{R}} - \frac{u-h}{(\ln{R})^2}+\frac{1}{2\ln{R}}\frac{\mathrm{d}u}{\mathrm{d}\ln{R}}\right]\epsilon^2\frac{1+\cos{4\theta}}{2} + \mathcal{O}(\epsilon^3) \\
    &= h(R) - \epsilon \frac{u-h}{\ln{R}}\cos{2\theta} + \epsilon^2 \left[\frac{u-h}{2\ln{R}} - \frac{u-h}{2(\ln{R})^2} + \frac{1}{4\ln{R}}\frac{\mathrm{d}u}{\mathrm{d}\ln{R}} \right] + \epsilon^2 \left[-\frac{u-h}{2\ln{R}} - \frac{u-h}{2(\ln{R})^2} + \frac{1}{4\ln{R}}\frac{\mathrm{d}u}{\mathrm{d}\ln{R}} \right] \cos{4\theta} + \mathcal{O}(\epsilon^3)  \\
    \end{aligned}
    \label{eq:power_expansion1}
\end{equation}

Now, we consider an elliptical distribution, $\Sigma(R,\theta) = \Sigma_{\rm sph}(\tilde{R})=\tilde{R}^{k}$, where $k = h(\tilde{R})$. We want to describe $\Sigma$ using $R$ to match the form of multipoles. In the end, we also use $u$ instead of $h$ to simplify the formulas. 
From Eq.~\ref{eq:deformed_radius_expansion}, using a Taylor expansion $(1+x)^\alpha = 1+\alpha x+{\alpha(\alpha-1)}x^2/{2}+\mathcal{O}(x^3)$ when $x\rightarrow0$, we have 
\begin{equation}
    \begin{aligned}
    \Sigma &= R^k \left[1 - k\epsilon\cos{2\theta} + k\frac{3-\cos{4\theta}}{4}\epsilon^2 + \frac{k(k-1)}{2}\epsilon^2\cos^2{2\theta} + \mathcal{O}(\epsilon^3) \right] \\
    &= R^k \left[1 - k\epsilon\cos{2\theta} + k\frac{3-\cos{4\theta}}{4}\epsilon^2 + \frac{k(k-1)}{2}\epsilon^2\frac{1+\cos{4\theta}}{2} + \mathcal{O}(\epsilon^3) \right] \\
    &= R^k \left[1 - k\epsilon\cos{2\theta} + \left(\frac{k}{2} + \frac{k^2}{4}\right)\epsilon^2 + \left(-\frac{k}{2}+\frac{k^2}{4}\right)\epsilon^2\cos{4\theta} + \mathcal{O}(\epsilon^3) \right].  \\
    \end{aligned}
    \label{eq:Sigma_expansion0}
\end{equation}
Using Eq.~\ref{eq:power_expansion1}, we have
\begin{equation}
    \begin{aligned}
        \Sigma/R^k &= 1 - h\epsilon\cos{2\theta} + \left(\frac{h}{2} + \frac{h^2}{4}\right)\epsilon^2 + \left(-\frac{h}{2}+\frac{h^2}{4}\right)\epsilon^2\cos{4\theta} + \epsilon^2\frac{u-h}{\ln{R}}\cos^2{2\theta} + \mathcal{O}(\epsilon^3) \\
        &= 1 - h\epsilon\cos{2\theta} + \left(\frac{h}{2} + \frac{h^2}{4} + \frac{u-h}{2\ln{R}}\right)\epsilon^2 + \left(-\frac{h}{2}+\frac{h^2}{4} + \frac{u-h}{2\ln{R}}\right)\epsilon^2\cos{4\theta} + \mathcal{O}(\epsilon^3)
    \end{aligned}
    \label{eq:Sigma_expansion1}
\end{equation}

Let $s\equiv k - h=h(\tilde{R})-h(R)$, so that $R^k=R^h R^s$. Using Eq.~\ref{eq:power_expansion1}
and the Taylor expansion $A^x=1+x\ln{A}+x^2{(\ln{A})^2}/{2}+\mathcal{O}(x^3)$ when $x\rightarrow0$,
we have
\begin{equation}
    \begin{aligned}
    R^s &= 1 + s\ln{R} + \epsilon^2\left(\frac{u-h}{\ln{R}}\right)^2\cos^2{2\theta}\frac{(\ln{R})^2}{2} + \mathcal{O}(\epsilon^3) \\
    &= 1 + s\ln{R} + \epsilon^2(u-h)^2\frac{\cos{4\theta}+1}{4} + \mathcal{O}(\epsilon^3) \\
    &= 1 + s\ln{R} + \epsilon^2\left(\frac{u-h}{2}\right)^2 + \epsilon^2\left(\frac{u-h}{2}\right)^2\cos{4\theta} + \mathcal{O}(\epsilon^3) \\
    & = 1 - \epsilon (u-h)\cos{2\theta} + \epsilon^2 \left[\frac{u-h}{2} + \left(\frac{u-h}{2}\right)^2 - \frac{u-h}{2\ln{R}} + \frac{1}{4}\frac{\mathrm{d}u}{\mathrm{d}\ln{R}} \right] + \\
    &~~~~ \epsilon^2 \left[-\frac{u-h}{2} + \left(\frac{u-h}{2}\right)^2 - \frac{u-h}{2\ln{R}} + \frac{1}{4}\frac{\mathrm{d}u}{\mathrm{d}\ln{R}} \right] \cos{4\theta} + \mathcal{O}(\epsilon^3)  \\
    \end{aligned}
    \label{eq:Sigma_expansion2}
\end{equation}

Now, combine Eq.~\ref{eq:Sigma_expansion1} and Eq.~\ref{eq:Sigma_expansion2}, check the power terms of $\epsilon$ and the terms related to the angle. 
\begin{equation}
    \begin{aligned}
    \Sigma/\Sigma_{\rm sph} &= \Sigma/R^h =  \\ 
    &= \left[ 1 - h\epsilon\cos{2\theta} + \left(\frac{h}{2} + \frac{h^2}{4} + \frac{u-h}{2\ln{R}}\right)\epsilon^2 + \left(-\frac{h}{2}+\frac{h^2}{4} + \frac{u-h}{2\ln{R}}\right)\epsilon^2\cos{4\theta} + \mathcal{O}(\epsilon^3) \right] \times \\
    &~~~~ \Bigg\{ 1 - \epsilon (u-h)\cos{2\theta} + \epsilon^2 \left[\frac{u-h}{2} + \left(\frac{u-h}{2}\right)^2 - \frac{u-h}{2\ln{R}} + \frac{1}{4}\frac{\mathrm{d}u}{\mathrm{d}\ln{R}} \right] + \\
    &~~~~ \epsilon^2 \left[-\frac{u-h}{2} + \left(\frac{u-h}{2}\right)^2 - \frac{u-h}{2\ln{R}} + \frac{1}{4}\frac{\mathrm{d}u}{\mathrm{d}\ln{R}} \right] \cos{4\theta} + \mathcal{O}(\epsilon^3) \Bigg\} \\
    &= 1 - \epsilon u \cos{2\theta} + \epsilon^2 \left(\frac{u}{2} + \frac{u^2-2uh+2h^2}{4} + \frac{1}{4}\frac{\mathrm{d}u}{\mathrm{d}\ln{R}}\right) + \epsilon^2 h(u-h) \cos^2{2\theta} + \\
    &~~~~ \epsilon^2\left(-\frac{u}{2} + \frac{u^2-2uh+2h^2}{4} + \frac{1}{4}\frac{\mathrm{d}u}{\mathrm{d}\ln{R}}\right)\cos{4\theta} + \mathcal{O}(\epsilon^3)\\
    &= 1 - \epsilon u \cos{2\theta} + \epsilon^2 \left(\frac{u}{2} + \frac{u^2-2uh+2h^2}{4} + \frac{1}{4}\frac{\mathrm{d}u}{\mathrm{d}\ln{R}}\right) + \epsilon^2 h(u-h) \frac{1+\cos{4\theta}}{2} + \\
    &~~~~ \epsilon^2\left(-\frac{u}{2} + \frac{u^2-2uh+2h^2}{4} + \frac{1}{4}\frac{\mathrm{d}u}{\mathrm{d}\ln{R}}\right)\cos{4\theta} + \mathcal{O}(\epsilon^3)\\
     &= 1 - \epsilon u \cos{2\theta} + \epsilon^2 \left(\frac{u}{2} + \frac{u^2}{4} + \frac{1}{4}\frac{\mathrm{d}u}{\mathrm{d}\ln{R}}\right) + \epsilon^2\left(-\frac{u}{2} + \frac{u^2}{4} + \frac{1}{4}\frac{\mathrm{d}u}{\mathrm{d}\ln{R}}\right)\cos{4\theta} + \mathcal{O}(\epsilon^3)
    \end{aligned}
\end{equation}

Hence, we obtain Eq.~\ref{eq:sigma0_sigma2}. Note that by definition $\tilde{R}(R,\theta;\epsilon)=\tilde{R}(R,\theta+\pi/2;-\epsilon)$~\citep{Umetsu2026}; putting this into $\Sigma=\Sigma_{\rm sph}(\tilde{R})=\Sigma_0(R;\epsilon)+\Sigma_2(R;\epsilon)\cos{2\theta}+\cdots$, we have $\Sigma_0(R;\epsilon)=\Sigma_0(R;-\epsilon)$ and $\Sigma_2(R;\epsilon)=-\Sigma_2(R;-\epsilon)$. Thus $\Sigma_0$ has no $\epsilon^3$ term and is accurate up to $\mathcal{O}(\epsilon^4)$.
In this work, we skip the $\cos{4\theta}$ term (octupole): $\Sigma_4=\epsilon^2\left(-\frac{u}{2} + \frac{u^2}{4} + \frac{1}{4}\frac{\mathrm{d}u}{\mathrm{d}\ln{R}}\right)\Sigma_{\rm sph}(R)+\mathcal{O}(\epsilon^4)$. 
The reason is that it will be canceled out after the integral in our quadrupole models $\Sigma_{\rm const}$ and $\Sigma_{\rm 4\theta}$, even when the misalignment angle $\theta_0\neq0$. 
However, the octupole can provide further constraints on the lens shape when there is sufficiently deep data, for example, the LSST Y10 data, with specifically constructed estimators.

\subsection{Models}

\subsubsection{General forms}
\label{app:derive_multipoles_models_general}

\citet{Adhikari2015} derived the monopole and quadrupole of the tangential and cross shear (Eq.~\ref{eq:models_tx}, ~\ref{eq:models_integral}), so that $\gamma_{\rm T/X} =\gamma_{\rm T/X; 0} + \gamma_{\rm T/X;2} + \cdots$. 
This is achieved by solving the lensing potential ($\phi=\sum_{m\geq0} \phi_m (R)\cos{m\theta}$) from $\kappa=\frac{1}{2}\nabla\phi$ through the multipole expansion of convergence ($\kappa=\sum_m \kappa_{m\geq0} (R)\cos{m\theta}$). Then, the tangential and cross components of shear are derived from $\phi$ in the polar coordinate system, and only the $m=0$ and $m=2$ multipoles are kept. 
Here we have converted the parameters to the ones used in this paper; the same is applied below. Note that $L_{1,2}$ has the same dimension as $\Sigma_2$, and $R$ can be a physical or angular length in $L_{1,2}(R)$.

\begin{equation}
    \left\{
    \begin{aligned}
    \Sigma_{\rm crit}\gamma_{\rm T;0} (R) &= \bar{\Sigma}_0 (<R) -  \Sigma_0 (R) = \Delta\Sigma_0(R) \\
     \Sigma_{\rm crit}\gamma_{\rm X;0} (R)&= 0 \\
     \Sigma_{\rm crit}\gamma_{\rm T;2} (R) &= [-\Sigma_2 + L_1 + L_2 ] \cos{2\theta} \\
     \Sigma_{\rm crit}\gamma_{\rm X;2} (R) &=  [ L_1 - L_2 ] \sin{2\theta}
    \end{aligned}
    \right.
    \label{eq:models_tx}
\end{equation}

\begin{equation}
    \left\{
    \begin{aligned}
    L_1(R) &= \frac{3}{R^4}\int_0^R R'^3 \Sigma_2(R')\mathrm{d}R' \\
    L_2(R) &= \int_R^{\infty} \frac{\Sigma_2(R')}{R'}\mathrm{d}R'
    \end{aligned}
    \right.
    \label{eq:models_integral}
\end{equation}

Next,~\citet{Clampitt2016} derived the shear quadrupoles in the Cartesian system (Eq.~\ref{eq:models_12}; see also Eq.~\ref{eq:shear_conversion_matrix}), so that $\gamma_{\rm 1/2} =\gamma_{\rm 1/2;0} + \gamma_{\rm 1/2;2} + \cdots$.

\begin{equation}
    \left\{
    \begin{aligned}
        \Sigma_{\rm crit}\gamma_{1;0} &=  (\Sigma_0 - \bar{\Sigma}_0)\cos{2\theta}  \\
        \Sigma_{\rm crit}\gamma_{2;0} &=  (\Sigma_0 - \bar{\Sigma}_0)\sin{2\theta}  \\
        \Sigma_{\rm crit}\gamma_{1;2} &= \frac{1}{2}[(\Sigma_2 - 2L_1)\cos{4\theta}  + \Sigma_2 -  2L_2] \\
        \Sigma_{\rm crit}\gamma_{2;2} &= \frac{1}{2}(\Sigma_2 - 2L_1)\sin{4\theta}
    \end{aligned}
    \right.
    \label{eq:models_12}
\end{equation}

Then,~\citet{Shin2018} re-organized the formulas to separate $L_1$ (internal, $<R$ term) and $L_2$ (external, $>R$ term). Using the periodicity of trigonometric functions and the axisymmetry of the monopole, we have Eq.~\ref{eq:models_c4}. 

\begin{equation}
    \left\{
    \begin{aligned}
        \Delta\Sigma_{\rm const} &= \frac{1}{2\pi} \int_0^{2\pi} \Sigma_{\rm crit}\gamma_{1} \mathrm{d}\theta = \frac{1}{2\pi} \int_0^{2\pi} \Sigma_{\rm crit}\gamma_{1;2}\mathrm{d}\theta = \frac{\Sigma_2}{2} - L_2 \\
        \Delta\Sigma_{4\theta} &= \frac{1}{2\pi} \int_0^{2\pi} (\Sigma_{\rm crit}\gamma_{1}\cos{4\theta} + \Sigma_{\rm crit}\gamma_{2}\sin{4\theta})\mathrm{d}\theta = \frac{1}{2\pi} \int_0^{2\pi} (\Sigma_{\rm crit}\gamma_{1;2}\cos{4\theta} + \Sigma_{\rm crit}\gamma_{2;2}\sin{4\theta})\mathrm{d}\theta \\
        &= \frac{\Sigma_2}{2} - L_1 \\
        &= \frac{1}{2}\frac{1}{2\pi} \int_0^{2\pi} \left(\frac{\Sigma_{\rm crit}\gamma_{1}}{\cos{4\theta}} + \frac{\Sigma_{\rm crit}\gamma_{2}}{\sin{4\theta}}\right)\mathrm{d}\theta
    \end{aligned}
    \right.
    \label{eq:models_c4}
\end{equation}

In the above equations, $\Sigma_2$ can be replaced by $\Sigma_2 = -\epsilon u \Sigma_0 + \mathcal{O}(\epsilon^3)$ or $\Sigma_2 = -\epsilon u \Sigma_{\rm sph} + \mathcal{O}(\epsilon^3)$, according to Eq.~\ref{eq:sigma0_sigma2}.

\subsubsection{Null tests}\label{app:null_test}

From Eq.~\ref{eq:models_tx} it is clear to see that the azimuthal average of $\Sigma_{\rm crit}\gamma_{\rm X}$ is zero (denoted as $\Delta\Sigma_{\rm x}$), which provides a common  null test for systematics~\citep{McClintock2019}. Similarly, following Eq.~\ref{eq:models_12},~\ref{eq:models_c4}, we consider two null tests for the quadruple measurements~\citep{Shin2018}: the azimuthal average of $\Sigma_{\rm crit}\gamma_2$ is zero (denoted as $\Delta\Sigma_{\rm const;x}$); the azimuthal average of $\Sigma_{\rm crit}\gamma_1\sin{4\theta}-\Sigma_{\rm crit}\gamma_2\cos{4\theta}$ is also zero (denoted as $\Delta\Sigma_{\rm 4\theta;x}$; or exchanging $\sin{4\theta}$ and $\cos{4\theta}$). In Section~\ref{sec:dis_null_test} we construct corresponding null estimators and the measurements are consistent with zero.

\subsubsection{Misalignment angle and dilution factor}
\label{app:derive_dilution_factor}

If the major axis is at a position angle of $\theta_0$ instead of 0, following~\citet{Clampitt2016} we have Eq.~\ref{eq:models_tx_t0} based on Eq.~\ref{eq:models_tx}.

\begin{equation}
    \left\{
    \begin{aligned}
     \Sigma_{\rm crit}\gamma_{\rm T;2} (R) &= [-\Sigma_2 + L_1 + L_2 ] \cos{(2\theta-2\theta_0)} \\
     &= [-\Sigma_2 + L_1 + L_2 ](\cos{2\theta}\cos{2\theta_0}+\sin{2\theta}\sin{2\theta_0})\\
     \Sigma_{\rm crit}\gamma_{\rm X;2} (R) &=  [ L_1 - L_2 ] \sin{(2\theta-2\theta_0)}\\
     &= [ L_1 - L_2 ] (\sin{2\theta}\cos{2\theta_0}-\cos{2\theta}\sin{2\theta_0})
    \end{aligned}
    \right.
    \label{eq:models_tx_t0}
\end{equation}

Then in the Cartesian system, one can derive Eq.~\ref{eq:models_12_t0}.

\begin{equation}
    \left\{
    \begin{aligned}
        \Sigma_{\rm crit}\gamma_{1;2} &= \frac{\cos{2\theta_0}}{2}[(\Sigma_2 - 2L_1)\cos{4\theta} + \Sigma_2 - 2L_2 ] + \frac{\sin{2\theta_0}}{2}(\Sigma_2 - 2L_1)\sin{4\theta}\\
        \Sigma_{\rm crit}\gamma_{2;2} &= \frac{\cos{2\theta_0}}{2}(\Sigma_2 - 2L_1)\sin{4\theta} + \frac{\sin{2\theta_0}}{2}[\Sigma_2-2L_2+(-\Sigma_2 + 2L_1)\cos{4\theta}]
    \end{aligned}
    \right.
    \label{eq:models_12_t0}
\end{equation}

Similarly to Eq.~\ref{eq:models_c4}, when we compute the quadrupole models using the integrals, we have Eq.~\ref{eq:models_c4_t0}, where both quadrupole models have an extra factor of $\cos{2\theta_0}$ compared to Eq.~\ref{eq:models_c4}. 

\begin{equation}
    \left\{
    \begin{aligned}
        \Delta\Sigma_{\rm const} &= \frac{1}{2\pi} \int_0^{2\pi} \Sigma_{\rm crit}\gamma_{1} \mathrm{d}\theta = \frac{1}{2\pi} \int_0^{2\pi} \Sigma_{\rm crit}\gamma_{1;2}\mathrm{d}\theta = \left(\frac{\Sigma_2}{2} - L_2\right)\cos{2\theta_0} \\
        \Delta\Sigma_{4\theta} &= \frac{1}{2\pi} \int_0^{2\pi} (\Sigma_{\rm crit}\gamma_{1}\cos{4\theta} + \Sigma_{\rm crit}\gamma_{2}\sin{4\theta})\mathrm{d}\theta = \frac{1}{2\pi} \int_0^{2\pi} (\Sigma_{\rm crit}\gamma_{1;2}\cos{4\theta} + \Sigma_{\rm crit}\gamma_{2;2}\sin{4\theta})\mathrm{d}\theta \\
        &= \left(\frac{\Sigma_2}{2} - L_1\right)\cos{2\theta_0} \\
        &= \frac{1}{2}\frac{1}{2\pi} \int_0^{2\pi} \left(\frac{\Sigma_{\rm crit}\gamma_{1}}{\cos{4\theta}} + \frac{\Sigma_{\rm crit}\gamma_{2}}{\sin{4\theta}}\right)\mathrm{d}\theta
    \end{aligned}
    \right.
    \label{eq:models_c4_t0}
\end{equation}

\subsubsection{Reduced shear}
\label{app:derive_reduced_shear}

In real observations, reduced shear is directly measured instead of shear. Replacing $\gamma$ with $g$, we find the models of the monopole and the quadrupoles both have an extra factor of $\sim (1+\Sigma_0(R)/\Sigma_{\rm crit})$ in Eq.~\ref{eq:mean_shear_excess_surface_density},~\ref{eq:models_tx},~\ref{eq:models_main},~\ref{eq:models_c4}. The $\Sigma_{\rm crit}$ inside that factor can not be directly determined, as it requires single-point representative redshifts for the lens and the source. It needs to be estimated by the photo-zs of the lens and the source, and it can be strongly influenced by the uncertainties of photo-zs. We thus exclude small radii, where the convergence is non-negligible, to avoid this factor. Also, in Section~\ref{sec:dis_multi_sys} we test and find that it has a minor impact on the halo shape measurement, because it affects the monopole and the quadrupoles simultaneously.

\subsubsection{Miscentering}\label{app:derive_miscentering}
 
\begin{figure}[htb!]
    \centering
    \includegraphics[width=0.5\linewidth]{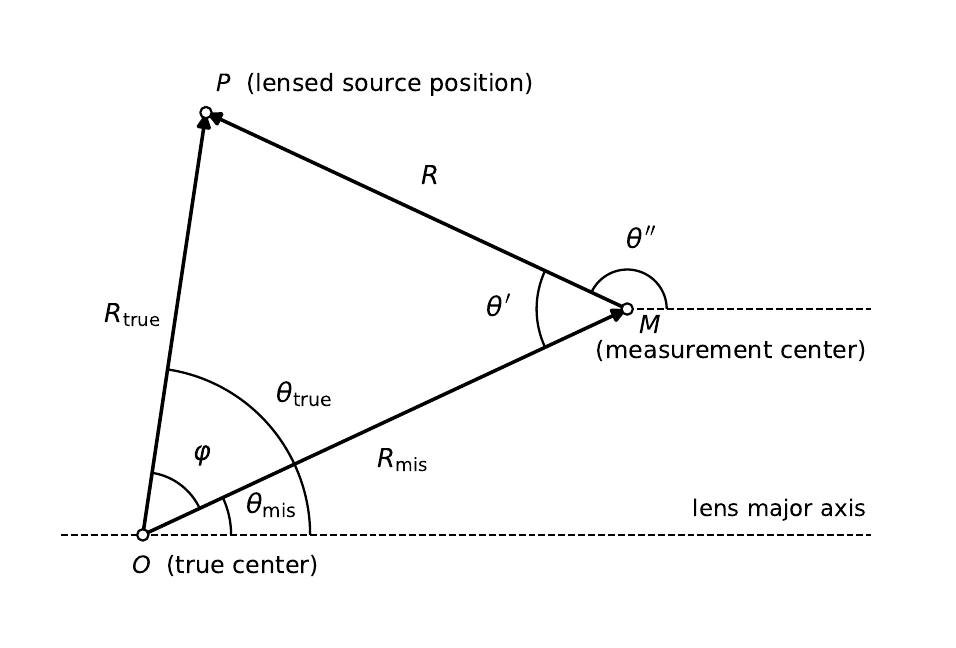}
    \caption{Illustration of miscentering for an elliptical lens.}
    \label{fig:miscenter_demo}
\end{figure}

If the center for measurement has an offset from the true center (``miscentering''; Figure~\ref{fig:miscenter_demo}), 
the Cartesian shears can be written schematically as $\Sigma_{\rm crit}\gamma_1
=
A(R_{\rm true})\cos{2\theta_{\rm true}} 
+ B(R_{\rm true})\cos{4\theta_{\rm true}} 
+ C(R_{\rm true})$ and $\Sigma_{\rm crit}\gamma_2
=
A(R_{\rm true})\sin{2\theta_{\rm true}}
 + B(R_{\rm true})\sin{4\theta_{\rm true}}$ (Eq.~\ref{eq:models_12}), where $R_{\rm true}^2
=
R_{\rm mis}^2+R^2-2R_{\rm mis}R\cos\theta'$ and $\theta_{\rm true}=\theta_{\rm mis}+\varphi$. In fact, $A=-\Delta\Sigma_0$, $B=\Delta\Sigma_{4\theta}$, $C=\Delta\Sigma_{\rm const}$, but for simplicity we use $A$, $B$, $C$ first.

The measured ``const'' quadrupole becomes the average of $\Sigma_{\rm crit} \gamma_1$ over $\theta'$, i.e., $\Delta\Sigma_{\rm const}^{\rm mis}=\langle \Sigma_{\rm crit} \gamma_1 \rangle|_{\theta'}$ (Eq.~\ref{eq:models_c4}). 
For the $A$ term, since $\cos{2\theta_{\rm true}} = \cos {2\theta_{\rm mis}} \cos{2\varphi} - \sin{2\theta_{\rm mis}} \sin{2\varphi}$, $R_{\rm true}(-\theta')=R_{\rm true}(\theta')$, and $\varphi(-\theta')=-\varphi(\theta')$, we have $\langle A(R_{\rm true})\cos{2\theta_{\rm true}}  \rangle=\cos {2\theta_{\rm mis}} \langle A(R_{\rm true})\cos{2\varphi}\rangle$. Similarly, $\langle B(R_{\rm true})\cos {4\theta_{\rm true}}  \rangle=\cos {4\theta_{\rm mis}} \langle B(R_{\rm true})\cos{4\varphi}\rangle$. 
Therefore, it can be written as Eq.~\ref{eq:Delta_Sigma_const_mis}, where $F_2$ scales with $\Sigma_0$, while $F_0$ and $F_4$ scale with $\Sigma_2\propto\epsilon$. 
Likewise, using the symmetry one can prove that $\langle \Sigma_{\rm crit} \gamma_2 \rangle=F_2\sin{2\theta_{\rm mis}}+F_4\sin{4\theta_{\rm mis}}$.

\begin{equation}
    \left\{
    \begin{aligned}
        \Delta\Sigma_{\rm const}^{\rm mis}&=F_0+F_2\cos{2\theta_{\rm mis}}+F_4\cos{4\theta_{\rm mis}}\\
        F_0(R,R_{\rm mis})&=\langle C(R_{\rm true})\rangle = O(\epsilon) \\
        F_2(R,R_{\rm mis})&=\langle A(R_{\rm true})\cos{2\varphi} \rangle = O(\epsilon^0) \\
        F_4(R,R_{\rm mis})&=\langle B(R_{\rm true})\cos{4\varphi} \rangle = O(\epsilon)
    \end{aligned}
    \right.
    \label{eq:Delta_Sigma_const_mis}
\end{equation}

For the measured ``$4\theta$'' quadrupole, it becomes $\Delta\Sigma_{\rm 4\theta}^{\rm mis}=\langle \Sigma_{\rm crit} \gamma_1 \cos{4\theta''} + \Sigma_{\rm crit} \gamma_2 \sin{4\theta''}\rangle$, as the position angle of $P$ with respect to $M$ is $\theta''=\pi-\theta'+\theta_{\rm mis}$  (Eq.~\ref{eq:models_c4}). Thus, $\Delta\Sigma_{\rm 4\theta}^{\rm mis}=\langle \Sigma_{\rm crit} \gamma_1 \cos{(4\theta_{\rm mis}-4\theta')} + \Sigma_{\rm crit} \gamma_2 \sin{(4\theta_{\rm mis}-4\theta')}\rangle$.
It can be divided into $A$, $B$, $C$ terms as well: $\langle A \cos{[2(\theta_{\rm mis}+\varphi) - (4\theta_{\rm mis}-4\theta')]}\rangle=\langle A \cos{(-2\theta_{\rm mis} + 2\varphi + 4\theta')}\rangle$; 
$\langle B \cos{[4(\theta_{\rm mis}+\varphi) - (4\theta_{\rm mis}-4\theta')]}\rangle=\langle B \cos{(4\varphi + 4\theta')}\rangle$; $\langle C \cos{(4\theta_{\rm mis}-4\theta')}\rangle$. 
Using the same symmetry, we have Eq.~\ref{eq:Delta_Sigma_4theta_mis}, where $G_2$ scales with $\Sigma_0$, while $G_0$ and $G_4$ scale with $\Sigma_2\propto\epsilon$.

\begin{equation}
    \left\{
    \begin{aligned}
        \Delta\Sigma_{\rm 4\theta}^{\rm mis}&=G_0+G_2\cos{2\theta_{\rm mis}}+G_4\cos{4\theta_{\rm mis}}\\
        G_0(R,R_{\rm mis})&=\langle B(R_{\rm true})\cos{(4\varphi+4\theta')} \rangle = O(\epsilon) \\
        G_2(R,R_{\rm mis})&=\langle A(R_{\rm true})\cos{(2\varphi+4\theta')} \rangle = O(\epsilon^0) \\
        G_4(R,R_{\rm mis})&=\langle C(R_{\rm true})\cos{4\theta'} \rangle = O(\epsilon)
    \end{aligned}
    \right.
    \label{eq:Delta_Sigma_4theta_mis}
\end{equation}

For the measured monopole, it can be derived from either the new tangential shear with respect to $M$ based on $\gamma_{1,2}$ (Eq.~\ref{eq:models_12}), or from the relation between the new tangential shear and the new excess surface density (Eq.~\ref{eq:mean_shear_excess_surface_density}); the latter is more straightforward. 
We define $\Sigma_M(R, R_{\rm mis})$ as the average of the true surface density $\Sigma(R_{\rm true},\theta_{\rm true})$ with respect to $M$ when $R$ is fixed and $\theta'$ varies. 
Since $\Sigma(R_{\rm true},\theta_{\rm true})=\Sigma_0(R_{\rm true})+\Sigma_2(R_{\rm true})\cos{2\theta_{\rm true}}$, we have $\Sigma_M(R,R_{\rm mis}) = \langle\Sigma_0(R_{\rm true})\rangle + \langle\Sigma_2(R_{\rm true})\cos{2(\theta_{\rm mis}+\varphi)}\rangle$, and the second term is $\cos{2\theta_{\rm mis}}
\left\langle
\Sigma_2(R_{\rm true})\cos{2\varphi}
\right\rangle$ because of the same symmetry. 
Thus, $\Sigma_M$ can be written as $\Sigma_M(R,R_{\rm mis})=S_0(R,R_{\rm mis})+S_2(R,R_{\rm mis})\cos{2\theta_{\rm mis}}$.
Then, by definition $\bar{\Sigma}_M(<R, R_{\rm mis})=
\bar{S}_0(<R,R_{\rm mis})
+
\bar{S}_2(<R,R_{\rm mis})
\cos{2\theta_{\rm mis}}$. Therefore we have Eq.~\ref{eq:Delta_Sigma_0_mis}.

\begin{equation}
    \left\{
    \begin{aligned}
        \Delta\Sigma_{0}^{\rm mis}&=H_0+H_2\cos{2\theta_{\rm mis}}\\
        H_0(R,R_{\rm mis})&=\bar{S}_0(<R,R_{\rm mis})-S_0(R,R_{\rm mis}) = O(\epsilon^0) \\
        H_2(R,R_{\rm mis})&=\bar{S}_2(<R,R_{\rm mis})-S_2(R,R_{\rm mis}) = O(\epsilon)
    \end{aligned}
    \right.
    \label{eq:Delta_Sigma_0_mis}
\end{equation}

At a fixed $R_{\rm mis}$, the measured ellipticity $\epsilon_{\rm fit}^{\rm mis}$ is obtained by jointly fitting $\Delta\Sigma_0^{\rm mis}$, $\Delta\Sigma_{\rm const}^{\rm mis}$, and $\Delta\Sigma_{\rm 4\theta}^{\rm mis}$. Since the quadrupole models depend linearly on $\epsilon$ at leading order, the fitted ellipticity is expected to inherit the same angular dependence approximately (Eq.~\ref{eq:epsilon_fit}) and to have strong dependence on $\cos{2\theta_{\rm mis}}$ (Eq.~\ref{eq:Delta_Sigma_const_mis},~\ref{eq:Delta_Sigma_4theta_mis}). 
Indeed, it is demonstrated with the mock in Section~\ref{sec:dis_miscenter}.

\begin{equation}
\epsilon_{\rm fit}(\theta_{\rm mis}) \simeq \epsilon_0+\epsilon_2\cos{2\theta_{\rm mis}}+\epsilon_4\cos{4\theta_{\rm mis}}
\label{eq:epsilon_fit}
\end{equation}

Hence, when the measurement centers are uniformly distributed on a circle after the axis-aligned stacking, the mean fitted ellipticity is close to the result at $\theta_{\rm mis}=45^\circ$, where increasing $R_{\rm mis}$ mainly dilutes the $\Delta\Sigma_{\rm 4\theta}$ signal (sensitive to internal matter) and reduces the fitted ellipticity for the mock. 
When the measurement centers are distributed along an ellipse after stacking, whose the major axis is aligned with the x-axis (same as the lens), the mean fitted ellipticity would be closer to the original one (more likely to increase the fitted ellipticity than to decrease), and the result would be the opposite when the major axis is aligned with the y-axis (perpendicular to the lens).


\bibliography{main}{}
\bibliographystyle{aasjournalv7}

\end{document}